\documentclass[english]{extarticle}
\usepackage[T1]{fontenc}
\usepackage[latin9]{inputenc}
\usepackage{mathrsfs}
\usepackage{amsmath}
\usepackage{amssymb}
\usepackage{graphicx}
\usepackage{esint}

\makeatletter

\providecommand{\tabularnewline}{\\}

\numberwithin{equation}{section}

\usepackage{geometry}
\usepackage{cite}
\usepackage{mathrsfs}
\usepackage{mnsymbol}

\allowdisplaybreaks[1]

\newif\ifContLineOne
\newif\ifContLineTwo
\newif\ifContLineThree

\def\conC#1{\vbox{\ialign{##\crcr
  \ifContLineThree\hrulefill\else\vphantom{\hrulefill}\fi\crcr
  \noalign{\kern3.2pt\nointerlineskip}
  \ifContLineTwo\hrulefill\else\vphantom{\hrulefill}\fi\crcr
  \noalign{\kern3.2pt\nointerlineskip}
  \ifContLineOne\hrulefill\else\vphantom{\hrulefill}\fi\crcr
  \noalign{\nointerlineskip}
  $\hfil\textstyle{\vbox to 14pt{}#1}\hfil$\crcr}}}

\def\DrawLeg#1#2{
  \kern-.2pt              
  \dimen2 =#1             
  \advance\dimen2 by 2pt  
  \dimen3 = 10.6pt        
  \dimen4 =3.6pt          
  \advance\dimen3 by -\dimen2 
  \multiply\dimen4 by #2
  \advance\dimen3 by \dimen4
  \raise\dimen2 \hbox{\vrule height\dimen3 width .4pt} 
  \kern-.2pt}             

\def\begC#1#2{\setbox0 =\hbox{$\textstyle{#2}$}
  \dimen0=.5\wd0 \dimen1=\ht0
  \conC{\hskip\dimen0}
  \count255=#1
  \ifnum\count255 =1 \ContLineOnetrue\else
  \ifnum\count255 =2 \ContLineTwotrue\else
  \ifnum\count255 =3 \ContLineThreetrue\fi\fi\fi
  \DrawLeg{\dimen1}{\count255}
  \conC{\hskip\dimen0}
  \kern-\dimen0\kern-\dimen0 \box0}

\def\endC#1#2{\setbox0 =\hbox{$\textstyle{#2}$}
  \dimen0=.5\wd0 \dimen1=\ht0
  \conC{\hskip\dimen0}
  \count255=#1
  \ifnum\count255 =1 \ContLineOnefalse\else
  \ifnum\count255 =2 \ContLineTwofalse\else
  \ifnum\count255 =3 \ContLineThreefalse\fi\fi\fi
  \DrawLeg{\dimen1}{\count255}
  \conC{\hskip\dimen0}
  \kern-\dimen0\kern-\dimen0 \box0}

\makeatother

\usepackage{babel}
\begin{document}
\begin{titlepage}

\global\long\def\thefootnote{\fnsymbol{footnote}}%

\begin{flushright}
\begin{tabular}{l}
UTHEP- 773\tabularnewline
\end{tabular}
\par\end{flushright}

\bigskip{}

\begin{center}
\textbf{\Large{}The Fokker-Planck formalism for closed bosonic strings}\bigskip{}
\par\end{center}

\begin{center}
{\large{}{}Nobuyuki Ishibashi}\footnote{e-mail: ishibashi.nobuyuk.ft@u.tsukuba.ac.jp}
{\large{}{}}{\large\par}
\par\end{center}

\begin{center}
\textit{Tomonaga Center for the History of the Universe,}\\
\textit{ University of Tsukuba}\\
\textit{ Tsukuba, Ibaraki 305-8571, JAPAN}\\
 
\par\end{center}

\bigskip{}

\bigskip{}

\bigskip{}

\begin{abstract}
Every Riemann surface with genus $g$ and $n$ punctures admits a
hyperbolic metric, if $2g-2+n>0$. Such a surface can be decomposed
into pairs of pants whose boundaries are geodesics. We construct a
string field theory for closed bosonic strings based on this pants
decomposition. In order to do so, we derive a recursion relation satisfied
by the off-shell amplitudes, using the Mirzakhani's scheme for computing
integrals over the moduli space of bordered Riemann surfaces. The
recursion relation can be turned into a string field theory via the
Fokker-Planck formalism. The Fokker-Planck Hamiltonian consists of
kinetic terms and three string vertices. Unfortunately, the worldsheet
BRST symmetry is not manifest in the theory thus constructed. We will
show that the invariance can be made manifest by introducing auxiliary
fields. 
\end{abstract}
\global\long\def\thefootnote{\arabic{footnote}}%

\end{titlepage}

\pagebreak{}

\section{Introduction}

For constructing a string field theory (SFT), we should specify a
rule to cut worldsheets into fundamental building blocks, i.e. propagators
and vertices. A few simple rules were proposed and SFTs for bosonic
strings were constructed following these rules \cite{Kaku1974,Witten1986,Kugo1992,Zwiebach1993}.
Construction of an SFT for superstrings is more complicated because
of the spurious singularities \cite{Lacroix2017}. 

The worldsheets of closed strings describing scattering amplitudes
are punctured Riemann surfaces. In mathematics, there exists a convenient
way to decompose them into fundamental building blocks. On a Riemann
surface with genus $g$ and $n$ boundaries or punctures, one can
introduce a metric with constant negative curvature, if $2g-2+n>0$.
Such a metric is called a hyperbolic metric and surfaces with hyperbolic
metrics are called hyperbolic surfaces. With a hyperbolic metric,
one can decompose the surface into pairs of pants with geodesic boundaries.
It may be possible to consider the pair of pants as the fundamental
building block of the surface. 

\begin{figure}
\begin{centering}
\includegraphics[scale=0.6]{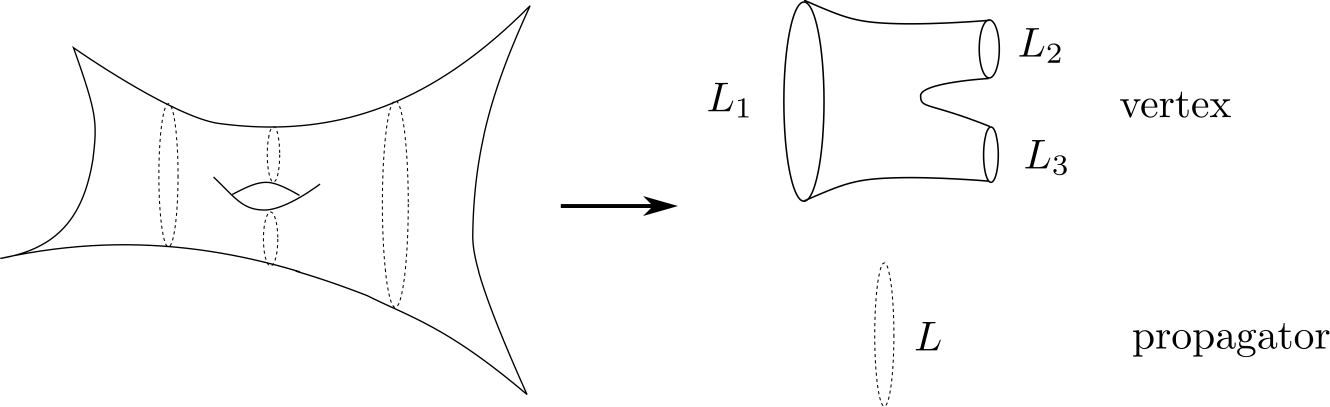}
\par\end{centering}
\caption{A pants decomposition.\label{fig:A-pants-decomposition.}}
\end{figure}

The hyperbolic metric was used to construct SFT in \cite{Moosavian2019,Moosavian2019a,Costello2022},
in which the kinetic term of the action was taken to be the conventional
one 
\begin{equation}
\int\Psi c_{0}^{-}Q\Psi\,,\label{eq:kinetic}
\end{equation}
so that the propagators correspond to cylinders. The theories include
infinitely many vertices besides the three string vertex and the Feynman
graphs have nothing to do with the pants decomposition. In this paper,
we would like to construct an SFT based on the pants decomposition.
Namely, we will construct an SFT for closed bosonic strings regarding
the pair of pants as the three string vertex and the cylinders with
vanishing heights as the propagator, as depicted in Figure \ref{fig:A-pants-decomposition.}. 

In such a theory, a state of string will correspond to a boundary
of a pair of pants. Accordingly, the string field should be labeled
by an element of the Hilbert space of the first quantized strings
and the length $L$ of the boundary. The external states of the scattering
amplitudes are regarded as the limit $L\to0$ of such states. The
off-shell amplitudes may correspond to Riemann surfaces which have
geodesic boundaries with fixed lengths and will be expressed by integrals
over the moduli spaces of such surfaces. 

\begin{figure}
\begin{centering}
\includegraphics[scale=0.6]{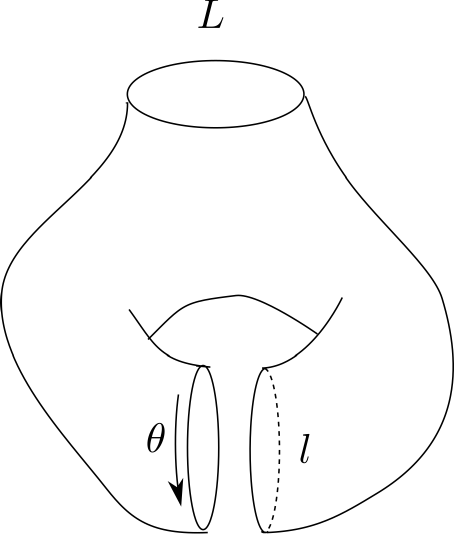}
\par\end{centering}
\caption{One loop one point function.\label{fig:One-loop-one}}
\end{figure}

Unfortunately, such an approach suffers from a problem addressed in
\cite{DHoker1988} (section IV. E). The three string vertex will be
given by the correlation function of the worldsheet theory on hyperbolic
pants with the boundary lengths specified. Suppose that one calculates
the one loop one point function following the conventional Feynman
rules. The amplitude corresponds to the worldsheet in Figure \ref{fig:One-loop-one}
and we should integrate over the length $l$ and the twist angle $\theta$.
By doing so, the fundamental domain of the modular group is covered
infinitely many times, as will be seen in section \ref{subsec:String-field-action}.
The same happens for all the other amplitudes. Therefore, the conventional
Feynman rule with the vertex and the propagator in Figure \ref{fig:A-pants-decomposition.}
does not yield the correct amplitudes. 

In order to overcome this problem, we formulate the theory using the
Mirzakhani's scheme \cite{Mirzakhani2006,Mirzakhani2007} for computing
integrals over moduli space of bordered Riemann surfaces. Mirzakhani
derived a recursion relation for the volume of the moduli space. Applying
her method to the off-shell amplitudes of closed bosonic strings,
we derive a recursion relation satisfied by these amplitudes. 

As was pointed out in \cite{Eynard2007,Saad2019a}, Mirzakhani's recursion
relation is related to the loop equation of minimal string theory.
On the other hand, the loop equations for minimal strings can be described
by an SFT via the Fokker-Planck formalism \cite{Ishibashi:1993nq,Jevicki1994}.
We will show that the recursion relation of the off-shell amplitudes
can be described by an SFT using the Fokker-Planck formalism. The
Fokker-Planck Hamiltonian consists of kinetic terms and three string
vertices. One can develop perturbation theory which does not suffer
from the above mentioned problem. Unfortunately, the worldsheet BRST
symmetry is not manifest in the SFT thus constructed. We will show
that we can make the invariance manifest by introducing auxiliary
fields. 

The organization of this paper is as follows. In section \ref{sec:Bordered-Riemann-surface},
we define the off-shell amplitudes of closed bosonic string theory
based on the moduli space of bordered Riemann surfaces. In section
\ref{sec:A-recursion-relation}, we derive recursion relations satisfied
by the off-shell amplitudes. In section \ref{sec:Fokker-Planck-Hamiltonian},
we prove that the off-shell amplitudes defined in section \ref{sec:Bordered-Riemann-surface}
can be derived from the Fokker-Planck formalism for string fields.
We show that the solution of the recursion relations in section \ref{sec:A-recursion-relation}
satisfies the Schwinger-Dyson equation derived from the Fokker-Planck
Hamiltonian. In section \ref{sec:BRST-invariant-formulation}, we
modify the theory by introducing auxiliary fields and make it manifestly
BRST invariant. Section \ref{sec:Discussions} is devoted to discussions
and comments. In Appendix \ref{sec:Hyperbolic-metric-on}, we present
formulas for the local coordinates on hyperbolic pants. In appendix
\ref{sec:BRST-identity}, we prove the BRST identity. 

\section{Off-shell amplitudes\label{sec:Bordered-Riemann-surface}}

The off-shell amplitudes of the theory we will study should correspond
to hyperbolic surfaces which have geodesic boundaries with fixed lengths.
In this section, we would like to define such amplitudes. The formulation
is a modification of the conventional ones \cite{Zwiebach1993,Sen2015,Erler2020,Erbin2021}. 

\subsection{The moduli space $\mathcal{M}_{g,n,\mathbf{L}}$}

Let $\Sigma_{g,n,\mathbf{L}}$ with $\mathbf{L}=(L_{1},\cdots,L_{n})$
be a genus $g$ hyperbolic surface with $n$ geodesic boundaries (labeled
by an index $a=1,\cdots,n$) whose lengths are $L_{1},\cdots,L_{n}$.
Cutting the surface $\Sigma_{g,n,\mathbf{L}}$ along non-peripheral
simple closed geodesics, we can decompose it into pairs of pants $S_{i}\ (i=1,\cdots2g-2+n)$.
There are many choices for such decomposition and here we pick one.
The hyperbolic structure of the surface is specified by the lengths
of the non-peripheral simple closed geodesics and the way how boundaries
of $S_{i}$'s are identified. Therefore the hyperbolic structure of
$\Sigma_{g,n,\mathbf{L}}$ can be parametrized by the Fenchel-Nielsen
coordinates $(l_{s};\tau_{s})\ (s=1,\cdots,3g-3+n)$, where $l_{s}$
are the lengths of the nonperipheral boundaries of $S_{i}$ and $\tau_{s}$
denotes the twist parameters which specify how boundaries of different
pairs of pants are identified. The Teichm\"{u}ller space $\mathcal{T}_{g,n,\mathbf{L}}$
corresponds to the region $0<l_{s}<\infty,-\infty<\tau_{s}<\infty$.
A volume form $\Omega_{g,n,\mathbf{L}}$ on $\mathcal{T}_{g,n,\mathbf{L}}$
called the Weil-Petersson volume form is given by
\[
\Omega_{g,n,\mathbf{L}}=\bigwedge_{s=1}^{3g-3+n}\left[dl_{s}\wedge d\tau_{s}\right]\,.
\]
$\Omega_{g,n,\mathbf{L}}$ does not depend on the choice of the pants
decomposition. The moduli space $\mathcal{M}_{g,n,\mathbf{L}}$ is
defined as
\[
\mathcal{M}_{g,n,\mathbf{L}}\equiv\mathcal{T}_{g,n,\mathbf{L}}/\Gamma\,,
\]
where $\Gamma$ denotes the mapping class group. The Fenchel-Nielsen
coordinates $(l_{s};\tau_{s})$ can be used as local coordinates on
$\mathcal{M}_{g,n,\mathbf{L}}$. We will define the off-shell amplitudes
as integrals over $\mathcal{M}_{g,n,\mathbf{L}}$. The space of all
inequivalent hyperbolic structures on a surface is the same as that
of the complex structures. Hence the definition of the off-shell amplitudes
here can be regarded as the traditional one for the case where the
lengths of the external strings are specified. 

\subsection{$b$-ghost insertions}

\begin{figure}
\begin{centering}
\includegraphics[scale=0.5]{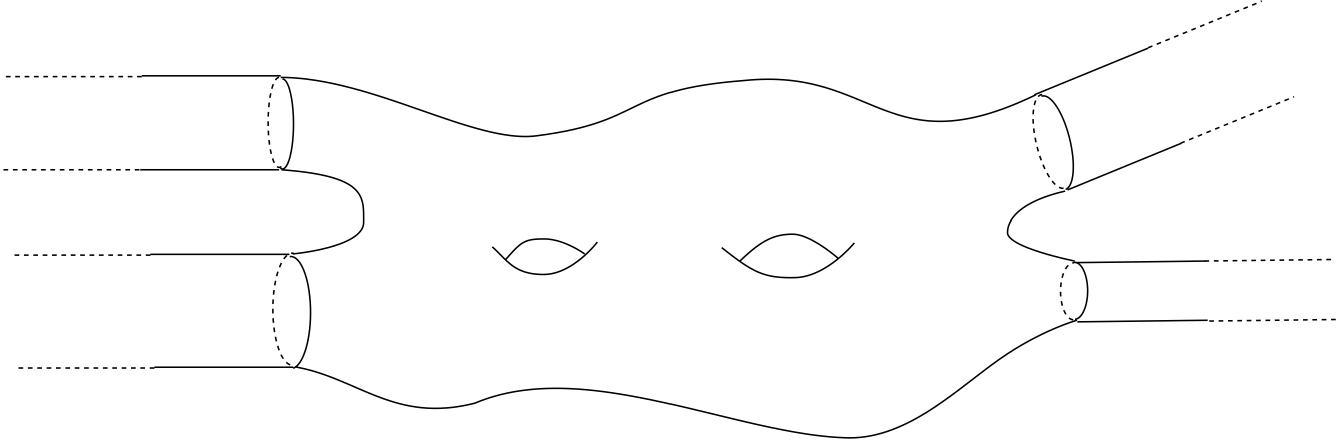}
\par\end{centering}
\caption{Attaching flat semi-infinite cylinders to $\Sigma_{g,n,\mathbf{L}}$.\label{fig:The-map-.}}
\end{figure}

Let us consider an element $\Sigma_{g,n,\mathbf{L}}$ of $\mathcal{M}_{g,n,\mathbf{L}}$.
One can attach a flat semi-infinite cylinder to each boundary \cite{Costello2022}
as depicted in Figure \ref{fig:The-map-.} and obtain a punctured
Riemann surface. The cylinder is conformally equivalent to a disk
with a puncture. Let $w_{a}\ (a=1,\cdots n)$ be a local coordinate
on the $a$-th disk $D_{a}$ such that $D_{a}$ corresponds to the
region $\left|w_{a}\right|\leq1$, the flat metric is given as
\[
ds^{2}=\frac{L_{a}^{2}}{(2\pi)^{2}}\frac{|dw_{a}|^{2}}{|w_{a}|^{2}}\,,
\]
and the $a$-th puncture corresponds to $w_{a}=0$. By these conditions,
$w_{a}$ is fixed up to a phase rotation. $w_{a}$ can be expressed
as a function $w_{a}(z)$ of a local coordinate $z$ on $\Sigma_{g,n,\mathbf{L}}$.
$w_{a}(z)$ is holomorphic in a neighborhood of $\partial D_{a}$. 

In this way, from $\Sigma_{g,n,\mathbf{L}}$, we obtain a punctured
Riemann surface $\Sigma_{g,n}$ with local coordinates around punctures,
which are specified up to phase rotations. To $\Sigma_{g,n}$ thus
obtained, one can associate a surface state, picking a local coordinate
$w_{a}$ as above for each $D_{a}$. Let us denote this surface state
by $\langle\Sigma_{g,n,\mathbf{L}}|$. By definition, we have
\begin{equation}
\langle\Sigma_{g,n,\mathbf{L}}|\Psi_{1}\rangle\cdots|\Psi_{n}\rangle=\langle\prod_{a=1}^{n}w_{a}^{-1}\circ\mathcal{O}_{\Psi_{a}}(0)\rangle_{\Sigma_{g,n}}\,,\label{eq:surface}
\end{equation}
where $\mathcal{O}_{\Psi_{a}}$ denotes the operator corresponding
to the state $|\Psi_{a}\rangle$ and $\left\langle \cdot\right\rangle _{\Sigma_{g,n}}$
denotes the correlation function on $\Sigma_{g,n}$. Under a phase
rotation $w_{a}\to e^{i\alpha_{a}}w_{a}$, $\langle\Sigma_{g,n,\mathbf{L}}|$
transforms as
\[
\langle\Sigma_{g,n,\mathbf{L}}|\to\langle\Sigma_{g,n,\mathbf{L}}|\prod_{a}e^{i\alpha_{a}(L_{0}^{(a)}-\bar{L}_{0}^{(a)})}\,.
\]
The correlation function $\langle\Sigma_{g,n,\mathbf{L}}|\Psi_{1}\rangle\cdots|\Psi_{n}\rangle$
is invariant under the phase rotation, if 
\[
(L_{0}-\bar{L}_{0})|\Psi_{a}\rangle=0\,.
\]

In order to define the amplitudes, we need to construct a top form
on the moduli space $\mathcal{M}_{g,n,\mathbf{L}}$ from the $b$-ghost.
A deformation of the hyperbolic structure of a surface induces that
of the complex structure. Therefore we can construct the $b$-ghost
insertion corresponding to a tangent vector of $\mathcal{M}_{g,n,\mathbf{L}}$,
following the procedure given in \cite{Zwiebach1993,Sen2015,Polchinski2007,Erbin2021}.
Let $z_{i}$ be a local coordinate on the pair of pants $S_{i}$,
such that the hyperbolic metric on $S_{i}$ is in the form 
\[
ds^{2}=e^{\varphi}|dz_{i}|^{2}\,.
\]
 Each boundary of $S_{i}$ is either shared by another pair of pants
$S_{j}\ (j\neq i)$ or is equal to one of the boundaries of $\Sigma_{g,n,\mathbf{L}}$.
In the former case, the local coordinate $z_{j}$ on $S_{j}$ and
$z_{i}$ are related by 
\begin{equation}
z_{i}=F_{ij}(z_{j})\,,\label{eq:Fij}
\end{equation}
in a neighborhood of $S_{i}\cap S_{j}=\mathrm{C}_{ij}$. If the boundary
of $S_{i}$ coincides with $\partial D_{a}$, $z_{i}$ and $w_{a}$
are related by
\begin{equation}
z_{i}=f_{ia}(w_{a})\,,\label{eq:fia}
\end{equation}
 in a neighborhood of $\partial D_{a}$. The transition functions
$F_{ij},f_{ia}$ describe the moduli of $\Sigma_{g,n,\mathbf{L}}$. 

Suppose that under an infinitesimal change of moduli, $z_{i},w_{a},F_{ij},f_{ia}$
change as 
\begin{eqnarray*}
z_{i} & \to & z_{i}+\varepsilon v_{i}\,,\\
w_{a} & \to & w_{a}\,,\\
F_{ij} & \to & F_{ij}+\delta F_{ij}\,,\\
f_{ia} & \to & f_{ia}+\delta f_{ia}\,.
\end{eqnarray*}
Eqs. (\ref{eq:Fij}), (\ref{eq:fia}) imply
\begin{eqnarray*}
 &  & z_{i}+\varepsilon v_{i}=\left(F_{ij}+\delta F_{ij}\right)(z_{j}+\varepsilon v_{j})\,.\\
 &  & z_{i}+\varepsilon v_{i}=\left(f_{ia}+\delta f_{ia}\right)(w_{a})\,,
\end{eqnarray*}
and we obtain
\begin{eqnarray*}
 &  & \varepsilon(v_{i}-\frac{dz_{i}}{dz_{j}}v_{j})=\delta F_{ij}(z_{j})\,,\\
 &  & \varepsilon v_{i}=\delta f_{ia}(w_{a})\,,
\end{eqnarray*}
in a neighborhood of $C_{ij},\partial D_{a}$ respectively. One can
take $v_{i}$ to be holomorphic in neighborhoods of boundaries of
$S_{i}$ and smooth inside. For such $v_{i}$, we define 
\begin{equation}
b(v)\equiv\sum_{i}\left[\oint_{\partial S_{i}}\frac{dz_{i}}{2\pi i}v_{i}(z_{i})b(z_{i})-\oint_{\partial S_{i}}\frac{d\bar{z}_{i}}{2\pi i}\bar{v}_{i}(\bar{z}_{i})\bar{b}(\bar{z}_{i})\right]\,.\label{eq:partialxs}
\end{equation}
Here the integration contours are taken so that they run along $\partial S_{i}$
keeping $S_{i}$ on the left for $z_{i}$. 

For our purpose, we need to make the formulas (\ref{eq:Fij}), (\ref{eq:fia})
and (\ref{eq:partialxs}) more explicit. $S_{i}$ itself is a hyperbolic
surface with three boundaries and by attaching flat semi-infinite
cylinders to the boundaries as above, we get a three punctured sphere
with local coordinates $W_{k}\ (k=1,2,3)$. Therefore $S_{i}$ is
conformally equivalent to $\mathbb{C}-\bigcup_{k=1}^{3}D_{k}$ where
$D_{k}$ are the disks corresponding to the cylinders. We choose the
local coordinate $z_{i}$ on $S_{i}$ to be the complex coordinate
$z$ on $\mathbb{C}$ such that the three punctures are at $z=0,1,\infty$.
The explicit forms of $W_{k}(z_{i})$ are given in \cite{Hadasz2004,Firat2021},
which are presented in appendix \ref{sec:Hyperbolic-metric-on}. There
is a freedom in choosing which of $\partial D_{k}$ corresponds to
each boundary of $S_{i}$, but (\ref{eq:sl2}) implies $W_{k}(z_{i})$'s
are related by $\mathrm{SL}(2,\mathbb{C})$ transformation of $z_{i}$
and a phase rotation and the choice does not change the result. If
the boundary $\partial D_{a}$ of $\Sigma_{g,n,\mathbf{L}}$ coincides
with $\left|W_{k}(z_{i})\right|=1$, we can take $w_{a}$ to be equal
to $W_{k}(z_{i})$. Then the explicit form of (\ref{eq:fia}) becomes
\begin{equation}
z_{i}=W_{k}^{-1}(w_{a})\,.\label{eq:fia2}
\end{equation}
$\mathrm{C}_{ij}$ should coincide with $\left|W_{k}(z_{i})\right|=1$
and $\left|W_{k^{\prime}}(z_{j})\right|=1$ for some $k,k^{\prime}$
and we obtain the explicit form of $\eqref{eq:Fij}$ as 
\begin{equation}
z_{i}=W_{k}^{-1}\left(\frac{e^{i\theta_{ij}}}{W_{k^{\prime}}(z_{j})}\right)\,,\label{eq:Fij2}
\end{equation}
where $\theta_{ij}$ is the twist angle. 

We take the Fenchel-Nielsen coordinates $l_{s},\tau_{s}\ (s=1,\cdots,3g-3+n)$
on $\mathcal{M}_{g,n,\mathbf{L}}$. Changes of the transition functions
(\ref{eq:fia2}) and (\ref{eq:Fij2}) under the variation $l_{s}\to l_{s}+\delta l_{s},\tau_{s}\to\tau_{s}+\delta\tau_{s}$
describe those of the hyperbolic structure of $\Sigma_{g,n,\mathbf{L}}$.
If $l_{s}$ is the length of $\mathrm{C}_{ij}$, $\tau_{s}=\frac{l_{s}}{2\pi}\theta_{ij}$.
For $l_{s}\to l_{s}+\varepsilon$, we can take 
\begin{eqnarray*}
v_{i} & = & -\frac{\partial W_{k}(z_{i})}{\partial l_{s}}\left(\frac{\partial W_{k}}{\partial z_{i}}\right)^{-1}\,,\\
v_{j} & = & -\frac{\partial W_{k}(z_{j})}{\partial l_{s}}\left(\frac{\partial W_{k}}{\partial z_{j}}\right)^{-1}\,,
\end{eqnarray*}
for $k=1,2,3$, in neighborhoods of boundaries $|W_{k}(z_{i})|=1,\ |W_{k}(z_{j})|=1$
of $S_{i},S_{j}$ respectively. Therefore we define 
\begin{eqnarray}
 &  & b(\partial_{l_{s}})\equiv b(v)=b_{S_{i}}(\partial_{l_{s}})+b_{S_{j}}(\partial_{l_{s}})\,,\nonumber \\
 &  & b_{S_{i}}(\partial_{l_{s}})=-\oint_{\partial S_{i}}\frac{dz_{i}}{2\pi i}\frac{\partial W_{k}}{\partial l_{s}}\left(\frac{\partial W_{k}}{\partial z_{i}}\right)^{-1}b(z_{i})+\oint_{\partial S_{i}}\frac{d\bar{z}_{i}}{2\pi i}\frac{\partial\bar{W}_{k}}{\partial l_{s}}\left(\frac{\partial\bar{W}_{k}}{\partial\bar{z}_{i}}\right)^{-1}\bar{b}(\bar{z}_{i})\,,\nonumber \\
 &  & b_{S_{j}}(\partial_{l_{s}})=-\oint_{\partial S_{j}}\frac{dz_{j}}{2\pi i}\frac{\partial W_{k}}{\partial l_{s}}\left(\frac{\partial W_{k}}{\partial z_{j}}\right)^{-1}b(z_{j})+\oint_{\partial S_{j}}\frac{d\bar{z}_{j}}{2\pi i}\frac{\partial\bar{W}_{k}}{\partial l_{s}}\left(\frac{\partial\bar{W}_{k}}{\partial\bar{z}_{i}}\right)^{-1}\bar{b}(\bar{z}_{j})\,.\label{eq:partialls}
\end{eqnarray}
Here $k\ (k=1,2,3)$ for $W_{k}$ in each term is chosen so that the
relevant component of the boundary corresponds to $\left|W_{k}\right|=1$.
For $\tau_{s}\to\tau_{s}+\varepsilon$, we define
\begin{equation}
b(\partial_{\tau_{s}})=-\frac{2\pi}{l_{s}}\left[\oint_{\mathrm{C}_{ij}}\frac{dz_{i}}{2\pi i}iW_{k}(z_{i})\left(\frac{\partial W_{k}}{\partial z_{i}}\right)^{-1}b(z_{i})+\oint_{\mathrm{C}_{ij}}\frac{d\bar{z}_{i}}{2\pi i}i\bar{W}_{k}\left(\frac{\partial\bar{W}_{k}}{\partial\bar{z}_{i}}\right)^{-1}\bar{b}(\bar{z}_{i})\right]\,,\label{eq:partialtaus}
\end{equation}
where $k$ for $W_{k}$ is chosen so that $\mathrm{C}_{ij}$ coincides
with $\left|W_{k}(z_{i})\right|=1$. The contours run along $\mathrm{C}_{ij}$
so that $S_{j}$ lies to its left for $z_{j}$. 

In the same way, for a pair of pants $S_{i}$ one of whose boundary
coincides with $\partial D_{a}$, we define 
\begin{equation}
b_{S_{i}}(\partial_{L_{a}})=-\oint_{\partial S_{i}}\frac{dz_{i}}{2\pi i}\frac{\partial W_{k}}{\partial L_{a}}\left(\frac{\partial W_{k}}{\partial z_{i}}\right)^{-1}b(z_{i})+\oint_{\partial S_{i}}\frac{d\bar{z}_{i}}{2\pi i}\frac{\partial\bar{W}_{k}}{\partial L_{a}}\left(\frac{\partial\bar{W}_{k}}{\partial\bar{z}_{i}}\right)^{-1}\bar{b}(\bar{z}_{i})\,.\label{eq:partialla}
\end{equation}

\subsection{Off-shell amplitudes}

Now we define the connected $g$ loop $n$ point amplitude $A_{g,n}\left((|\Psi_{1}\rangle,L_{1}),\cdots,(|\Psi_{n}\rangle,L_{n})\right)$
by
\begin{equation}
A_{g,n}\left((|\Psi_{1}\rangle,L_{1}),\cdots,(|\Psi_{n}\rangle,L_{n})\right)=2^{-\delta_{g,1}\delta_{n,1}}\int_{\mathcal{M}_{g,n,\mathbf{L}}}(2\pi i)^{-3g+3-n}\langle\Sigma_{g,n,\mathbf{L}}|B_{6g-6+2n}|\Psi_{1}\rangle\cdots|\Psi_{n}\rangle\,.\label{eq:Agn}
\end{equation}
Here $\langle\Sigma_{g,n,\mathbf{L}}|B_{6g-6+2n}$ is defined so that
\[
\langle\Sigma_{g,n,\mathbf{L}}|B_{6g-6+2n}|\Psi_{1}\rangle\cdots|\Psi_{n}\rangle=\langle B_{6g-6+2n}\prod_{a=1}^{n}w_{a}^{-1}\circ\mathcal{O}_{\Psi_{a}}(0)\rangle_{\Sigma_{g,n}}\,,
\]
holds for any $|\Psi_{a}\rangle$, with
\begin{equation}
B_{6g-6+2n}=\prod_{s=1}^{3g-3+n}\left[b(\partial_{l_{s}})b(\partial_{\tau_{s}})\right]\bigwedge_{s=1}^{3g-3+n}\left[dl_{s}\wedge d\tau_{s}\right]\,.\label{eq:B6g-6+2n}
\end{equation}
The factor $2^{-\delta_{g,1}\delta_{n,1}}$ is due to the fact that
$\Sigma_{1,1,L}$ has a $\mathbb{Z}_{2}$ symmetry. The state $|\Psi_{a}\rangle$
is taken to be an element of $\mathcal{H}_{0}$ which consists of
the states $|\Psi\rangle$ satisfying 
\begin{equation}
b_{0}^{-}|\Psi\rangle=(L_{0}-\bar{L}_{0})|\Psi\rangle=0\,,\label{eq:b0-}
\end{equation}
where $b_{0}^{\pm}\equiv b_{0}\pm\bar{b}_{0}$. 

$B_{6g-6+2n}$ is defined by using the Fenchel-Nielsen coordinate
$l_{s},\tau_{s}$ associated to a pants decomposition of $\Sigma_{g,n,\mathbf{L}}$.
We should check if the amplitude (\ref{eq:Agn}) does not depend on
the choice of the pants decomposition. Suppose that we have two pants
decompositions, in which $\Sigma_{g,n,\mathbf{L}}$ is decomposed
into pairs of pants $S_{i}\ (i=1,\cdots,2g-2+n)$ and $S_{j}^{\prime}\ (j=1,\cdots,2g-2+n)$.
Let $z_{i}$ and $z_{j}^{\prime}$ be the local coordinates on $S_{i},S_{j}^{\prime}$
respectively. There should be a function $G_{ij}$ holomorphic on
$S_{i}\cap S_{j}^{\prime}$ such that 
\[
z_{i}=G_{ij}(z_{j}^{\prime})\,.
\]
If a boundary of $S_{i}\cap S_{j}^{\prime}$ coincides with $\partial D_{a}$,
we have functions $g_{ia},g_{ja}^{\prime}$ such that 
\begin{eqnarray}
z_{i} & = & g_{ia}(w_{a})\,,\nonumber \\
z_{j}^{\prime} & = & g_{ja}^{\prime}(w_{a})\,,\label{eq:gprime0}
\end{eqnarray}
 in a neighborhood of $\partial D_{a}$. Suppose that under an infinitesimal
change of moduli, $z_{i},z_{j}^{\prime},w_{a},G_{ij},g_{ia},g_{ia}^{\prime}$
change as
\begin{eqnarray*}
z_{i} & \to & z_{i}+\varepsilon v_{i}\,,\\
z_{j}^{\prime} & \to & z_{j}^{\prime}+\varepsilon v_{j}^{\prime}\,,\\
w_{a} & \to & w_{a}\,,\\
G_{ij} & \to & G_{ij}+\delta G_{ij}\,,\\
g_{ia} & \to & g_{ia}+\delta g_{ia}\,,\\
g_{ja}^{\prime} & \to & g_{ja}^{\prime}+\delta g_{ja}^{\prime}\,.
\end{eqnarray*}
We can derive 
\begin{eqnarray}
\varepsilon(v_{i}-\frac{\partial z_{i}}{\partial z_{j}^{\prime}}v_{j}^{\prime}) & = & \delta G_{ij}(z_{j}^{\prime})\,,\label{eq:Gij}\\
\varepsilon v_{i} & = & \delta g_{ia}(w_{a})\,,\nonumber \\
\varepsilon v_{j}^{\prime} & = & \delta g_{ja}^{\prime}(w_{a})\,.\label{eq:gprime}
\end{eqnarray}
(\ref{eq:Gij}) implies 
\[
\oint_{\partial(S_{i}\cap S_{j}^{\prime})}\frac{dz_{i}}{2\pi i}(v_{i}-\frac{\partial z_{i}}{\partial z_{j}^{\prime}}v_{j}^{\prime})b(z_{i})=0\,.
\]
If we take $v_{i},v_{j}^{\prime}$ to be holomorphic in neighborhoods
of the boundaries of $S_{i},S_{j}^{\prime}$ respectively, we get
\begin{eqnarray*}
0 & = & \sum_{i,j}\oint_{\partial(S_{i}\cap S_{j}^{\prime})}\frac{dz_{i}}{2\pi i}(v_{i}-\frac{\partial z_{i}}{\partial z_{j}^{\prime}}v_{j}^{\prime})b(z_{i})\\
 & = & \sum_{i}\oint_{\partial S_{i}}\frac{dz_{i}}{2\pi i}v_{i}(z_{i})b(z_{i})-\sum_{j}\oint_{\partial S_{j}^{\prime}}\frac{dz_{j}^{\prime}}{2\pi i}v_{j}^{\prime}(z_{j}^{\prime})b(z_{j}^{\prime})\,.
\end{eqnarray*}
Therefore the $b$-ghost insertion (\ref{eq:partialxs}) satisfies
\begin{equation}
b(v)=b(v^{\prime})\,,\label{eq:bvbvprime}
\end{equation}
if $v$ and $v^{\prime}$ corresponds to the same change of moduli. 

Let $(l_{s};\tau_{s}),(l_{t}^{\prime};\tau_{t}^{\prime})$ be the
Fenchel-Nielsen coordinates associated to the two different pants
decompositions. Using (\ref{eq:bvbvprime}), we may be able to express
$b(\partial_{l_{s}}),b(\partial_{\tau_{s}})$ in terms of $b(\partial_{l_{t}^{\prime}}),b(\partial_{\tau_{t}^{\prime}})$.
In doing so, there is one thing one should be careful about. In defining
$b(\partial_{l_{s}})$, we have taken the coordinate on $D_{a}$ to
be $W_{k}(z_{i})$, if $\partial D_{a}$ coincides with a boundary
of $S_{i}$. If one of boundaries of $S_{j}^{\prime}$ coincides with
$\partial D_{a}$, 
\[
W_{k}(z_{i})=e^{i\alpha_{a}}W_{k^{\prime}}(z_{j}^{\prime})\,,
\]
should hold with some $k^{\prime}$. Here $\alpha_{a}$ is a real
function of moduli. If we fix $w_{a}$ in (\ref{eq:gprime0}) to be
$W_{k}(z_{i})$, (\ref{eq:gprime}) implies
\[
\varepsilon\oint_{\partial D_{a}}\frac{dz_{j}^{\prime}}{2\pi i}v_{j}^{\prime}(z_{j}^{\prime})b(z_{j}^{\prime})=-\oint_{\partial D_{a}}\frac{dz_{j}^{\prime}}{2\pi i}\delta W_{k^{\prime}}(z_{j}^{\prime})\left(\frac{\partial W_{k^{\prime}}}{\partial z_{j}^{\prime}}\right)^{-1}b(z_{j}^{\prime})-i\delta\alpha_{a}\oint_{\partial D_{a}}\frac{dw_{a}}{2\pi i}w_{a}b(w_{a})\,.
\]
Therefore the relations between $b(\partial_{l_{s}}),b(\partial_{\tau_{s}})$
and $b(\partial_{l_{t}^{\prime}}),b(\partial_{\tau_{t}^{\prime}})$
should be
\begin{eqnarray*}
b(\partial_{l_{s}}) & = & \sum_{t}\left[\frac{\partial l_{t}^{\prime}}{\partial l_{s}}b(\partial_{l_{t}^{\prime}})+\frac{\partial\tau_{t}^{\prime}}{\partial l_{s}}b(\partial_{\tau_{t}^{\prime}})\right]-i\sum_{a=1}^{n}\frac{\partial\alpha_{a}}{\partial l_{s}}b_{0}^{-(a)}\,,\\
b(\partial_{\tau_{s}}) & = & \sum_{t}\left[\frac{\partial l_{t}^{\prime}}{\partial\tau_{s}}b(\partial_{l_{t}^{\prime}})+\frac{\partial\tau_{t}^{\prime}}{\partial\tau_{s}}b(\partial_{\tau_{t}^{\prime}})\right]-i\sum_{a=1}^{n}\frac{\partial\alpha_{a}}{\partial\tau_{s}}b_{0}^{-(a)}\,.
\end{eqnarray*}
Here $b_{0}^{-(a)}$ denotes $b_{0}^{-}$ acting on the $a$-th Hilbert
space. Substituting these into the amplitude (\ref{eq:Agn}), we can
see that it is independent of the pants decomposition, if $|\Psi_{a}\rangle\ (a=1,\cdots,n)$
satisfy the condition (\ref{eq:b0-}). 

By the BRST identity proved in appendix \ref{sec:BRST-identity},
we have
\begin{equation}
\langle\Sigma_{g,n,\mathbf{L}}|B_{6g-6+2n}\sum_{a}Q^{(a)}|\Psi_{1}\rangle\cdots|\Psi_{n}\rangle=d\left[\langle\Sigma_{g,n,\mathbf{L}}|B_{6g-7+2n}|\Psi_{1}\rangle\cdots|\Psi_{n}\rangle\right]\,,\label{eq:BRST}
\end{equation}
 and the amplitude $A_{g,n}\left((|\Psi_{1}\rangle,L_{1}),\cdots,(|\Psi_{n}\rangle,L_{n})\right)$
is BRST invariant if one treats the boundary contributions appropriately.
By construction, $A_{g,n}\left((|\Psi_{1}\rangle,L_{1}),\cdots,(|\Psi_{n}\rangle,L_{n})\right)$
exists for $2g-2+n>0$.

The amplitude (\ref{eq:Agn}) is not something we usually deal with
in string theory. In the limit $L_{a}\to0$, $\mathcal{M}_{g,n,\mathbf{L}}$
coincides with the moduli space $\mathcal{M}_{g,n}$ of punctured
Riemann surfaces and $(l_{s};\tau_{s})$ become the Fenchel-Nielsen
coordinates on $\mathcal{M}_{g,n}$. Therefore
\begin{equation}
\lim_{L_{a}\to0}A_{g,n}\left((|\Psi_{1}\rangle,L_{1}),\cdots,(|\Psi_{n}\rangle,L_{n})\right)\label{eq:A1}
\end{equation}
is equal to the on-shell amplitude when $|\Psi_{a}\rangle$ are taken
to be on-shell physical states. In section \ref{sec:BRST-invariant-formulation},
we will show that the off-shell amplitudes of the kind studied in
\cite{Cohen1986,Jaskolski1991,Bolte1991} can also be derived in our
formalism. 

\section{A recursion relation of the off-shell amplitudes\label{sec:A-recursion-relation}}

Given a propagator, one can construct the string field action which
reproduces the off-shell amplitudes defined in the previous section
order by order in the string coupling constant $g_{\mathrm{s}}.$
If we take the propagator to be the one depicted in Figure \ref{fig:A-pants-decomposition.},
we run into the difficulty mentioned in introduction. In this paper,
as a workaround, we construct an SFT by studying equations satisfied
by the off-shell amplitudes.

In order to calculate the right hand side of (\ref{eq:Agn}), we need
to specify the integration region in terms of the Fenchel-Nielsen
coordinates. However no concrete description of the fundamental domain
of the mapping class group in $\mathcal{T}_{g,n,\mathbf{L}}$ is known
in general. Mathematicians were trying to calculate the the Weil-Petersson
volume $V_{g,n}(\mathbf{L})$ of $\mathcal{M}_{g,n,\mathbf{L}}$ defined
by\footnote{There are two conventions for $V_{1,1}(L)$ due to the presence of
$\mathbb{Z}_{2}$ symmetry. Here we adopt (\ref{eq:Vgn}) so as to
make (\ref{eq:Mirzakhani}) look simple.}
\begin{equation}
V_{g,n}(\mathbf{L})\equiv2^{-\delta_{g,1}\delta_{n,1}}\int_{\mathcal{M}_{g,n,\mathbf{L}}}\Omega_{g,n,\mathbf{L}}\,,\label{eq:Vgn}
\end{equation}
and encounter the same problem. Mirzakhani discovered \cite{Mirzakhani2006,Mirzakhani2007}
a way to overcome this difficulty. In this section, we would like
to explain her method (for reviews, see for instance \cite{Moosavian2019a,Do2011,Huang2016})
and apply it to the off-shell amplitudes. 

\subsection{Mirzakhani's scheme}

Mirzakhani's idea is to transform an integral over $\mathcal{M}_{g,n,\mathbf{L}}$
into the one over its covering space. Suppose $X_{1}$ are $X_{2}$
are manifolds and 
\[
\pi:X_{1}\to X_{2}\,,
\]
is a covering map. Let $dv_{2}$ be a volume form on $X_{2}$, and
we define $dv_{1}$ to be the pull back, i.e. 
\[
dv_{1}=\pi^{*}dv_{2}\,.
\]
For a function $f$ on $X_{1}$, one can define the push forward $\pi_{*}f$
by
\[
(\pi_{*}f)(x)=\sum_{y\in\pi^{-1}(x)}f(y)\,.
\]
Then
\begin{equation}
\int_{X_{2}}(\pi_{*}f)dv_{2}=\int_{X_{1}}fdv_{1}\,,\label{eq:coverformula}
\end{equation}
holds. 

Eq. (\ref{eq:coverformula}) can be used to calculate the volume of
the moduli space $\mathcal{M}_{1,1,0}$, for example. We take $X_{2}$
to be $\mathcal{M}_{1,1,0}$ and $X_{1}$ to be the following space
of pairs
\[
\{(\Sigma_{1,1,0},\gamma)\,|\Sigma_{1,1,0}\in\mathcal{M}_{1,1,0}\text{ and }\gamma\text{ is a simple closed geodesic on }\Sigma_{1,1,0}\}\,.
\]
The set of simple closed geodesics $\gamma$ on $\Sigma_{1,1,0}$
is a discrete set with infinitely many elements and a mapping class
group orbit. $X_{1}$ can be described by the pair $(l_{\gamma},\tau_{\gamma})$
where $l_{\gamma}$ is the length of $\gamma$ and $\tau_{\gamma}$
is the twist parameter corresponding to it. $X_{1}$ corresponds to
the region 
\[
0<l_{\gamma}<\infty,\ 0\leq\tau_{\gamma}\leq l_{\gamma}\,,
\]
with $(l_{\gamma},0)\sim(l_{\gamma},l_{\gamma})$. The projection
$\pi$ can be defined by 
\[
\pi(\Sigma_{1,1,0},\gamma)=\Sigma_{1,1,0}\,,
\]
and for $dv_{2}=\Omega_{1,1,0}$, we have
\[
dv_{1}=\pi^{*}dv_{2}=dl_{\gamma}\wedge d\tau_{\gamma}\,.
\]

If one takes the function $f$ to be a function of $l_{\gamma}$,
the value of $\pi_{*}f$ at $\Sigma_{1,1,0}\in\mathcal{M}_{1,1,0}$
becomes 
\[
\sum_{\gamma}f(l_{\gamma})\,,
\]
where the sum is over the set of simple geodesics on $\Sigma_{1,1,0}$.
In \cite{McShane1991}, McShane proved that for $f(l)=\frac{2}{1+e^{l}}$,
\begin{equation}
\sum_{\gamma}f(l_{\gamma})=1\,,\label{eq:McShane}
\end{equation}
holds. (\ref{eq:McShane}) is called the McShane identity. For this
choice of $f$, (\ref{eq:coverformula}) becomes
\begin{equation}
\int_{\mathcal{M}_{1,1,0}}\Omega_{1,1,0}=\int_{X_{2}}(\pi_{*}f)dv_{2}=\int_{X_{1}}fdv_{1}=\int_{0}^{\infty}dl_{\gamma}\frac{2l_{\gamma}}{1+e^{l_{\gamma}}}=\frac{\pi^{2}}{6}\,,\label{eq:V11}
\end{equation}
and we get the volume of $\mathcal{M}_{1,1,0}$.

Mirzakhani generalized this procedure to general $(g,n)$, by discovering
a generalization of the McShane identity. For $\Sigma_{g,n,\mathbf{L}}\in\mathcal{\mathcal{M}}_{g,n,\mathbf{L}}$,
let $\beta_{1},\cdots,\beta_{n}$ be the boundaries so that the lengths
of $\beta_{1},\cdots,\beta_{n}$ are $L_{1},\cdots,L_{n}$ respectively.
The generalized McShane identity derived in \cite{Mirzakhani2006}
is
\begin{equation}
L_{1}=\sum_{\left\{ \gamma,\delta\right\} \in\mathscr{C}_{1}}\mathsf{D}_{L_{1}l_{\gamma}l_{\delta}}+\sum_{a=2}^{n}\sum_{\gamma\in\mathscr{C}_{a}}(\mathsf{T}_{L_{1}L_{a}l_{\gamma}}+\mathsf{D}_{L_{1}L_{a}l_{\gamma}})\,,\label{eq:gMS}
\end{equation}
where
\begin{eqnarray}
\mathsf{D}_{LL^{\prime}L^{\prime\prime}} & = & 2\left(\log(e^{\frac{L}{2}}+e^{\frac{L^{\prime}+L^{\prime\prime}}{2}})-\log(e^{-\frac{L}{2}}+e^{\frac{L^{\prime}+L^{\prime\prime}}{2}})\right)\,,\label{eq:D}\\
\mathsf{T}_{LL^{\prime}L^{\prime\prime}} & = & \log\frac{\cosh\frac{L^{\prime\prime}}{2}+\cosh\frac{L+L^{\prime}}{2}}{\cosh\frac{L^{\prime\prime}}{2}+\cosh\frac{L-L^{\prime}}{2}}\,,\label{eq:T}
\end{eqnarray}
\begin{eqnarray*}
\mathscr{C}_{1} & \equiv & \left\{ \begin{array}{l}
\text{the collection of unordered pairs of nonperipheral simple closed geodesics }\left\{ \gamma,\delta\right\} \\
\text{on }\Sigma_{g,n,\mathbf{L}}\text{ which bounds a pair of pants along with the bondary }\beta_{1}
\end{array}\right\} \,,\\
\mathscr{C}_{a} & \equiv & \left\{ \begin{array}{l}
\text{the collection of simple closed geodesics }\gamma\text{ on }\Sigma_{g,n,\mathbf{L}}\\
\text{which bounds a pair of pants along with the bondaries }\beta_{1}\text{ and }\beta_{a}
\end{array}\right\} \,,
\end{eqnarray*}
and $l_{\gamma},l_{\delta}$ are the lengths of $\gamma,\delta$ respectively.
For $L,L^{\prime},L^{\prime\prime}>0$, $\mathsf{D}_{LL^{\prime}L^{\prime\prime}},\mathsf{T}_{LL^{\prime}L^{\prime\prime}}$>0
and
\begin{eqnarray}
 &  & \mathsf{D}_{LL^{\prime}L^{\prime\prime}}=\mathsf{D}_{LL^{\prime\prime}L^{\prime}}\,,\nonumber \\
 &  & \mathsf{T}_{LL^{\prime}L^{\prime\prime}}=\mathsf{T}_{L^{\prime}LL^{\prime\prime}\,,}\nonumber \\
 &  & \mathsf{D}_{LL^{\prime}L^{\prime\prime}}+\mathsf{T}_{LL^{\prime}L^{\prime\prime}}+\mathsf{T}_{LL^{\prime\prime}L^{\prime}}=L\,.\label{eq:DTidentity}
\end{eqnarray}

Multiplying (\ref{eq:gMS}) by $\Omega_{g,n,\mathbf{L}}$ and integrating
it over $\mathcal{M}_{g,n,\mathbf{L}}$, one obtains the Mirzakhani's
recursion relation:
\begin{eqnarray}
LV_{g,n+1}(L,\mathbf{L}) & = & \frac{1}{2}\int_{0}^{\infty}dL^{\prime}L^{\prime}\int_{0}^{\infty}dL^{\prime\prime}L^{\prime\prime}\mathsf{D}_{LL^{\prime}L^{\prime\prime}}\left(V_{g-1,n+2}(L^{\prime},L^{\prime\prime},\mathbf{L})+\sum_{\text{stable}}V_{g_{1},n_{1}}(L^{\prime},\mathbf{L}_{1})V_{g_{2},n_{2}}(L^{\prime\prime},\mathbf{L}_{2})\right)\nonumber \\
 &  & +\sum_{a=1}^{n}\int_{0}^{\infty}dL^{\prime}L^{\prime}\left(\mathsf{T}_{L_{1}L_{a}L^{\prime}}+\mathsf{D}_{L_{1}L_{a}L^{\prime}}\right)V_{g,n}(L,\mathbf{L}\backslash L_{a})\,,\label{eq:Mirzakhani}
\end{eqnarray}
which holds for $2g-2+n>0$. The sum $\sum_{\text{stable}}$ here
means the sum over $g_{1},g_{2},n_{1},n_{2},\mathbf{L}_{1},\mathbf{L}_{2}$
such that\footnote{Here we consider $\mathbf{L}_{1},\mathbf{L}_{2}$ as unordered subsets
of $\mathbf{L}=\left\{ L_{1},\cdots,L_{n}\right\} $. }
\begin{eqnarray}
g_{1}+g_{2} & = & g\,,\nonumber \\
n_{1}+n_{2} & = & n+2\,,\nonumber \\
\mathbf{L}_{1}\cup\mathbf{L}_{2} & = & \left\{ L_{1},\cdots,L_{n}\right\} \,,\nonumber \\
\mathbf{L}_{1}\cap\mathbf{L}_{2} & = & \phi\,,\nonumber \\
2g_{1}-2+n_{1} & > & 0\,,\nonumber \\
2g_{2}-2+n_{2} & > & 0\,.\label{eq:stable-2}
\end{eqnarray}
With the information $V_{0,3}(L_{1},L_{2},L_{3})=1$, $V_{1,1}(L)=\frac{\pi^{2}}{12}+\frac{L^{2}}{48}$,
it is possible to calculate $V_{g,n}(\mathbf{L})$ for all the other
$g,n$ by the recursion relation (\ref{eq:Mirzakhani}). 

\subsection{Recursion relation of the off-shell amplitudes}

A recursion relation of the off-shell amplitudes (\ref{eq:Agn}) is
derived in the same way as the Mirzakhani's recursion relation. Multiplying
(\ref{eq:gMS}) by $(2\pi i)^{-3g+3-n}\langle\Sigma_{g,n,\mathbf{L}}|B_{6g-6+2n}|\Psi_{1}\rangle\cdots|\Psi_{n}\rangle$
and integrating it over $\mathcal{M}_{g,n,\mathbf{L}}$, we obtain
\begin{eqnarray}
 &  & L_{1}\int_{\mathcal{M}_{g,n,\mathbf{L}}}(2\pi i)^{-3g+3-n}\langle\Sigma_{g,n,\mathbf{L}}|B_{6g-6+2n}|\Psi_{1}\rangle\cdots|\Psi_{n}\rangle\nonumber \\
 &  & \quad=\int_{\mathcal{M}_{g,n,\mathbf{L}}}\sum_{\left\{ \gamma,\delta\right\} \in\mathscr{C}_{1}}\mathsf{D}_{L_{1}l_{\gamma}l_{\delta}}\cdot(2\pi i)^{-3g+3-n}\langle\Sigma_{g,n,\mathbf{L}}|B_{6g-6+2n}|\Psi_{1}\rangle\cdots|\Psi_{n}\rangle\nonumber \\
 &  & \hphantom{\quad=}+\sum_{a=2}^{n}\int_{\mathcal{M}_{g,n,\mathbf{L}}}\sum_{\gamma\in\mathscr{C}_{a}}(\mathsf{T}_{L_{1}L_{a}l_{\gamma}}+\mathsf{D}_{L_{1}L_{a}l_{\gamma}})\cdot(2\pi i)^{-3g+3-n}\langle\Sigma_{g,n,\mathbf{L}}|B_{6g-6+2n}|\Psi_{1}\rangle\cdots|\Psi_{n}\rangle\,.\label{eq:intomega}
\end{eqnarray}
The left hand side yields 
\[
L_{1}A_{g,n}\left((|\Psi_{1}\rangle,L_{1}),\cdots,(|\Psi_{n}\rangle,L_{n})\right)\,.
\]

Here we restrict ourselves to the case $2g-2+n>1$. We will rewrite
the terms on the right hand side by using the formula (\ref{eq:coverformula}).
Let us first consider the integral
\begin{equation}
\int_{\mathcal{M}_{g,n,\mathbf{L}}}\sum_{\gamma\in\mathscr{C}_{a}}(\mathsf{T}_{L_{1}L_{a}l_{\gamma}}+\mathsf{D}_{L_{1}L_{a}l_{\gamma}})\cdot(2\pi i)^{-3g+3-n}\langle\Sigma_{g,n,\mathbf{L}}|B_{6g-6+2n}|\Psi_{1}\rangle\cdots|\Psi_{n}\rangle\,.\label{eq:int1}
\end{equation}
In order to unfold this integral, we take $X_{2}$ to be $\mathcal{M}_{g,n,\mathbf{L}}$
and $X_{1}$ to be the space of pairs
\[
\left\{ \vphantom{\begin{array}{c}
\Sigma_{g,n,\mathbf{L}}\in\mathcal{M}_{g,n,\mathbf{L}}\text{ and }\gamma\text{ is a simple closed geodesic on }\Sigma_{g,n,\mathbf{L}}\\
\text{which bounds a pair of pants along with the bondary }\beta_{1}\text{ and }\beta_{a}
\end{array}}(\Sigma_{g,n,\mathbf{L}},\gamma)\right.\left|\begin{array}{c}
\Sigma_{g,n,\mathbf{L}}\in\mathcal{M}_{g,n,\mathbf{L}}\text{ and }\gamma\text{ is a simple closed geodesic on }\Sigma_{g,n,\mathbf{L}}\\
\text{which bounds a pair of pants along with the boundaries }\beta_{1}\text{ and }\beta_{a}
\end{array}\right\} \,.
\]
The set of possible $\gamma$ on $\Sigma_{g,n,\mathbf{L}}$ is exactly
$\mathscr{C}_{a}$ and is a mapping class group orbit. $X_{1}$ can
be described by the triple $(l_{\gamma},\tau_{\gamma},\Sigma_{g,n-1,\mathbf{L^{\prime}}}^{\prime})$
where $\tau_{\gamma}$ is the twist parameter corresponding to $\gamma$
and $\Sigma_{g,n-1,\mathbf{L^{\prime}}}^{\prime}$ is the complement
of the pair of pants bounded by $\beta_{1},\beta_{a},\gamma$ with
$\mathbf{L^{\prime}}=(l_{\gamma},L_{2},\cdots,\hat{L}_{a},\cdots,L_{n})$.
$X_{1}$ corresponds to the region 
\[
0<l_{\gamma}<\infty,\ 0\leq\tau_{\gamma}\leq l_{\gamma}\,,
\]
with $(l_{\gamma},0,\Sigma_{g,n-1,\mathbf{L^{\prime}}}^{\prime})\sim(l_{\gamma},l_{\gamma},\Sigma_{g,n-1,\mathbf{L^{\prime}}}^{\prime})$.
The projection $\pi$ can be defined by 
\[
\pi(\Sigma_{g,n,\mathbf{L}},\gamma)=\Sigma_{g,n,\mathbf{L}}\,,
\]
and for $dv_{2}=\Omega_{g,n,\mathbf{L}}$, we have
\[
dv_{1}=\pi^{*}dv_{2}=dl_{\gamma}\wedge d\tau_{\gamma}\wedge\Omega_{g,n-1,\mathbf{L^{\prime}}}^{\prime}\,,
\]
where $\Omega_{g,n-1,\mathbf{L^{\prime}}}^{\prime}$ is the Weil-Petersson
volume form on $\Sigma_{g,n-1,\mathbf{L^{\prime}}}^{\prime}$. 

Now, if one takes 
\[
fdv_{1}=(\mathsf{T}_{L_{1}L_{a}l_{\gamma}}+\mathsf{D}_{L_{1}L_{a}l_{\gamma}})\cdot(2\pi i)^{-3g+3-n}\langle\Sigma_{g,n,\mathbf{L}}|B_{6g-6+2n}|\Psi_{1}\rangle\cdots|\Psi_{n}\rangle\,,
\]
(\ref{eq:coverformula}) becomes
\begin{eqnarray}
 &  & \int_{X_{1}}(\mathsf{T}_{L_{1}L_{a}l_{\gamma}}+\mathsf{D}_{L_{1}L_{a}l_{\gamma}})\cdot(2\pi i)^{-3g+3-n}\langle\Sigma_{g,n,\mathbf{L}}|B_{6g-6+2n}|\Psi_{1}\rangle\cdots|\Psi_{n}\rangle\nonumber \\
 &  & \quad=\int_{\mathcal{M}_{g,n,\mathbf{L}}}\sum_{\gamma\in\mathscr{C}_{a}}(\mathsf{T}_{L_{1}L_{a}l_{\gamma}}+\mathsf{D}_{L_{1}L_{a}l_{\gamma}})\cdot(2\pi i)^{-3g+3-n}\langle\Sigma_{g,n,\mathbf{L}}|B_{6g-6+2n}|\Psi_{1}\rangle\cdots|\Psi_{n}\rangle\,.\label{eq:2nd-1}
\end{eqnarray}
Therefore (\ref{eq:int1}) is obtained by evaluating the left hand
side of (\ref{eq:2nd-1}). 

\begin{figure}
\begin{centering}
\includegraphics[scale=0.7]{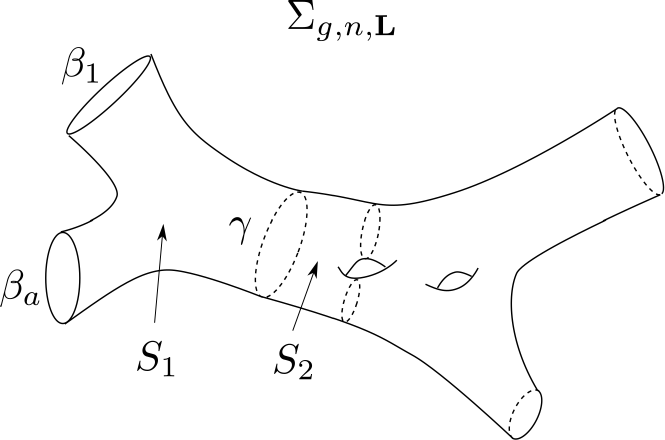}
\par\end{centering}
\caption{$\Sigma_{g,n,\mathbf{L}}$ and $\gamma$.\label{fig:-and-.}}
\end{figure}

Let us consider a pants decomposition of $\Sigma_{g,n,\mathbf{L}}$
such that one pair of the pants is with boundaries $\beta_{1},\beta_{a},\gamma$
(Figure \ref{fig:-and-.}). We denote this pair of pants by $S_{1}$
and the adjacent one by $S_{2}$. Based on the pants decomposition
we define the Fenchel-Nielsen coordinates $l_{s},\tau_{s}\ (s=1,\cdots,3g-3+n)$
such that $(l_{1},\tau_{1})=(l_{\gamma},\tau_{\gamma})$. Cutting
$\Sigma_{g,n,\mathbf{L}}$ along $\gamma$, we get a three holed sphere
$\Sigma_{0,3,(L_{1},L_{a},l_{1})}$ and $\Sigma_{g,n-1,\mathbf{L^{\prime}}}^{\prime}$.
$\Sigma_{g,n-1,\mathbf{L^{\prime}}}^{\prime}$ inherits the pants
decomposition of $\Sigma_{g,n,\mathbf{L}}$. Then $(2\pi i)^{-3g+3-n}\langle\Sigma_{g,n,\mathbf{L}}|B_{6g-6+2n}|\Psi_{1}\rangle\cdots|\Psi_{n}\rangle$
can be expressed as
\begin{eqnarray}
 &  & (2\pi i)^{-3g+3-n}\langle\Sigma_{g,n,\mathbf{L}}|B_{6g-6+2n}|\Psi_{1}\rangle\cdots|\Psi_{n}\rangle\nonumber \\
 &  & \quad=(2\pi i)^{-3g+3-n}dl_{1}\wedge d\tau_{1}\langle\Sigma_{g,n,\mathbf{L}}|\left[b_{S_{1}}(\partial_{l_{1}})+b_{S_{2}}(\partial_{l_{1}})\right]b(\partial_{\tau_{1}})B_{6g-8+2n}^{\prime}|\Psi_{1}\rangle\cdots|\Psi_{n}\rangle\,,\label{eq:pistaromega}
\end{eqnarray}
where $B_{6g-8+2n}^{\prime}$ denotes the $6g-6+2(n-1)$ form on $\Sigma_{g,n-1,\mathbf{L^{\prime}}}^{\prime}$
defined through (\ref{eq:B6g-6+2n}). 

\begin{figure}
\begin{centering}
\includegraphics[scale=0.7]{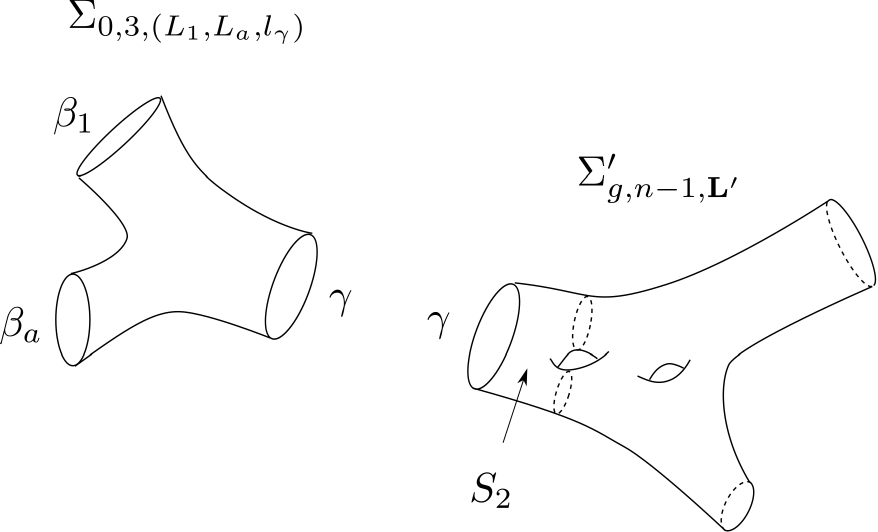}
\par\end{centering}
\caption{The decomposition of $\Sigma_{g,n,\mathbf{L}}$ corresponding to (\ref{eq:2ndfactor}).
\label{fig:The-decomposition-of}}
\end{figure}

$\Sigma_{g,n,\mathbf{L}}$ can be generated by gluing a pair of pants
$\Sigma_{0,3,(L_{1},L_{a},l_{1})}$ and $\Sigma_{g,n-1,\mathbf{L^{\prime}}}^{\prime}$
(Figure \ref{fig:The-decomposition-of}) using the plumbing fixture
relation (\ref{eq:Fij2}). Hence the correlation function on the right
hand side of (\ref{eq:pistaromega}) can be factorized into those
on $\Sigma_{0,3,(L_{1},L_{a},l_{1})}$ and $\Sigma_{g,n-1,\mathbf{L^{\prime}}}^{\prime}$.
Let $|\varphi_{i}\rangle$ be a basis of the Hilbert space $\mathcal{H}$
of the worldsheet theory of the strings and $\langle\varphi_{i}^{c}|$
be the conjugate state of $|\varphi_{i}\rangle$ such that 
\begin{eqnarray*}
\langle\varphi_{i}^{c}|\varphi_{j}\rangle & = & \delta_{ij}\,,\\
\langle\varphi_{j}|\varphi_{i}^{c}\rangle & = & (-1)^{n_{\varphi_{i}}}\delta_{ij}\,,\\
\sum_{i}|\varphi_{i}\rangle\langle\varphi_{i}^{c}| & = & \sum_{i}|\varphi_{i}^{c}\rangle\langle\varphi_{i}|(-1)^{n_{\varphi_{i}}}=\mathbf{1}\,.
\end{eqnarray*}
Here $\langle\varphi_{i}|$ is the BPZ conjugate of $|\varphi_{i}\rangle$
and $n_{\varphi_{i}}$ is the ghost number of $|\varphi_{i}\rangle$.
Then we have
\begin{eqnarray}
 &  & \langle\Sigma_{g,n,\mathbf{L}}|\left[b_{S_{1}}(\partial_{l_{1}})+b_{S_{2}}(\partial_{l_{1}})\right]b(\partial_{\tau_{1}})B_{6g-6+2n}|\Psi_{1}\rangle\cdots|\Psi_{n}\rangle\nonumber \\
 &  & \quad=-\frac{2\pi i}{l_{1}}\varepsilon_{a}\sum_{i,j}\left[\langle\Sigma_{0,3,(L_{1},L_{a},l_{1})}|b_{\Sigma_{0,3,(L_{1},L_{a},l_{1})}}(\partial_{l_{1}})b_{0}^{-(0)}|\Psi_{1}\rangle_{1}|\Psi_{a}\rangle_{a}e^{i\theta(L_{0}-\bar{L}_{0})}|\varphi_{i}\rangle_{0}\right.\nonumber \\
 &  & \hphantom{\quad=-\frac{2\pi i}{l_{1}}\varepsilon_{a}\sum_{i}\quad\quad}\times2^{-\delta_{g,1}\delta_{n,2}}\langle\Sigma_{g,n-1,\mathbf{L^{\prime}}}^{\prime}|B_{6g-8+2n}^{\prime}|\varphi_{j}\rangle|\Psi_{2}\rangle\cdots\widehat{|\Psi_{a}\rangle}\cdots|\Psi_{n}\rangle\nonumber \\
 &  & \hphantom{\quad=-\frac{2\pi i}{l_{1}}\varepsilon_{a}\sum_{i}\quad}+\langle\Sigma_{0,3,(L_{1},L_{a},l_{1})}|\Psi_{1}\rangle|\Psi_{a}\rangle|\varphi_{i}\rangle\nonumber \\
 &  & \hphantom{\quad=-\frac{2\pi i}{l_{1}}\varepsilon_{a}\sum_{i}\quad+\quad}\left.\times2^{-\delta_{g,1}\delta_{n,2}}\langle\Sigma_{g,n-1,\mathbf{L^{\prime}}}^{\prime}|B_{6g-8+2n}^{\prime}b_{S_{2}}(\partial_{l_{1}})b_{0}^{-}e^{i\theta(L_{0}-\bar{L}_{0})}|\varphi_{j}\rangle|\Psi_{2}\rangle\cdots\widehat{|\Psi_{a}\rangle}\cdots|\Psi_{n}\rangle\right]\nonumber \\
 &  & \hphantom{\quad=-\frac{2\pi i}{l_{1}}\varepsilon_{a}\sum_{i}\quad}\times\langle\varphi_{i}^{c}|\varphi_{j}^{c}\rangle(-1)^{n_{\varphi_{j}}}\,,\label{eq:2ndfactor}
\end{eqnarray}
where
\[
\varepsilon_{a}=(-1)^{n_{a}(n_{2}+\cdots+n_{a-1})}\,,
\]
 $n_{b}$ denotes the ghost number of $|\Psi_{b}\rangle$ and $\theta$
denotes the twist angle. The factor $2^{-\delta_{g,1}\delta_{n,2}}$
is due to the fact that $\Sigma_{g,n-1,\mathbf{L^{\prime}}}^{\prime}$
has a $\mathbb{Z}_{2}$ symmetry for $g=1,n=2$. 

Substituting (\ref{eq:pistaromega}) and (\ref{eq:2ndfactor}) into
(\ref{eq:2nd-1}), we can see that the second term on the right hand
side of (\ref{eq:intomega}) becomes
\begin{eqnarray}
 &  & \sum_{a=2}^{n}\int_{\mathcal{M}_{g,n,\mathbf{L}}}\sum_{\gamma\in\mathscr{C}_{a}}(\mathsf{T}_{L_{1}L_{a}l_{\gamma}}+\mathsf{D}_{L_{1}L_{a}l_{\gamma}})\cdot(2\pi i)^{-3g+3-n}\langle\Sigma_{g,n,\mathbf{L}}|B_{6g-6+2n}|\Psi_{1}\rangle\cdots|\Psi_{n}\rangle\nonumber \\
 &  & \quad=-\sum_{a=2}^{n}\sum_{i,j}\varepsilon_{a}\left[\int_{0}^{\infty}dl_{1}(\mathsf{T}_{L_{1}L_{a}l_{1}}+\mathsf{D}_{L_{1}L_{a}l_{1}})\langle\Sigma_{0,3,(L_{1},L_{a},l_{1})}|b_{\Sigma_{0,3,(L_{1},L_{a},l_{1})}}(\partial_{l_{1}})b_{0}^{-(0)}P^{(0)}|\Psi_{1}\rangle_{1}|\Psi_{a}\rangle_{a}|\varphi_{i}\rangle_{0}\right.\nonumber \\
 &  & \hphantom{\quad=-\sum_{a=2}^{n}\sum_{i}\varepsilon_{a}\quad\quad}\times2^{-\delta_{g,1}\delta_{n,2}}\int_{\mathcal{M}_{g,n-1,\mathbf{L^{\prime}}}}(2\pi i)^{-3g+4-n}\langle\Sigma_{g,n-1,\mathbf{L^{\prime}}}^{\prime}|B_{6g-8+2n}^{\prime}|\varphi_{j}\rangle|\Psi_{2}\rangle\cdots\widehat{|\Psi_{a}\rangle}\cdots|\Psi_{n}\rangle\nonumber \\
 &  & \hphantom{\quad=-\sum_{a=2}^{n}\sum_{i}\varepsilon_{a}\quad}+\int_{0}^{\infty}dl_{1}(\mathsf{T}_{L_{1}L_{a}l_{1}}+\mathsf{D}_{L_{1}L_{a}l_{1}})\langle\Sigma_{0,3,(L_{1},L_{a},l_{1})}|\Psi_{1}\rangle|\Psi_{a}\rangle|\varphi_{i}\rangle\nonumber \\
 &  & \hphantom{\quad=-\sum_{a=2}^{n}\sum_{i}\varepsilon_{a}\quad\quad}\left.\times2^{-\delta_{g,1}\delta_{n,2}}\int_{\mathcal{M}_{g,n-1,\mathbf{L^{\prime}}}}(2\pi i)^{-3g+4-n}\langle\Sigma_{g,n-1,\mathbf{L^{\prime}}}^{\prime}|B_{6g-8+2n}^{\prime}b_{S_{2}}(\partial_{l_{1}})b_{0}^{-}P|\varphi_{j}\rangle|\Psi_{2}\rangle\cdots\widehat{|\Psi_{a}\rangle}\cdots|\Psi_{n}\rangle\right]\nonumber \\
 &  & \hphantom{\quad=-\sum_{a=2}^{n}\sum_{i}\varepsilon_{a}\quad}\times\langle\varphi_{i}^{c}|\varphi_{j}^{c}\rangle(-1)^{n_{\varphi_{j}}}\,,\label{eq:2nd-2}
\end{eqnarray}
with $P=\int_{0}^{2\pi}\frac{d\theta}{2\pi}e^{i\theta(L_{0}-\bar{L}_{0})}$.

(\ref{eq:2nd-2}) implies that it will be convenient to consider the
recursion relation of the amplitudes of the form
\begin{eqnarray}
 &  & A_{g,n}\left((|\varphi_{i_{1}}\rangle,\alpha_{1},L_{1}),\cdots,(|\varphi_{i_{n}}\rangle,\alpha_{n},L_{n})\right)\nonumber \\
 &  & \quad=2^{-\delta_{g,1}\delta_{n,1}}\int_{\mathcal{M}_{g,n,\mathbf{L}}}(2\pi i)^{-3g+3-n}\langle\Sigma_{g,n,\mathbf{L}}|B_{6g-6+2n}B_{\alpha_{1}}^{1}\cdots B_{\alpha_{n}}^{n}|\varphi_{i_{1}}\rangle_{1}\cdots|\varphi_{i_{n}}\rangle_{n}\,.\label{eq:Agn-1}
\end{eqnarray}
Here the indices $\alpha_{a}\ (a=1,\cdots,n)$ take values $\pm$
and
\begin{equation}
B_{\alpha_{a}}^{a}\equiv\begin{cases}
1 & \alpha_{a}=+\\
b_{0}^{-(a)}b_{S_{a}}(\partial_{L_{a}})P^{(a)} & \alpha_{a}=-
\end{cases}\,.\label{eq:alpha}
\end{equation}
 $S_{a}$ for $b_{S_{a}}(\partial_{L_{a}})$ in (\ref{eq:alpha})
denotes the pair of pants which has a boundary corresponding to the
$a$-th external line in a pants decomposition of $\Sigma_{g,n,\mathbf{L}}$.
$b_{S_{a}}(\partial_{L_{a}})$ depends on the choice of the pants
decomposition, because it corresponds to the variation $L_{a}\to L_{a}+\varepsilon$
with $l_{s},\tau_{s}$ fixed. However, $b_{S_{a}}(\partial_{L_{a}})B_{6g-6+2n}$
and the amplitude in (\ref{eq:Agn-1}) is independent of the choice
of $S_{a}$. 

(\ref{eq:2nd-2}) can be recast into 
\begin{eqnarray}
 &  & \sum_{a=2}^{n}\int_{\mathcal{M}_{g,n,\mathbf{L}}}\sum_{\gamma\in\mathscr{C}_{a}}(\mathsf{T}_{L_{1}L_{a}l_{\gamma}}+\mathsf{D}_{L_{1}L_{a}l_{\gamma}})\cdot(2\pi i)^{-3g+3-n}\langle\Sigma_{g,n,\mathbf{L}}|B_{6g-6+2n}B_{\alpha_{1}}^{1}\cdots B_{\alpha_{n}}^{n}|\varphi_{i_{1}}\rangle_{1}\cdots|\varphi_{i_{n}}\rangle_{n}\nonumber \\
 &  & \quad=\sum_{a=2}^{n}\int_{0}^{\infty}dL(\mathsf{T}_{L_{1}L_{a}L}+\mathsf{D}_{L_{1}L_{a}L})\nonumber \\
 &  & \hphantom{\quad=\sum_{a=2}^{n}\int_{0}^{\infty}dl}\times\sum_{i,j,\alpha}\varepsilon_{a}\left[\langle\Sigma_{0,3,(L_{1},L_{a},L)}|B_{\alpha_{1}}^{1}B_{\alpha_{a}}^{a}B_{-\alpha}^{0}|\varphi_{i_{1}}\rangle_{1}|\varphi_{i_{a}}\rangle_{a}|\varphi_{i}\rangle_{0}\langle\varphi_{i}^{c}|\varphi_{j}^{c}\rangle(-1)^{n_{\varphi_{j}}}\right.\nonumber \\
 &  & \hphantom{\quad=\sum_{a=2}^{n}\int_{0}^{\infty}dl\sum_{i,j}\varepsilon_{a}\quad\quad\quad}\times A_{g,n-1}\left((|\varphi_{j}\rangle,\alpha,L),(|\varphi_{i_{2}}\rangle,\alpha_{2},L_{2}),\cdots,\widehat{(|\varphi_{i_{a}}\rangle,\alpha_{a},L_{a})},\cdots,(|\varphi_{i_{n}}\rangle,\alpha_{n},L_{n})\right)\nonumber \\
\label{eq:2nd-3}
\end{eqnarray}
We simplify the formula by introducing the following notation. The
external states are labeled by $i$ (for $|\varphi_{i}\rangle$),
$\alpha$, and $L$. We denote these collectively by $I$ and rewrite
(\ref{eq:2nd-3}) in the following way:
\begin{eqnarray}
 &  & \sum_{a=2}^{n}\int_{\mathcal{M}_{g,n,\mathbf{L}}}\sum_{\gamma\in\mathscr{C}_{a}}(\mathsf{T}_{L_{1}L_{a}l_{\gamma}}+\mathsf{D}_{L_{1}L_{a}l_{\gamma}})\cdot(2\pi i)^{-3g+3-n}\langle\Sigma_{g,n,\mathbf{L}}|B_{6g-6+2n}B_{\alpha_{1}}^{1}\cdots B_{\alpha_{n}}^{n}|\varphi_{i_{1}}\rangle\cdots|\varphi_{i_{n}}\rangle\nonumber \\
 &  & \quad=\sum_{a=2}^{n}\varepsilon_{a}(T^{I_{1}I_{a}J}+D^{I_{1}I_{a}J})G_{JI}A_{g,n-1}^{II_{2}\cdots\hat{I}_{a}\cdots I_{n}}\,,\label{eq:2nd-4}
\end{eqnarray}
where
\begin{eqnarray*}
T^{I_{1}I_{2}I_{3}} & \equiv & \mathsf{T}_{L_{1}L_{2}L_{3}}\langle\Sigma_{0,3,(L_{1},L_{2},L_{3})}|B_{\alpha_{1}}^{1}B_{\alpha_{2}}^{2}B_{\alpha_{3}}^{3}|\varphi_{i_{1}}\rangle_{1}|\varphi_{i_{2}}\rangle_{2}|\varphi_{i_{3}}\rangle_{3}\,,\\
D^{I_{1}I_{2}I_{3}} & \equiv & \mathsf{D}_{L_{1}L_{2}L_{3}}\langle\Sigma_{0,3,(L_{1},L_{2},L_{3})}|B_{\alpha_{1}}^{1}B_{\alpha_{2}}^{2}B_{\alpha_{3}}^{3}|\varphi_{i_{1}}\rangle_{1}|\varphi_{i_{2}}\rangle_{2}|\varphi_{i_{3}}\rangle_{3}\,,\\
G_{I_{1}I_{2}} & \equiv & \langle\varphi_{i_{1}}^{c}|\varphi_{i_{2}}^{c}\rangle(-1)^{n_{\varphi_{i_{2}}}}\delta(L_{1}-L_{2})\delta_{\alpha_{1},-\alpha_{2}}\,,\\
A_{g,n}^{I_{1}\cdots I_{n}} & \equiv & A_{g,n}\left((|\varphi_{i_{1}}\rangle,\alpha_{1},L_{1}),\cdots,(|\varphi_{i_{n}}\rangle,\alpha_{n},L_{n})\right)\,,
\end{eqnarray*}
and for $X_{I}=X(i,\alpha,L)$ and $Y^{I}=Y(i,\alpha,L)$
\[
X_{I}Y^{I}=\sum_{i}\sum_{\alpha=\pm}\int_{0}^{\infty}dLX(i,\alpha,L)Y(i,\alpha,L)\,.
\]

\begin{figure}
\begin{centering}
\includegraphics[scale=0.6]{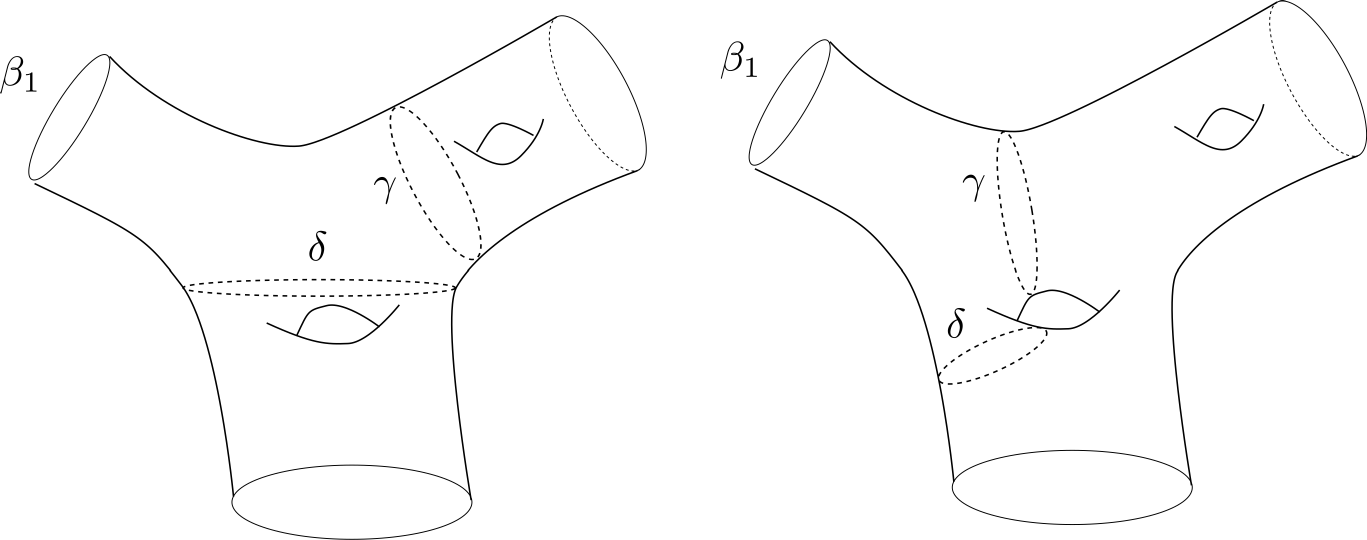}
\par\end{centering}
\caption{Examples of $\{\gamma,\delta\}$ in $\mathscr{C}_{1}$.\label{fig:Examples-of-}}

\end{figure}

\begin{figure}
\begin{centering}
\includegraphics[scale=0.6]{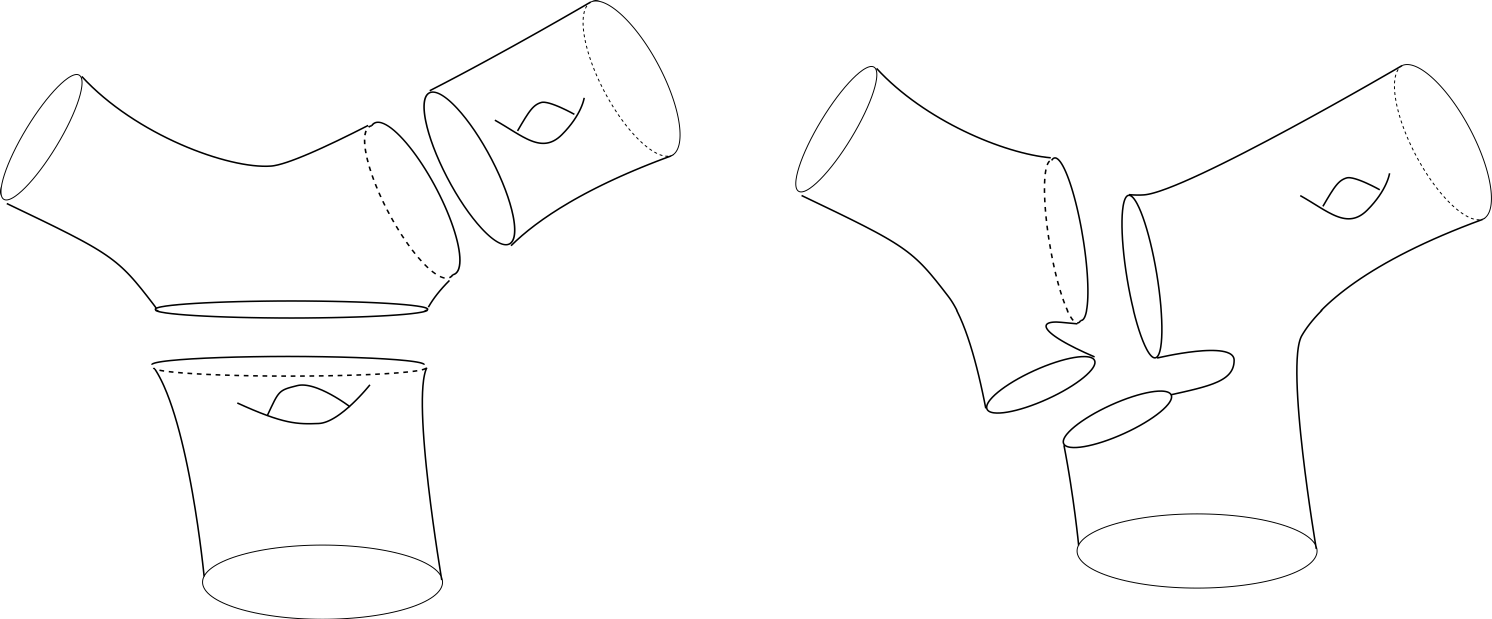}
\par\end{centering}
\caption{Factorizations of the surfaces in Figure \ref{fig:Examples-of-}.\label{fig:Factorizations-of-the}}

\end{figure}

The integral
\begin{equation}
\int_{\mathcal{M}_{g,n,\mathbf{L}}}\sum_{\left\{ \gamma,\delta\right\} \in\mathscr{C}_{1}}\mathsf{D}_{L_{1}l_{\gamma}l_{\delta}}\cdot(2\pi i)^{-3g+3-n}\langle\Sigma_{g,n,\mathbf{L}}|B_{6g-6+2n}|\Psi_{1}\rangle\cdots|\Psi_{n}\rangle\,,\label{eq:1st}
\end{equation}
 on the right hand side of (\ref{eq:intomega}) can be dealt with
in the same way. In this case, there can be topologically distinct
configurations of the pair $\{\gamma,\delta\}$ in $\mathscr{C}_{1}$
as depicted in Figure \ref{fig:Examples-of-}. They belong to different
mapping class group orbits. For each orbit, we take $X_{1}$ to be
the space of the combination $(\Sigma_{g,n,\mathbf{L}}$,$\gamma,\delta)$
where $(\gamma,\delta)$ is in the orbit and express the contribution
to (\ref{eq:1st}) by an integral over $X_{1}$. The amplitudes are
factorized as in Figure \ref{fig:Factorizations-of-the}. A formula
similar to (\ref{eq:2nd-4}) can be derived for each contribution. 

Putting everything together, we can see that (\ref{eq:intomega})
is transformed into
\begin{eqnarray}
L_{1}A_{g,n}^{I_{1}\cdots I_{n}} & = & \frac{1}{2}D^{I_{1}J^{\prime}J}G_{JI}G_{J^{\prime}I^{\prime}}\left[A_{g-1,n+1}^{II^{\prime}I_{2}\cdots I_{n}}+\sum_{\text{stable}}\frac{\varepsilon_{\mathcal{I}_{1}\mathcal{I}_{2}}}{(n_{1}-1)!(n_{2}-1)!}A_{g_{1},n_{1}}^{I\mathcal{I}_{1}}A_{g_{2},n_{2}}^{I^{\prime}\mathcal{I}_{2}}\right]\nonumber \\
 &  & +\sum_{a=2}^{n}\varepsilon_{a}(T^{I_{1}I_{a}J}+D^{I_{1}I_{a}J})G_{JI}A_{g,n-1}^{II_{2}\cdots\hat{I}_{a}\cdots I_{n}}\,,\label{eq:recursion}
\end{eqnarray}
which holds for $2g-2+n>1$. Here $\mathcal{I}_{1},\mathcal{I}_{2}$
are ordered sets of indices with $n_{1}-1,n_{2}-1$ elements respectively.
The sum $\sum_{\text{stable}}$ means the sum over $g_{1},g_{2},n_{1},n_{2},\mathcal{I}_{1},\mathcal{I}_{2}$
such that
\begin{eqnarray}
g_{1}+g_{2} & = & g\,,\nonumber \\
n_{1}+n_{2} & = & n+1\,,\nonumber \\
\mathcal{I}_{1}\cup\mathcal{I}_{2} & = & \left\{ I_{2},\cdots,I_{n}\right\} \,,\nonumber \\
\mathcal{I}_{1}\cap\mathcal{I}_{2} & = & \phi\,,\nonumber \\
2g_{1}-2+n_{1} & > & 0\,,\nonumber \\
2g_{2}-2+n_{2} & > & 0\,.\label{eq:stable}
\end{eqnarray}
$\varepsilon_{\mathcal{I}_{1}\mathcal{I}_{2}}=\pm1$ is the sign which
will appear when we change the order of the product $II^{\prime}I_{2}\cdots I_{n}$
to $I\mathcal{I}_{1}I^{\prime}\mathcal{I}_{2}$, if we regard the
indices as Grassmann numbers with Grassmannality of the corresponding
string state. 

Eq.(\ref{eq:recursion}) can be made more tractable by introducing
$A_{0,2}^{I_{1}I_{2}}$. Since we define the amplitudes for surfaces
with $2g-2+n>0$, amplitudes for $g=0,n=2$ do not exist. We here
introduce a fictitious amplitude
\[
A_{0,2}^{I_{1}I_{2}}=G^{I_{1}I_{2}}\,,
\]
where 
\[
G^{I_{1}I_{2}}\equiv\langle\varphi_{i_{1}}|\varphi_{i_{2}}\rangle\delta(L_{1}-L_{2})\delta_{\alpha_{1},-\alpha_{2}}\,,
\]
which satisfies
\[
G_{I_{1}I_{2}}G^{I_{2}I_{3}}=\delta_{I_{1}}^{\ I_{3}}=\delta_{i_{1},i_{3}}\delta_{\alpha_{1},\alpha_{3}}\delta(L_{1}-L_{3})\,.
\]
Taking this into account, we can turn (\ref{eq:recursion}) into 
\begin{eqnarray}
L_{1}A_{g,n}^{I_{1}\cdots I_{n}} & = & L_{1}G^{I_{1}I_{2}}\delta_{g,0}\delta_{n,2}\nonumber \\
 &  & +\frac{1}{2}D^{I_{1}J^{\prime}J}G_{JI}G_{J^{\prime}I^{\prime}}\left[A_{g-1,n+1}^{II^{\prime}I_{2}\cdots I_{n}}+\sum\hspace{0bp}^{\prime}\frac{\varepsilon_{\mathcal{I}_{1}\mathcal{I}_{2}}}{(n_{1}-1)!(n_{2}-1)!}A_{g_{1},n_{1}}^{I\mathcal{I}_{1}}A_{g_{2},n_{2}}^{I^{\prime}\mathcal{I}_{2}}\right]\nonumber \\
 &  & +\sum_{a=2}^{n}\varepsilon_{a}T^{I_{1}I_{a}J}G_{JI}A_{g,n-1}^{II_{2}\cdots\hat{I}_{a}\cdots I_{n}}\,,\label{eq:recursion2}
\end{eqnarray}
which holds for $2g-2+n>0$ or $g=0,n=2$. Here the summation $\sum\hspace{0bp}^{\prime}$
is over $g_{1},g_{2},n_{1},n_{2},\mathcal{I}_{1},\mathcal{I}_{2}$
such that
\begin{eqnarray*}
g_{1}+g_{2} & = & g\,,\\
n_{1}+n_{2} & = & n+1\,,\\
\mathcal{I}_{1}\cup\mathcal{I}_{2} & = & \left\{ I_{2},\cdots,I_{n}\right\} \,,\\
\mathcal{I}_{1}\cap\mathcal{I}_{2} & = & \phi\,,\\
2g_{1}-2+n_{1} & \geq & 0\,,\\
2g_{2}-2+n_{2} & \geq & 0\,.
\end{eqnarray*}

Let us check if (\ref{eq:recursion2}) is valid for $(g,n)=(0,2),(0,3),(1,1)$.
For $g=0,n=2$, (\ref{eq:recursion2}) becomes\footnote{Notice that $A_{0,1}^{I}$ does not exist. }
\begin{equation}
L_{1}A_{0,2}^{I_{1}I_{2}}=L_{1}G^{I_{1}I_{2}}\,.\label{eq:A02}
\end{equation}
The first term on the right hand side of (\ref{eq:recursion2}) is
introduced so that $A_{0,2}^{I_{1}I_{2}}=G^{I_{1}I_{2}}$ holds. For
$g=0,n=3$, we have
\begin{eqnarray*}
L_{1}A_{0,3}^{I_{1}I_{2}I_{3}} & = & \frac{1}{2}D^{I_{1}J^{\prime}J}G_{JI}G_{J^{\prime}I^{\prime}}\left[(-1)^{\left|I_{2}\right|\left|I_{3}\right|}G^{II_{2}}G^{I^{\prime}I_{3}}+(-1)^{\left|I_{3}\right|(\left|I_{2}\right|+\left|I_{2}\right|)}G^{II_{3}}G^{I^{\prime}I_{2}}\right]\\
 &  & +T^{I_{1}I_{2}J}G_{JI}G^{II_{3}}+(-1)^{\left|I_{2}\right|\left|I_{3}\right|}T^{I_{1}I_{3}J}G_{JI}G^{II_{2}}\\
 & = & (\frac{1}{2}\mathsf{D}_{L_{1}L_{2}L_{3}}+\frac{1}{2}\mathsf{D}_{L_{1}L_{3}L_{2}}+\mathsf{T}_{L_{1}L_{2}L_{3}}+\mathsf{T}_{L_{1}L_{3}L_{2}})\\
 &  & \quad\times\langle\Sigma_{0,3,(L_{1},L_{2},L_{3})}|B_{\alpha_{1}}^{1}B_{\alpha_{2}}^{2}B_{\alpha_{3}}^{3}|\varphi_{i_{1}}\rangle_{1}|\varphi_{i_{2}}\rangle_{2}|\varphi_{i_{3}}\rangle_{3}\,,
\end{eqnarray*}
where $\left|I\right|$ denotes the Grassmannality of $|\varphi_{i}\rangle$.
Substituting (\ref{eq:DTidentity}) into this, we obtain
\begin{equation}
L_{1}A_{0,3}^{I_{1}I_{2}I_{3}}=L_{1}\langle\Sigma_{0,3,(L_{1},L_{2},L_{3})}|B_{\alpha_{1}}^{1}B_{\alpha_{2}}^{2}B_{\alpha_{3}}^{3}|\varphi_{i_{1}}\rangle_{1}|\varphi_{i_{2}}\rangle_{2}|\varphi_{i_{3}}\rangle_{3}\,.\label{eq:A03}
\end{equation}
Notice that $\mathcal{M}_{0,3,\mathbf{L}}$ is a point and (\ref{eq:Agn-1})
implies
\begin{equation}
A_{0,3}^{I_{1}I_{2}I_{3}}=\langle\Sigma_{0,3,(L_{1},L_{2},L_{3})}|B_{\alpha_{1}}^{1}B_{\alpha_{2}}^{2}B_{\alpha_{3}}^{3}|\varphi_{i_{1}}\rangle_{1}|\varphi_{i_{2}}\rangle_{2}|\varphi_{i_{3}}\rangle_{3}\,,\label{eq:A03A}
\end{equation}
which is consistent with the above equation. For $g=1,n=1$, (\ref{eq:recursion2})
becomes
\begin{eqnarray}
L_{1}A_{1,1}^{I} & = & \frac{1}{2}D^{IJ^{\prime}J}G_{J^{\prime}J}\nonumber \\
 & = & -\frac{1}{2}\int dl_{\gamma}\mathsf{D}_{Ll_{\gamma}l_{\gamma}}\sum_{j}\left.\langle\Sigma_{0,3,(L,L_{2},L_{3})}|B_{\alpha}^{1}(b(\partial_{L_{2}})+b(\partial_{L_{3}}))b_{0}^{-(2)}P^{(2)}|\varphi_{i}\rangle_{1}|\varphi_{j}\rangle_{2}|\varphi_{j}^{c}\rangle_{3}\right|_{L_{2}=L_{3}=l_{\gamma}}\,.\label{eq:g1n1}
\end{eqnarray}
On the other hand, $A_{1,1}^{I}$ can be given as
\begin{eqnarray*}
A_{1,1}^{I} & = & \frac{1}{2}\int_{\mathcal{M}_{1,1,L}}\langle\Sigma_{1,1,L}|B_{\alpha}B_{2}|\varphi_{i}^{\alpha}\rangle(2\pi i)^{-1}\\
 & = & -\frac{1}{2}\int_{\mathcal{M}_{1,1,L}}\sum_{j}\left.\langle\Sigma_{0,3,(L,L_{2},L_{3})}|B_{\alpha}^{1}(b(\partial_{L_{2}})+b(\partial_{L_{3}}))b_{0}^{-(2)}\frac{1}{2\pi}e^{i\theta_{\gamma}(L_{0}^{(2)}-\bar{L}_{0}^{(29})}|\varphi_{i}\rangle_{1}|\varphi_{j}\rangle_{2}|\varphi_{j}^{c}\rangle_{3}\right|_{L_{2}=L_{3}=l_{\gamma}}dl_{\gamma}\wedge d\theta_{\gamma}\,.
\end{eqnarray*}
The integral on the last line can be unfolded by using the McShane
identity and we get exactly (\ref{eq:g1n1}). 

\subsection{The solution of the recursion relation}

The recursion relation (\ref{eq:recursion}) is derived from the properties
of the off-shell amplitudes $A_{g,n}^{I_{1}\cdots I_{n}}$ . Conversely,
$A_{g,n}^{I_{1}\cdots I_{n}}$ can be derived by solving the equation
(\ref{eq:recursion}). 

$A_{g,n}^{I_{1}\cdots I_{n}}$ is the order $g_{\mathrm{s}}^{2g-2+n}$
contribution to the $n$ point amplitude. (\ref{eq:recursion2}) can
be solved order by order in $g_{\mathrm{s}}$, because the right hand
side of (\ref{eq:recursion2}) consists of lower order terms compared
with the $A_{g,n}^{I_{1}\cdots I_{n}}$ on the left hand side. For
example, the equation for $g=0,n=3$ becomes (\ref{eq:A03}) and the
solution is (\ref{eq:A03A}) because $A_{0,3}^{I_{1}I_{2}I_{3}}$
is defined for $L_{1}>0$. (\ref{eq:recursion2}) can be solved in
the same way for general $g,n$. The solution is unique, because $A_{g,n}^{I_{1}\cdots I_{n}}$
is defined for $L_{1}>0$. This unique solution should coincide with
the $A_{g,n}^{I_{1}\cdots I_{n}}$ in (\ref{eq:Agn-1}). Therefore
the equation (\ref{eq:recursion2}) can be used to derive the off-shell
amplitudes of closed bosonic string theory. 

For later convenience, let us define the generating functional of
the off-shell amplitudes:
\begin{equation}
W_{A}[J]\equiv\sum_{g=0}^{\infty}\sum_{n=2}^{\infty}g_{\mathrm{s}}^{2g-2+n}\frac{1}{n!}J_{I_{n}}\cdots J_{I_{1}}A_{g,n}^{I_{1}\cdots I_{n}}\,.\label{eq:WA}
\end{equation}
$J_{I}$ is taken to have the same Grassmannality as that of $\phi^{I}$.
It is straightforward to show that the recursion relation (\ref{eq:recursion2})
is equivalent to the following identity:
\begin{eqnarray}
L\frac{\delta W_{A}[J]}{\delta J_{I}} & = & LJ_{I^{\prime}}G^{I^{\prime}I}\nonumber \\
 &  & +\frac{1}{2}g_{\mathrm{s}}D^{II^{\prime}I^{\prime\prime}}G_{I^{\prime\prime}K^{\prime\prime}}G_{I^{\prime}K^{\prime}}\left[\frac{\delta^{2}W_{A}[J]}{\delta J_{K^{\prime\prime}}\delta J_{K^{\prime}}}+\frac{\delta W_{A}[J]}{\delta J_{K^{\prime\prime}}}\frac{\delta W_{A}[J]}{\delta J_{K^{\prime}}}\right]\nonumber \\
 &  & +g_{\mathrm{s}}T^{II^{\prime}I^{\prime\prime}}G_{I^{\prime\prime}K^{\prime\prime}}J_{I^{\prime}}\frac{\delta W_{A}[J]}{\delta J_{K^{\prime\prime}}}(-1)^{\left|I\right|\left|I^{\prime}\right|}\,.\label{eq:WArecursion}
\end{eqnarray}
Here all the functional derivatives are the left derivatives.

\section{The Fokker-Planck formalism\label{sec:Fokker-Planck-Hamiltonian}}

In this section, we would like to develop the Fokker-Planck formalism
for the string theory from which we can derive the recursion relation
(\ref{eq:recursion2}) through the Schwinger-Dyson equation. 

\subsection{The Fokker-Planck formalism for conventional field theory}

Let $\phi(x)$ be a scalar field with action S{[}$\phi]$. The Euclidean
correlation functions are defined by 
\begin{equation}
\langle\phi(x_{1})\cdots\phi(x_{n})\rangle=\int[d\phi]P[\phi]\phi(x_{1})\cdots\phi(x_{n})\,,\label{eq:scalarcorr}
\end{equation}
where
\begin{equation}
P[\phi]=\frac{e^{-S[\phi]}}{\int[d\phi]e^{-S[\phi]}}\,.\label{eq:Pphi}
\end{equation}
In order to describe this quantum field theory, we consider a system
governed by the following Fokker-Planck equation:
\begin{equation}
-\frac{\partial}{\partial\tau}P[\phi,\tau]=H_{\mathrm{FP}}P[\phi,\tau]\,.\label{eq:FPeq}
\end{equation}
Here $H_{\mathrm{FP}}$ is the Fokker-Planck Hamiltonian defined by
\begin{equation}
H_{\mathrm{FP}}=-\int dx\frac{\delta}{\delta\phi(x)}\left(\frac{\delta}{\delta\phi(x)}+\frac{\delta S[\phi]}{\delta\phi(x)}\right)\,.\label{eq:HFPf}
\end{equation}
It is possible to show that for a solution of (\ref{eq:FPeq}) with
an appropriate initial condition, 
\[
\lim_{\tau\to\infty}P[\phi,\tau]=P[\phi]\,,
\]
holds. The Fokker-Planck equation with the Fokker-Planck Hamiltonian
(\ref{eq:HFPf}) appears in the context of stochastic quantization
\cite{Parisi:1980ys} where $\tau$ coincides with the fictitious
time. 

The Fokker-Planck Hamiltonian can be realized as an operator acting
on a Hilbert space. Let $\hat{\pi}(x),\hat{\phi}(x)$ be operators
satisfying the commutation relations
\begin{eqnarray*}
[\hat{\pi}(x),\hat{\phi}(y)] & = & \delta(x-y)\,,\\{}
[\hat{\pi}(x),\hat{\pi}(y)] & = & [\hat{\phi}(x),\hat{\phi}(y)]=0\,,
\end{eqnarray*}
and $|0\rangle,\langle0|$ be states satisfying 
\begin{eqnarray*}
\hat{\pi}(x)|0\rangle & = & \langle0|\hat{\phi}(x)=0\,,\\
\langle0|0\rangle & = & 1\,.
\end{eqnarray*}
Then 
\[
P[\phi,\tau]=\langle0|e^{-\tau\hat{H}_{\mathrm{FP}}}\prod_{x}\delta(\hat{\phi}(x)-\phi(x))|0\rangle\,,
\]
with 
\[
\hat{H}_{\mathrm{FP}}=-\int dx\left(\hat{\pi}(x)-\frac{\delta S}{\delta\phi(x)}[\hat{\phi}]\right)\hat{\pi}(x)\,,
\]
gives a solution to (\ref{eq:FPeq}) with initial condition $P[\phi,0]=\prod_{x}\delta(\phi(x))$.
Assuming that this is a good initial condition, we get 
\begin{equation}
P[\phi]=\lim_{\tau\to\infty}\langle0|e^{-\tau\hat{H}_{\mathrm{FP}}}\prod_{x}\delta(\hat{\phi}(x)-\phi(x))|0\rangle\,.\label{eq:Pphi2}
\end{equation}
The correlation function in (\ref{eq:scalarcorr}) is given by 
\[
\lim_{\tau\to\infty}\langle0|e^{-\tau\hat{H}_{\mathrm{FP}}}\hat{\phi}(x_{1})\cdots\hat{\phi}(x_{n})|0\rangle\,.
\]

In \cite{Ishibashi:1993nq}, a string field theory for the $(2,3)$
minimal string theory using this kind of operator formalism was proposed.
The string fields are labeled by the length $l$ of the string and
we define the operators $\hat{\pi}(l),\hat{\phi}(l)$ accordingly.
The Fokker-Planck Hamiltonian is given by
\begin{eqnarray*}
\hat{H}_{\mathrm{FP}} & = & 2\int_{0}^{\infty}dl_{1}\int_{0}^{\infty}dl_{2}\hat{\phi}(l_{1})w(l_{2})\hat{\pi}(l_{1}+l_{2})(l_{1}+l_{2})+\int_{0}^{\infty}dl_{1}\int_{0}^{\infty}dl_{2}w(l_{1}+l_{2})\hat{\pi}(l_{1})l_{1}\hat{\pi}((l_{2})l_{2}\\
 &  & +g_{\mathrm{s}}\int_{0}^{\infty}dl_{1}\int_{0}^{\infty}dl_{2}\hat{\phi}(l_{1})\hat{\phi}(l_{2})\hat{\pi}(l_{1}+l_{2})(l_{1}+l_{2})+g_{\mathrm{s}}\int_{0}^{\infty}dl_{1}\int_{0}^{\infty}dl_{2}\hat{\phi}(l_{1}+l_{2})\hat{\pi}(l_{1})l_{1}\hat{\pi}((l_{2})l_{2}\,,
\end{eqnarray*}
where $w(l)$ is the disk amplitude for the $(2,3)$ minimal string
theory\footnote{The correspondence between our notation and that in \cite{Ishibashi:1993nq}
is given by
\begin{eqnarray*}
\frac{1}{g_{\mathrm{s}}}w(l)+\hat{\phi}(l) & \leftrightarrow & \Psi^{\dagger}(l)\,,\\
\hat{\pi}(l) & \leftrightarrow & \Psi(l)\,.
\end{eqnarray*}
}. The correlation functions of the string fields are given by 
\begin{equation}
\lim_{\tau\to\infty}\langle0|e^{-\tau\hat{H}_{\mathrm{FP}}}\hat{\phi}(l_{1})\cdots\hat{\phi}(l_{n})|0\rangle\,.\label{eq:c=00003D0SFT}
\end{equation}
One can prove that the correlation functions thus defined coincide
with the loop amplitudes of the $(2,3)$ minimal string theory, in
the following way. In order for the limit (\ref{eq:c=00003D0SFT})
to exist, 
\begin{equation}
\lim_{\tau\to\infty}\partial_{\tau}\langle0|e^{-\tau\hat{H}_{\mathrm{FP}}}\hat{\phi}(l_{1})\cdots\hat{\phi}(l_{n})|0\rangle=-\lim_{\tau\to\infty}\langle0|e^{-\tau\hat{H}_{\mathrm{FP}}}\hat{H}_{\mathrm{FP}}\hat{\phi}(l_{1})\cdots\hat{\phi}(l_{n})|0\rangle=0\,,\label{eq:c=00003D0SD}
\end{equation}
should hold. (\ref{eq:c=00003D0SD}) yields the Schwinger-Dyson equation
satisfied by the correlation functions of the minimal string theory.
It can be shown that the loop equation of the minimal string theory
is equivalent to this Schwinger Dyson equation. Moreover this string
field theory can be derived from the stochastic quantization of the
one matrix model \cite{Jevicki1994}. The Fokker-Planck formalism
was applied to construct string field theories for general $(p,q)$
minimal string theories in \cite{Ishibashi:1993nqz,Ikehara1994,Ikehara1995}. 

\subsection{The Fokker-Planck Hamiltonian for closed bosonic strings\label{subsec:Fokker-Planck-Hamiltonian-for}}

In \cite{Eynard2007}, it was shown that Mirzakhani's recursion relation
(\ref{eq:Mirzakhani}) is a special case of random matrix recursion
relations. In \cite{Saad2019a}, (\ref{eq:Mirzakhani}) is identified
with a limit $p\to\infty$ of the loop equation of the $(2,p)$ minimal
string theory. Since $(2,p)$ minimal string theory is a close cousin
of the $(2,3)$ one, it is possible to develop the Fokker-Planck formalism
of string field theory corresponding to (\ref{eq:Mirzakhani}). The
recursion relation in (\ref{eq:recursion}) is a (not so close) cousin
of (\ref{eq:Mirzakhani}), it is conceivable that the same approach
is applicable to this equation. In this subsection, we would like
to show that this is the case. 

We introduce operators $\hat{\phi}^{I},\hat{\pi}_{I}$ which satisfy
the commutation relations
\begin{eqnarray*}
[\hat{\pi}_{I},\hat{\phi}^{K}] & = & \delta_{I}^{\ K}\,,\\{}
[\hat{\pi}_{I},\hat{\pi}_{K}] & = & [\hat{\phi}^{I},\hat{\phi}^{K}]=0\,.
\end{eqnarray*}
Here we define 
\[
[X^{I},Y^{K}]\equiv X^{I}Y^{K}-(-1)^{|I||K|}Y^{K}X^{I}\,.
\]
 Let $|0\rrangle,\llangle0|$ be states which satisfy
\begin{equation}
\llangle0|\hat{\phi}^{I}=\hat{\pi}_{I}|0\rrangle=0\,.\label{eq:0braket}
\end{equation}
We define the correlation functions of $\phi^{I}$'s as
\begin{equation}
\llangle\phi^{I_{1}}\cdots\phi^{I_{n}}\rrangle\equiv\lim_{\tau\to\infty}\llangle0|e^{-\tau\hat{H}}\hat{\phi}^{I_{1}}\cdots\hat{\phi}^{I_{n}}|0\rrangle\,,\label{eq:corr}
\end{equation}
with Hamiltonian
\begin{eqnarray}
\hat{H} & = & -L\hat{\pi}_{I}\hat{\pi}_{I^{\prime}}G^{I^{\prime}I}+L\hat{\phi}^{I}\hat{\pi}_{I}\nonumber \\
 &  & -\frac{1}{2}g_{\mathrm{s}}D^{II^{\prime}I^{\prime\prime}}G_{I^{\prime\prime}K^{\prime\prime}}G_{I^{\prime}K^{\prime}}\hat{\phi}^{K^{\prime\prime}}\hat{\phi}^{K^{\prime}}\hat{\pi}_{I}\nonumber \\
 &  & -g_{\mathrm{s}}T^{II^{\prime}I^{\prime\prime}}G_{I^{\prime\prime}K^{\prime\prime}}\hat{\phi}^{K^{\prime\prime}}\hat{\pi}_{I^{\prime}}\hat{\pi}_{I}\,.\label{eq:FPH}
\end{eqnarray}
As we will see, the right hand side of (\ref{eq:corr}) can be calculated
perturbatively with respect to $g_{\mathrm{s}}$. We define the connected
correlation functions $\llangle\phi^{I_{1}}\cdots\phi^{I_{n}}\rrangle^{\mathrm{c}}$
in the usual way and they can be expanded as
\[
\llangle\phi^{I_{1}}\cdots\phi^{I_{n}}\rrangle^{\mathrm{c}}=\sum_{g=0}^{\infty}g_{\mathrm{s}}^{2g-2+n}\llangle\phi^{I_{1}}\cdots\phi^{I_{n}}\rrangle_{g}^{\mathrm{c}}\,.
\]
 It is possible to show that 
\begin{equation}
\llangle\phi^{I_{1}}\cdots\phi^{I_{n}}\rrangle_{g}^{\mathrm{c}}=A_{g,n}^{I_{1}\cdots I_{n}}\,,\label{eq:corr=00003DA}
\end{equation}
holds.

In order to prove (\ref{eq:corr=00003DA}), we define the generating
functional $W[J]$ of the connected correlation functions
\begin{equation}
W[J]=\sum_{n=2}^{\infty}\frac{1}{n!}J_{I_{n}}\cdots J_{I_{1}}\llangle\phi^{I_{1}}\cdots\phi^{I_{n}}\rrangle^{\mathrm{c}}\,.\label{eq:WJ}
\end{equation}
 such that
\[
e^{W[J]}=\lim_{\tau\to\infty}\llangle0|e^{-\tau\hat{H}}e^{J_{I}\hat{\phi}^{I}}|0\rrangle\,.
\]
Since the limit $\tau\to\infty$ of $\llangle0|e^{-\tau\hat{H}}\hat{\phi}^{I_{1}}\cdots\hat{\phi}^{I_{n}}|0\rrangle$
exists, we have\footnote{(\ref{eq:SD}) can be proved perturbatively in $g_{\mathrm{s}}$.}
\begin{equation}
0=\lim_{\tau\to\infty}\partial_{\tau}\llangle0|e^{-\tau\hat{H}}e^{J_{I}\hat{\phi}^{I}}|0\rrangle=-\lim_{\tau\to\infty}\llangle0|e^{-\tau\hat{H}}\hat{H}e^{J_{I}\hat{\phi}^{I}}|0\rrangle\,.\label{eq:SD}
\end{equation}
 Using (\ref{eq:0braket}), we get the following equation from (\ref{eq:SD}):
\begin{eqnarray}
0 & = & J_{I}\left\{ L\frac{\delta W[J]}{\delta J_{I}}-LJ_{I^{\prime}}G^{I^{\prime}I}\right.\nonumber \\
 &  & \hphantom{J_{I}\quad}-\frac{1}{2}g_{\mathrm{s}}D^{II^{\prime}I^{\prime\prime}}G_{I^{\prime\prime}K^{\prime\prime}}G_{I^{\prime}K^{\prime}}\left[\frac{\delta^{2}W[J]}{\delta J_{K^{\prime\prime}}\delta J_{K^{\prime}}}+\frac{\delta W[J]}{\delta J_{K^{\prime\prime}}}\frac{\delta W[J]}{\delta J_{K^{\prime}}}\right]\nonumber \\
 &  & \hphantom{J_{I}\quad}\left.-g_{\mathrm{s}}T^{II^{\prime}I^{\prime\prime}}G_{I^{\prime\prime}K^{\prime\prime}}J_{I^{\prime}}\frac{\delta W[J]}{\delta J_{K^{\prime\prime}}}(-1)^{\left|I\right|\left|I^{\prime}\right|}\right\} \label{eq:Wrecursion}
\end{eqnarray}
It is possible to solve (\ref{eq:Wrecursion}) order by order in $g_{\mathrm{s}}$
and obtain $\llangle\phi^{I_{1}}\cdots\phi^{I_{n}}\rrangle_{g}^{\mathrm{c}}$.
For example, at $\mathcal{O}(g_{\mathrm{s}}^{0})$, (\ref{eq:Wrecursion})
implies
\[
J_{I}J_{I^{\prime}}(L+L^{\prime})\left(\llangle\phi^{I^{\prime}}\phi^{I}\rrangle_{0}^{\mathrm{c}}-G^{I^{\prime}I}\right)=0\,.
\]
Since $\llangle\phi^{I^{\prime}}\phi^{I}\rrangle_{0}^{\mathrm{c}}$
is defined for $L,L^{\prime}>0$, we obtain the unique solution 
\begin{equation}
\llangle\phi^{I^{\prime}}\phi^{I}\rrangle_{0}^{\mathrm{c}}=G^{I^{\prime}I}\,.\label{eq:propagator}
\end{equation}
In general, (\ref{eq:Wrecursion}) implies an equation in which $(L_{1}+\cdots+L_{n})\llangle\phi^{I_{1}}\cdots\phi^{I_{n}}\rrangle_{g}^{\mathrm{c}}$
is expressed in terms of lower order correlation functions. Since
$\llangle\phi^{I_{1}}\cdots\phi^{I_{n}}\rrangle_{g}^{c}$ is defined
for $L_{1},\cdots,L_{n}>0$, one can solve the equation and the solution
is unique. Hence all the coefficients of the expansion (\ref{eq:WJ})
is uniquely fixed by (\ref{eq:Wrecursion}). On the other hand, 
\begin{equation}
W[J]=W_{A}[J]\,,\label{eq:W=00003DWA}
\end{equation}
yields a solution to (\ref{eq:Wrecursion}) because $W_{A}[J]$ satisfies
(\ref{eq:WArecursion}). Since the solution of (\ref{eq:Wrecursion})
should be unique, we obtain (\ref{eq:corr=00003DA}). 

(\ref{eq:W=00003DWA}) implies that $W[J]$ satisfies the equation
(\ref{eq:WArecursion}), which can be expressed as 
\begin{equation}
\lim_{\tau\to\infty}\llangle0|e^{-\tau\hat{H}}\hat{\mathcal{T}}^{I}e^{J_{I}\hat{\phi}^{I}}|0\rrangle=0\,,\label{eq:SD2}
\end{equation}
 in the Fokker-Planck formalism. Here
\begin{eqnarray}
\hat{\mathcal{T}}^{I} & \equiv & -L\hat{\pi}_{I^{\prime}}G^{II^{\prime}}+L\hat{\phi}^{I}\nonumber \\
 &  & -\frac{1}{2}g_{\mathrm{s}}D^{II^{\prime}I^{\prime\prime}}G_{I^{\prime\prime}K^{\prime\prime}}G_{I^{\prime}K^{\prime}}\hat{\phi}^{K^{\prime\prime}}\hat{\phi}^{K^{\prime}}\nonumber \\
 &  & -g_{\mathrm{s}}T^{II^{\prime}I^{\prime\prime}}G_{I^{\prime\prime}K^{\prime\prime}}\hat{\phi}^{K^{\prime\prime}}\hat{\pi}_{I^{\prime}}\,,\label{eq:TI}
\end{eqnarray}
and we have 
\[
\hat{H}=\hat{\mathcal{T}}^{I}\hat{\pi}_{I}\,.
\]
Since every ket vector is expressed as a linear combination of states
of the form
\[
\llangle0|\hat{\pi}^{I_{1}}\cdots\hat{\pi}^{I_{n}}\,,
\]
(\ref{eq:corr}) means that $\lim_{\tau\to\infty}\llangle0|e^{-\tau\hat{H}}$
is expressed as
\[
\lim_{\tau\to\infty}\llangle0|e^{-\tau\hat{H}}=\sum_{n=0}^{\infty}\frac{1}{n!}\llangle\phi^{I_{1}}\cdots\phi^{I_{n}}\rrangle\llangle0|\hat{\pi}_{I_{n}}\cdots\hat{\pi}_{I_{1}}\,.
\]
In the same way, we can deduce from (\ref{eq:SD}), (\ref{eq:SD2})
\begin{eqnarray}
\left[\lim_{\tau\to\infty}\llangle0|e^{-\tau\hat{H}}\right]\hat{H} & = & 0\,,\nonumber \\
\left[\lim_{\tau\to\infty}\llangle0|e^{-\tau\hat{H}}\right]\hat{\mathcal{T}}^{I} & = & 0\,.\label{eq:SD3}
\end{eqnarray}

\subsection{String field action $S[\phi^{I}]$\label{subsec:String-field-action}}

In the case of conventional field theory, the Fokker-Planck formalism
is an alternative to the path integral formalism. Let us discuss if
the theory we have can be formulated using a path integral with action
$S[\phi^{I}]$. It is possible to define the weight $P[\phi^{I}]$
following (\ref{eq:Pphi2}):
\[
P[\phi^{I}]=\frac{e^{-S[\phi^{I}]}}{\int[d\phi^{I}]e^{-S[\phi^{I}]}}=\lim_{\tau\to\infty}\llangle0|e^{-\tau\hat{H}}\prod_{I}\delta(\hat{\phi}^{I}-\phi^{I})|0\rrangle\,.
\]
From (\ref{eq:SD3}), we obtain an equation for $S[\phi^{I}]$:
\begin{eqnarray}
 &  & [LG^{IJ}+g_{\mathrm{s}}T^{IJI^{\prime}}G_{I^{\prime}J^{\prime}}\phi^{J^{\prime}}]\frac{\delta S}{\delta\phi^{J}}\nonumber \\
 &  & \quad=L\phi^{I}-\frac{1}{2}g_{\mathrm{s}}D^{II^{\prime}I^{\prime\prime}}G_{I^{\prime}J^{\prime}}G_{I^{\prime\prime}J^{\prime\prime}}\phi^{J^{\prime\prime}}\phi^{J^{\prime}}+g_{\mathrm{s}}T^{II^{\prime}I^{\prime\prime}}G_{I^{\prime}I^{\prime\prime}}\,.\label{eq:Seq}
\end{eqnarray}
 Using (\ref{eq:DTidentity}), the last term on the right hand side
of (\ref{eq:Seq}) is expressed as
\begin{eqnarray*}
T^{II^{\prime}I^{\prime\prime}}G_{I^{\prime}I^{\prime\prime}} & = & \frac{1}{2}L\int_{0}^{\infty}dl_{\gamma}\sum_{j}\left.\langle\Sigma_{0,3,(L,L_{2},L_{3})}|B_{\alpha}^{1}(b(\partial_{L_{2}})+b(\partial_{L_{3}}))b_{0}^{-(2)}P^{(2)}|\varphi_{i}\rangle_{1}|\varphi_{j}\rangle_{2}|\varphi_{j}^{c}\rangle_{3}\right|_{L_{2}=L_{3}=l_{\gamma}}\\
 &  & -\frac{1}{2}\int_{0}^{\infty}dl_{\gamma}\mathsf{D}_{Ll_{\gamma}l_{\gamma}}\sum_{j}\left.\langle\Sigma_{0,3,(L,L_{2},L_{3})}|B_{\alpha}^{1}(b(\partial_{L_{2}})+b(\partial_{L_{3}}))b_{0}^{-(2)}P^{(2)}|\varphi_{i}\rangle_{1}|\varphi_{j}\rangle_{2}|\varphi_{j}^{c}\rangle_{3}\right|_{L_{2}=L_{3}=l_{\gamma}}\,.
\end{eqnarray*}
The integrand of the first term on the right hand side coincides with
that of $A_{1,1}^{I}$, but the integration region includes infinitely
many fundamental domains of the mapping class group. The second term
is equal to $-LA_{1,1}^{I}$. Therefore the last term on the right
hand side of (\ref{eq:Seq}) may be given as
\[
L\left[\infty\times A_{1,1}^{I}-A_{1,1}^{I}\right]\,,
\]
and divergent. Hence the equation (\ref{eq:Seq}) is not well-defined.

Still, (\ref{eq:Seq}) can be solved formally order by order in $g_{\mathrm{s}}$
and we get
\begin{equation}
S[\phi^{I}]=\frac{1}{2}G_{IJ}\phi^{I}\phi^{J}-\frac{g_{\mathrm{s}}}{6}A_{0,3}^{II^{\prime}I^{\prime\prime}}G_{IJ}G_{I^{\prime}J^{\prime}}G_{I^{\prime\prime}J^{\prime\prime}}\phi^{J^{\prime\prime}}\phi^{J^{\prime}}\phi^{J}+\frac{g_{\mathrm{s}}}{L}T^{II^{\prime}I^{\prime\prime}}G_{I^{\prime}I^{\prime\prime}}G_{IJ}\phi^{J}+\mathcal{O}(g_{\mathrm{s}}^{2})\,.\label{eq:Sphi}
\end{equation}
This should be the action which yields the off-shell amplitudes with
the propagator $G^{IJ}$. Let us compute the one loop one point function
using the path integral formalism. The contribution from the three
string vertex becomes
\[
\frac{g_{\mathrm{s}}}{2}\int dl_{\gamma}\sum_{j}\left.\langle\Sigma_{0,3,(L,L_{2},L_{3})}|B_{\alpha}^{1}(b(\partial_{L_{2}})+b(\partial_{L_{3}}))b_{0}^{-(2)}P^{(2)}|\varphi_{i}\rangle_{1}|\varphi_{j}\rangle_{2}|\varphi_{j}^{c}\rangle_{3}\right|_{L_{2}=L_{3}=l_{\gamma}}\,,
\]
and diverges. The contribution from divergent term $\frac{g_{\mathrm{s}}}{L}T^{II^{\prime}I^{\prime\prime}}G_{I^{\prime}I^{\prime\prime}}G_{IJ}\phi^{J}$
in the action cancels this divergence and we get the correct answer
$g_{\mathrm{s}}A_{1,1}^{I}$. This pattern seems to continue forever.
If one computes the four point amplitude using the three string vertex
in (\ref{eq:Sphi}), one gets a divergent result. The four string
vertex cancels the divergence and make the amplitude finite. 

Therefore, the Fokker-Planck formalism is necessary for a well-defined
formulation of the theory in our setup. On the other hand, the formally
defined action (\ref{eq:Sphi}) will be useful in studying various
aspects of our formulation. 

\subsection{SFT notation}

In order to discuss various properties of the theory, it is more convenient
to express the Fokker-Planck Hamiltonian (\ref{eq:FPH}) in terms
of the variables in the Hilbert space of strings. Let us define 

\begin{eqnarray}
|\phi^{\alpha}(L)\rangle & \equiv & \sum_{i}\hat{\phi}^{I}|\varphi_{i}^{c}\rangle\,,\label{eq:phialpha}\\
|\pi_{\alpha}(L)\rangle & \equiv & \sum_{i}|\varphi_{i}\rangle\hat{\pi}_{I}\,,\label{eq:pialpha}
\end{eqnarray}
The string fields $|\phi^{\alpha}(L)\rangle,|\pi_{\alpha}(L)\rangle$
are taken to satisfy
\begin{eqnarray}
 &  & |\pi_{+}(L)\rangle,|\varphi^{-}(L)\rangle\in\mathcal{H}_{0}\,,\nonumber \\
 &  & |\pi_{-}(L)\rangle,|\varphi^{+}(L)\rangle\in\mathcal{H}_{0}^{c}\,,\label{eq:H_0}
\end{eqnarray}
where $\mathcal{H}_{0}^{c}$ consists of the states $|\Psi\rangle$
satisfying
\[
c_{0}^{-}|\Psi\rangle=(L_{0}-\bar{L}_{0})|\Psi\rangle=0\,,
\]
where $c_{0}^{\pm}=c_{0}\pm\bar{c}_{0}$. We also impose the reality
condition \cite{Zwiebach1993,Sen2016}
\begin{eqnarray}
|\phi^{+}(L)\rangle^{\dagger} & = & \langle\phi^{+}(L)|\,,\nonumber \\
|\phi^{-}(L)\rangle^{\dagger} & = & -\langle\phi^{-}(L)|\,.\label{eq:reality}
\end{eqnarray}
The reality condition for $|\pi_{\alpha}(L)\rangle$ will not be so
simple, as is always the case in the Fokker-Planck formalism. Conditions
(\ref{eq:H_0}) and (\ref{eq:reality}) have been implicitly assumed
in the previous subsection. 

Notice that $|\phi^{\alpha}(L)\rangle,|\pi_{\alpha}(L)\rangle$ are
Grassmann even. They satisfy the canonical commutation relation
\begin{equation}
[|\pi_{\alpha}(L)\rangle_{1},|\phi^{\alpha^{\prime}}(L^{\prime})\rangle_{2}]=\delta_{\alpha}^{\alpha^{\prime}}\delta(L-L^{\prime})P_{\alpha}^{(1)}|R_{12}\rangle\,,\label{eq:ccr}
\end{equation}
where
\[
|R_{12}\rangle=\sum_{i}|\varphi_{i}\rangle_{1}|\varphi_{i}^{c}\rangle_{2}=\sum_{i}|\varphi_{i}\rangle_{2}|\varphi_{i}^{c}\rangle_{1}=|R_{21}\rangle=|R\rangle\,,
\]
is the reflector and
\[
P_{\alpha}=\begin{cases}
\frac{1}{2}b_{0}^{-}c_{0}^{-}P & \alpha=+\\
\frac{1}{2}c_{0}^{-}b_{0}^{-}P & \alpha=-
\end{cases}\,.
\]
 The states $|0\rrangle,\llangle0|$ satisfy
\[
|\pi_{\alpha}(L)\rangle\left|0\vphantom{\phi^{I}}\right\rrangle =\left\llangle 0\vphantom{\phi^{I}}\right||\phi^{\alpha}(L)\rangle=0\,.
\]

In terms of these string fields, the Fokker-Planck Hamiltonian (\ref{eq:FPH})
is expressed as
\begin{eqnarray}
\hat{H} & = & \int_{0}^{\infty}dLL\left[\langle R|\phi^{\alpha}(L)\rangle|\pi_{\alpha}(L)\rangle-\langle R|\pi_{\alpha}(L)\rangle|\pi_{-\alpha}(L)\rangle\right]\nonumber \\
 &  & -g_{\mathrm{s}}\int dL_{1}dL_{2}dL_{3}\langle T_{L_{2}L_{3}L_{1}}|B_{-\alpha_{1}}^{1}B_{\alpha_{2}}^{2}B_{\alpha_{3}}^{3}|\phi^{\alpha_{1}}(L_{1})\rangle_{1}|\pi_{\alpha_{2}}(L_{2})\rangle_{2}|\pi_{\alpha_{3}}(L_{3})\rangle_{3}\nonumber \\
 &  & -\frac{1}{2}g_{\mathrm{s}}\int dL_{1}dL_{2}dL_{3}\langle D_{L_{3}L_{1}L_{2}}|B_{-\alpha_{1}}^{1}B_{-\alpha_{2}}^{2}B_{\alpha_{3}}^{3}|\phi^{\alpha_{1}}(L_{1})\rangle_{1}|\phi^{\alpha_{2}}(L_{2})\rangle_{2}|\pi_{\alpha_{3}}(L_{3})\rangle_{3}\,,\label{eq:SFTFKH}
\end{eqnarray}
where
\begin{eqnarray*}
\langle T_{L_{2}L_{3}L_{1}}| & \equiv & \mathsf{T}_{L_{2}L_{3}L_{1}}\langle\Sigma_{0,3,(L_{1},L_{2},L_{3})}|\,,\\
\langle D_{L_{3}L_{1}L_{2}}| & \equiv & \mathsf{D}_{L_{3}L_{1}L_{2}}\langle\Sigma_{0,3,(L_{1},L_{2},L_{3})}|\,,
\end{eqnarray*}
and the sum over repeated indices $\alpha_{1},\alpha_{2},\alpha_{3}$
is understood. $\phi^{I}$ and $\pi_{I}$ are given by
\begin{eqnarray*}
\hat{\phi}^{I} & = & \langle\varphi_{i}|\phi^{\alpha}(L)\rangle\,,\\
\hat{\pi}_{I} & = & \langle\varphi_{i}^{c}|\pi_{\alpha}(L)\rangle\,,
\end{eqnarray*}
and the correlation functions of $|\phi^{\alpha}(L)\rangle$ are expressed
as
\begin{equation}
\llangle|\phi^{\alpha_{1}}(L_{1})\rangle_{1}\cdots|\phi^{\alpha_{n}}(L_{n})\rangle_{n}\rrangle_{g}^{\mathrm{c}}=\int_{\mathcal{M}_{g,n,\mathbf{L}}}{}_{1^{\prime}\cdots n^{\prime}}\langle\Sigma_{g,n,\mathbf{L}}|B_{6g-6+2n}B_{\alpha_{1}}^{1^{\prime}}\cdots B_{\alpha_{n}}^{n^{\prime}}P_{-\alpha_{1}}^{(1)}|R_{1^{\prime}1}\rangle\cdots P_{-\alpha_{n}}^{(n)}|R_{n^{\prime}n}\rangle\,.\label{eq:SFTamp}
\end{equation}

The correlation functions can be calculated perturbatively. The Euclidean
action corresponding to the Fokker-Planck Hamiltonian (\ref{eq:SFTFKH})
becomes
\begin{equation}
I=\int_{0}^{\infty}d\tau\left[-\int_{0}^{\infty}dL\langle R|\pi_{\alpha}(\tau,L)\rangle\frac{\partial}{\partial\tau}|\phi^{\alpha}(\tau,L)\rangle+H(\tau)\right]\,,\label{eq:Eucaction}
\end{equation}
where 
\begin{eqnarray*}
H(\tau) & = & \int_{0}^{\infty}dLL\left[\langle R|\phi^{\alpha}(\tau,L)\rangle|\pi_{\alpha}(\tau,L)\rangle-\langle R|\pi_{\alpha}(\tau,L)\rangle|\pi_{-\alpha}(\tau,L)\rangle\right]\\
 &  & -g_{\mathrm{s}}\int dL_{1}dL_{2}dL_{3}\langle T_{L_{2}L_{3}L_{1}}|B_{-\alpha_{1}}^{1}B_{\alpha_{2}}^{2}B_{\alpha_{3}}^{3}|\phi^{\alpha_{1}}(\tau,L_{1})\rangle_{1}|\pi_{\alpha_{2}}(\tau,L_{2})\rangle_{2}|\pi_{\alpha_{3}}(\tau,L_{3})\rangle_{3}\\
 &  & -\frac{1}{2}g_{\mathrm{s}}\int dL_{1}dL_{2}dL_{3}\langle D_{L_{3}L_{1}L_{2}}|B_{-\alpha_{1}}^{1}B_{-\alpha_{2}}^{2}B_{\alpha_{3}}^{3}|\phi^{\alpha_{1}}(\tau,L_{1})\rangle_{1}|\phi^{\alpha_{2}}(\tau,L_{2})\rangle_{2}|\pi_{\alpha_{3}}(\tau,L_{3})\rangle_{3}\,.
\end{eqnarray*}
 $|\phi^{\alpha}(\tau,L)\rangle,|\pi_{\alpha}(\tau,L)\rangle$ satisfy
the boundary conditions 
\[
\lim_{\tau\to\infty}|\phi^{\alpha}(\tau,L)\rangle=|\pi_{\alpha}(0,L)\rangle=0\,.
\]
The correlation functions are expressed as 
\begin{equation}
\llangle|\phi^{\alpha_{1}}(L_{1})\rangle\cdots|\phi^{\alpha_{n}}(L_{n})\rangle\rrangle=\frac{\int[d\pi d\phi]e^{-I}|\phi^{\alpha_{1}}(0,L_{1})\rangle\cdots|\phi^{\alpha_{n}}(0,L_{n})\rangle}{\int[d\pi d\phi]e^{-I}}\,,\label{eq:Icorr}
\end{equation}
using the path integral. To develop the perturbation theory, we decompose
the action as $I=I_{0}+g_{\mathrm{s}}V$, and we get propagators by
Wick's theorem:
\begin{eqnarray*}
|\begC1{\phi^{\alpha}}\conC{(\tau,L)\rangle_{1}|}\endC1{\phi^{\alpha^{\prime}}}(\tau^{\prime},L^{\prime})\rangle_{2} & = & e^{-|\tau-\tau^{\prime}|L}\delta(L-L^{\prime})\delta_{\alpha.-\alpha^{\prime}}P_{-\alpha}^{(1)}|R_{12}\rangle\,,\\
|\begC1{\pi_{\alpha}}\conC{(\tau,L)\rangle_{1}|}\endC1{\pi_{\alpha^{\prime}}}(\tau^{\prime},L^{\prime})\rangle_{2} & = & 0\,,\\
|\begC1{\pi_{\alpha}}\conC{(\tau,L)\rangle_{1}|}\endC1{\phi^{\alpha^{\prime}}}(\tau^{\prime},L^{\prime})\rangle_{2} & = & e^{-(\tau-\tau^{\prime})L}\theta(\tau-\tau^{\prime})\delta(L-L^{\prime})\delta_{\alpha}^{\alpha^{\prime}}P_{\alpha}^{(1)}|R_{12}\rangle\,.
\end{eqnarray*}
With the propagators and the vertex, it is straightforward to compute
(\ref{eq:Icorr}). By construction, the results are given by the integral
\[
\int_{\mathcal{M}_{g,n,\mathbf{L}}}{}_{1^{\prime}\cdots n^{\prime}}\langle\Sigma_{g,n,\mathbf{L}}|B_{6g-6+2n}B_{\alpha_{1}}^{1}\cdots B_{\alpha_{n}}^{n}P_{-\alpha_{1}}^{(1)}|R_{1^{\prime}1}\rangle\cdots P_{-\alpha_{n}}^{(n)}|R_{n^{\prime}n}\rangle\,,
\]
unfolded by Mirzakhani's method. The integrations can be done taking
care of the contributions from the boundaries of the moduli space
appropriately. 

\section{BRST invariant formulation\label{sec:BRST-invariant-formulation}}

With the Fokker-Planck formalism developed in the previous section,
one can express the off-shell amplitudes of the bosonic string theory.
In order to describe the string theory, we need the BRST symmetry
on the worldsheet to specify which states of strings are physical.
Unfortunately, the Fokker-Planck Hamiltonian (\ref{eq:SFTFKH}) and
the action (\ref{eq:Eucaction}) are not invariant under the BRST
symmetry, although the amplitudes are. We will modify the action (\ref{eq:Eucaction})
so that the BRST symmetry becomes manifest in our formalism.

\subsection{BRST transformation}

As is proved in appendix \ref{sec:BRST-identity}, $\langle\Sigma_{g,n,\mathbf{L}}|$
satisfies the BRST identity
\begin{eqnarray}
 &  & \langle\Sigma_{g,n,\mathbf{L}}|B_{6g-6+2n}B_{\alpha_{1}}^{1}\cdots B_{\alpha_{n}}^{n}\sum_{a=1}^{n}Q^{(a)}\nonumber \\
 &  & \quad=d\left(\langle\Sigma_{g,n,\mathbf{L}}|B_{6g-7+2n}B_{\alpha_{1}}^{1}\cdots B_{\alpha_{n}}^{n}\right)\nonumber \\
 &  & \hphantom{\quad=}-\sum_{a=1}^{n}\delta_{\alpha_{a},-}\partial_{L_{a}}\left(\langle\Sigma_{g,n,\mathbf{L}}|B_{6g-6+2n}B_{\alpha_{1}}^{1}\cdots b_{0}^{-(a)}P^{(a)}\cdots B_{\alpha_{n}}^{n}\right)\,.\label{eq:BRST1}
\end{eqnarray}
Integrating this over $\mathcal{M}_{g,n,\mathbf{L}}$, we obtain
\begin{eqnarray}
 &  & \sum_{a=1}^{n}P_{-\alpha_{a}}^{(a)}Q^{(a)}\llangle|\phi^{\alpha_{1}}(L_{1})\rangle_{1}\cdots|\phi^{\alpha_{n}}(L_{n})\rangle_{n}\rrangle_{g}^{\mathrm{c}}\nonumber \\
 &  & \quad=\sum_{a=1}^{n}\delta_{\alpha_{a},-}\llangle|\phi^{\alpha_{1}}(L_{1})\rangle_{1}\cdots b_{0}^{-(a)}P^{(a)}\partial_{L_{a}}|\phi^{+}(L_{a})\rangle_{a}\cdots|\phi^{\alpha_{n}}(L_{n})\rangle_{n}\rrangle_{g}^{\mathrm{c}}\,.\label{eq:BRSTcorr}
\end{eqnarray}
(\ref{eq:BRSTcorr}) implies that the correlation functions of $|\phi^{\alpha}(L)\rangle$
is invariant under 
\begin{eqnarray}
\delta_{\epsilon}|\phi^{+}(L)\rangle & = & \epsilon P_{-}Q|\phi^{+}(L)\rangle\,,\nonumber \\
\delta_{\epsilon}|\phi^{-}(L)\rangle & = & \epsilon Q|\phi^{-}(L)\rangle-\epsilon b_{0}^{-}P\partial_{L}|\phi^{+}(L)\rangle\,,\label{eq:BRSTphi}
\end{eqnarray}
with a Grassmann odd parameter $\epsilon$. This can be identified
with the BRST transformation of $|\phi^{\alpha}(L)\rangle$. It is
easily checked that the two point function
\[
\llangle|\phi^{\alpha_{1}}(L_{1})\rangle_{1}|\phi^{\alpha_{2}}(L_{2})\rangle_{2}\rrangle_{0}^{\mathrm{c}}=\delta(L_{1}-L_{2})\delta_{\alpha_{1},-\alpha_{2}}P_{-\alpha_{1}}^{(1)}|R_{12}\rangle\,,
\]
is also BRST invariant. The transformation of the $|\pi_{\alpha}(L)\rangle$
is fixed by requiring that the commutation relation (\ref{eq:ccr})
is invariant and we obtain
\begin{eqnarray}
\delta_{\epsilon}|\pi_{+}(L)\rangle & = & \epsilon Q|\pi_{+}(L)\rangle-\epsilon b_{0}^{-}P\partial_{L}|\pi_{-}(L)\rangle\,,\nonumber \\
\delta_{\epsilon}|\pi_{-}(L)\rangle & = & \epsilon P_{-}Q|\pi_{-}(L)\rangle\,.\label{eq:BRSTpi}
\end{eqnarray}

The generator of the transformation is given by
\begin{eqnarray*}
\hat{Q} & = & \int dL\left[\langle R|Q|\phi^{+}(L)\rangle|\pi_{+}(L)\rangle+\langle R|Q|\phi^{-}(L)\rangle|\pi_{-}(L)\rangle\right.\\
 &  & \hphantom{\int dL\quad\langle R|Q|\phi^{+}(L)\rangle|\pi_{+}(L)\rangle}\left.-\langle R|b_{0}^{-}P\partial_{L}|\phi^{+}(L)\rangle|\pi_{-}(L)\rangle\right]\,,
\end{eqnarray*}
which satisfies 
\begin{eqnarray}
 &  & \hat{Q}|0\rrangle=\llangle0|\hat{Q}=0\,,\nonumber \\
 &  & \left[\lim_{\tau\to\infty}\llangle0|e^{-\tau\hat{H}}\right]\hat{Q}=0\,.\label{eq:Qhat}
\end{eqnarray}

\subsection{The BRST variation of $\hat{H}$}

Although the correlation functions are invariant under (\ref{eq:BRSTphi}),
the Hamiltonian (\ref{eq:SFTFKH}) is not. The BRST variations of
the correlation functions which appear on the right hand side of (\ref{eq:SFTFKH})
yield total derivatives with respect to the length variables, but
they come with the coefficients $\mathsf{T}_{LL^{\prime}L^{\prime\prime}},\mathsf{D}_{LL^{\prime}L^{\prime\prime}}$
and do not vanish upon integration.

The Hamiltonian (\ref{eq:SFTFKH}) can be expressed as
\begin{equation}
\hat{H}=\int_{0}^{\infty}dL\langle R|\mathcal{T}^{\alpha}(L)\rangle|\pi_{\alpha}(L)\rangle\,,\label{eq:Hhat}
\end{equation}
where
\begin{eqnarray*}
|\mathcal{T}^{\alpha}(L)\rangle_{3^{\prime}} & = & L|\phi^{\alpha}(L)\rangle_{3^{\prime}}-L|\pi_{-\alpha}(L)\rangle_{3^{\prime}}\\
 &  & -g_{\mathrm{s}}\int dL_{1}dL_{2}\langle T_{L_{2}LL_{1}}|B_{-\alpha_{1}}^{1}B_{\alpha_{2}}^{2}B_{\alpha}^{3}|\phi^{\alpha_{1}}(L_{1})\rangle_{1}|\pi_{\alpha_{2}}(L_{2})\rangle_{2}|R_{33^{\prime}}\rangle\\
 &  & -\frac{1}{2}g_{\mathrm{s}}\int dL_{1}dL_{2}\langle D_{LL_{1}L_{2}}|B_{-\alpha_{1}}^{1}B_{-\alpha_{2}}^{2}B_{\alpha}^{3}|\phi^{\alpha_{1}}(L_{1})\rangle_{1}|\phi^{\alpha_{2}}(L_{2})\rangle_{2}|R_{33^{\prime}}\rangle\,.
\end{eqnarray*}
$|\mathcal{T}^{\alpha}(L)\rangle$ is the SFT version of $\mathcal{T}^{I}$
and we have 
\begin{equation}
\left[\lim_{\tau\to\infty}\llangle0|e^{-\tau\hat{H}}\right]|\mathcal{T}^{\alpha}(L)\rangle=0\,.\label{eq:Talpha}
\end{equation}
The BRST variation of $\hat{H}$ is given by
\[
[\hat{Q},\hat{H}]=\int_{0}^{\infty}dL\left(\langle R|\mathcal{Q}^{\alpha}(L)\rangle|\pi_{\alpha}(L)\rangle+\langle R|\mathcal{T}^{\alpha}(L)\rangle[\hat{Q},|\pi_{\alpha}(L)\rangle]\right)\,,
\]
where 
\[
|\mathcal{Q}^{\alpha}(L)\rangle\equiv[\hat{Q},|\mathcal{T}^{\alpha}(L)\rangle]\,.
\]
From (\ref{eq:Talpha}) and (\ref{eq:Qhat}), we obtain
\begin{equation}
\left[\lim_{\tau\to\infty}\llangle0|e^{-\tau\hat{H}}\right]|\mathcal{Q}^{\alpha}(L)\rangle=0\,.\label{eq:Qalpha}
\end{equation}

\subsection{BRST invariant formulation}

Although $[\hat{Q},\hat{H}]$ does not vanish, (\ref{eq:Talpha})
and (\ref{eq:Qalpha}) implies that it consists of ``null'' quantities.
Using this fact, we will make the theory manifestly invariant under
the BRST transformation by introducing auxiliary fields.

We modify the Euclidean action (\ref{eq:Eucaction}) by adding terms
involving auxiliary fields $|\lambda_{\alpha}^{\mathcal{T}}(\tau,L)\rangle,|\lambda_{\alpha}^{\mathcal{Q}}(\tau,L)\rangle$
as follows:
\begin{eqnarray}
I_{\mathrm{BRST}} & = & \int_{0}^{\infty}d\tau\left[-\int_{0}^{\infty}dL\langle R|\pi_{\alpha}(\tau,L)\rangle\frac{\partial}{\partial\tau}|\phi^{\alpha}(\tau,L)\rangle+H(\tau)\right.\nonumber \\
 &  & \hphantom{\int_{0}^{\infty}d\tau\quad}\left.+\int_{0}^{\infty}dL\left(\langle R|\mathcal{Q}^{\alpha}(\tau,L)\rangle|\lambda_{\alpha}^{\mathcal{Q}}(\tau,L)\rangle+\langle R|\mathcal{T}^{\alpha}(\tau,L)\rangle|\lambda_{\alpha}^{\mathcal{T}}(\tau,L)\rangle\right)\vphantom{\int_{0}^{\infty}dL\langle R|\pi_{\alpha}(\tau,L)\rangle\frac{\partial}{\partial\tau}|\phi^{\alpha}(\tau,L)\rangle}\right]\,.\label{eq:IBRST}
\end{eqnarray}
Here $|\mathcal{Q}^{\alpha}(\tau,L)\rangle(|\mathcal{T}^{\alpha}(\tau,L)\rangle)$
is equal to $|\mathcal{Q}^{\alpha}(L)\rangle(|\mathcal{T}^{\alpha}(L)\rangle$)
with $|\phi^{\alpha}(L)\rangle,|\pi_{\alpha}(L)\rangle$ replaced
by classical fields $|\phi^{\alpha}(\tau,L)\rangle,|\pi_{\alpha}(\tau,L)\rangle$
respectively. $|\lambda_{\alpha}^{\mathcal{T}}(\tau,L)\rangle$ and
$|\lambda_{\alpha}^{\mathcal{Q}}(\tau,L)\rangle$ are taken to satisfy
the boundary conditions
\[
|\lambda_{\alpha}^{\mathcal{T}}(0,L)\rangle=|\lambda_{\alpha}^{\mathcal{Q}}(0,L)\rangle=0\,.
\]
$I_{\mathrm{BRST}}$ is invariant under the BRST transformation 
\begin{eqnarray*}
\delta_{\epsilon}|\phi^{+}(\tau,L)\rangle & = & \epsilon P_{-}Q|\phi^{+}(\tau,L)\rangle\,,\\
\delta_{\epsilon}|\phi^{-}(\tau,L)\rangle & = & \epsilon Q|\phi^{-}(\tau,L)\rangle-\epsilon b_{0}^{-}P\partial_{L}|\phi^{+}(\tau,L)\rangle\,,\\
\delta_{\epsilon}|\pi_{+}(\tau,L)\rangle & = & \epsilon Q|\pi_{+}(\tau,L)\rangle-\epsilon b_{0}^{-}P\partial_{L}|\pi_{-}(\tau,L)\rangle\,,\\
\delta_{\epsilon}|\pi_{-}(\tau,L)\rangle & = & \epsilon P_{-}Q|\pi_{-}(\tau,L)\rangle\,,\\
\delta_{\epsilon}|\lambda_{\alpha}^{\mathcal{Q}}(\tau,L)\rangle & = & \epsilon\left[|\pi_{\alpha}(\tau,L)\rangle+|\lambda_{\alpha}^{\mathcal{T}}(\tau,L)\rangle\right]\,,\\
\delta_{\epsilon}|\lambda_{\alpha}^{\mathcal{T}}(\tau,L)\rangle & = & -\delta_{\epsilon}|\pi_{\alpha}(\tau,L)\rangle\,.
\end{eqnarray*}
The correlation functions are defined by 
\begin{equation}
\frac{\int[d\pi d\phi d\lambda^{\mathcal{Q}}d\lambda^{\mathcal{T}}]e^{-I_{\mathrm{BRST}}}|\phi^{\alpha_{1}}(0,L_{1})\rangle\cdots|\phi^{\alpha_{n}}(0,L_{n})\rangle}{\int[d\pi d\phi d\lambda^{\mathcal{Q}}d\lambda^{\mathcal{T}}]e^{-I_{\mathrm{BRST}}}}\,.\label{eq:IBRSTcorr}
\end{equation}

We would like to show that the correlation functions in this BRST
invariant theory coincide with those given in (\ref{eq:Icorr}). The
numerator of (\ref{eq:IBRSTcorr}) is computed as 
\begin{eqnarray*}
 &  & \int[d\pi d\phi d\lambda^{\mathcal{Q}}d\lambda^{\mathcal{T}}]e^{-I_{\mathrm{BRST}}}|\phi^{\alpha_{1}}(0,L_{1})\rangle\cdots|\phi^{\alpha_{n}}(0,L_{n})\rangle\\
 &  & \quad=\int[d\lambda^{\mathcal{Q}}d\lambda^{\mathcal{T}}]\int[d\pi d\phi]e^{-I}\sum_{n=0}^{\infty}\frac{1}{n!}\left[-\int_{0}^{\infty}d\tau\int_{0}^{\infty}dL\left(\langle R|\mathcal{Q}^{\alpha}(\tau,L)\rangle|\lambda_{\alpha}^{\mathcal{Q}}(\tau,L)\rangle+\langle R|\mathcal{T}^{\alpha}(\tau,L)\rangle|\lambda_{\alpha}^{\mathcal{T}}(\tau,L)\rangle\right)\right]^{n}\\
 &  & \hphantom{\quad=\int[d\pi d\phi d\lambda^{\mathcal{Q}}d\lambda^{\mathcal{T}}]e^{-I}\quad}\times|\phi^{\alpha_{1}}(0,L_{1})\rangle\cdots|\phi^{\alpha_{n}}(0,L_{n})\rangle\,.
\end{eqnarray*}
The $n\neq0$ terms on the right hand side vanish because of (\ref{eq:Talpha})
and (\ref{eq:Qalpha}), and this becomes
\[
\left[\lim_{\tau\to\infty}\llangle0|e^{-\tau\hat{H}}\right]|\phi^{\alpha_{1}}(0,L_{1})\rangle\cdots|\phi^{\alpha_{n}}(0,L_{n})\rangle|0\rrangle\int[d\pi d\phi d\lambda^{\mathcal{Q}}d\lambda^{\mathcal{T}}]e^{-I}\,.
\]
The denominator is evaluated in the same way and we obtain
\[
\int[d\pi d\phi d\lambda^{\mathcal{Q}}d\lambda^{\mathcal{T}}]e^{-I_{\mathrm{BRST}}}=\int[d\pi d\phi d\lambda^{\mathcal{Q}}d\lambda^{\mathcal{T}}]e^{-I}\,.
\]
Therefore the correlation function (\ref{eq:IBRSTcorr}) coincides
with (\ref{eq:Icorr}).

Now the theory is invariant under the BRST symmetry, we regard the
BRST invariant quantities as physical. One type of BRST invariant
observable is of the form
\[
\langle\varphi|\phi^{+}(L)\rangle\,,
\]
for $|\varphi\rangle$ satisfying
\[
Q|\varphi\rangle=0\,.
\]
As we mentioned in section \ref{sec:Bordered-Riemann-surface}, 
\[
\lim_{L_{a}\to0}\langle\varphi_{1}|\cdots\langle\varphi_{n}|\llangle|\phi^{+}(L_{1})\rangle\cdots|\phi^{+}(L_{n})\rangle\rrangle\,,
\]
 gives the on-shell amplitude if we take $|\varphi_{a}\rangle$ to
be on-shell physical states.

Another type of BRST invariant observable would be of the form 
\begin{equation}
\int_{0}^{\infty}dL\langle\varphi|\phi^{-}(L)\rangle\,,\label{eq:observable2}
\end{equation}
with $Q|\varphi\rangle=0$. The amplitude
\begin{equation}
\int_{0}^{\infty}dL_{1}\cdots\int_{0}^{\infty}dL_{n}\langle\varphi_{1}|\cdots\langle\varphi_{n}|\llangle|\phi^{-}(L_{1})\rangle\cdots|\phi^{-}(L_{n})\rangle\rrangle\,.\label{eq:boundary}
\end{equation}
for these observables is in the form of an integration over the moduli
space of complex structures of Riemann surfaces with boundaries. Therefore
it is natural to take $\langle\varphi|$ to be 
\[
\langle\varphi|=\langle B|(c_{0}-\bar{c}_{0})\,,
\]
where $\langle B|$ is the boundary state corresponding to some D-brane
configuration. For example, taking the states $|\varphi_{a}\rangle$
to be the point like string state with appropriate ghost part, we
obtain the off-shell amplitudes of the kind studied in \cite{Cohen1986,Jaskolski1991,Bolte1991}.
Such amplitudes involve the external leg contributions coming from
the integration region $L_{a}\sim0$. Indeed, using (\ref{eq:L0}),
the contribution of $\langle\varphi_{a}|\phi^{-}(L_{a})\rangle$ for
$L_{a}\sim0$ can be approximated as
\begin{eqnarray*}
\int_{0}dL_{a}b_{0}^{-}b(\partial_{L_{a}})P|\varphi_{a}\rangle & \sim & \int_{0}dL_{a}\frac{\pi^{2}}{L_{a}^{2}}b_{0}^{+}b_{0}^{-}Pe^{-(c+\frac{\pi^{2}}{L_{a}})(L_{0}+\bar{L}_{0})}|\varphi_{a}\rangle\\
 & \sim & \frac{b_{0}^{+}b_{0}^{-}}{L_{0}+\bar{L}_{0}}Pe^{-c(L_{0}+\bar{L}_{0})}|\varphi_{a}\rangle\,.
\end{eqnarray*}
This type of observable is suitable for studying mass renormalization
\cite{Pius2014a,Pius2014}. 

\section{Discussions\label{sec:Discussions}}

In this paper, we have constructed a string field theory for closed
bosonic strings based on the pants decomposition of hyperbolic surfaces.
In such a setup, the Fokker-Planck formalism is indispensable as discussed
in section \ref{subsec:String-field-action}. We have introduced auxiliary
fields to make the theory manifestly BRST invariant. The action (\ref{eq:IBRST})
we obtain consists of kinetic terms and three string vertices.

There are many interesting points that deserve further study. The
most obvious one would be to construct an SFT for superstrings based
on the same idea. It is straightforward to generalize our formalism
to Type 0 superstring case, using the supersymmetric version of (\ref{eq:gMS})
derived in \cite{Stanford2020}. We will present the results elsewhere.

The formulation we get in section \ref{sec:BRST-invariant-formulation}
is invariant under the worldsheet BRST symmetry. We should clarify
the meaning of this symmetry from the point of view of string fields.
In ordinary formulation of SFT, the worldsheet BRST symmetry is utilized
to define the gauge or BRST transformation for string fields which
is nonlinear with respect to these fields. In our case, the worldsheet
BRST symmetry will not be related to the gauge or BRST symmetry of
the string fields in the usual way, because the theory is not based
on a triangulation of the moduli space. The similarity between the
structure of our formalism and that of the covariantized light-cone
SFT \cite{Kugo1987} may provide a clue to this problem.

Our formalism is based on the one constructed for minimal strings.
In the minimal string case, the operator corresponding to $\hat{\mathcal{T}}^{I}$
in (\ref{eq:TI}) becomes
\begin{eqnarray*}
\hat{T}(l) & = & -2\int_{0}^{l}dl^{\prime}w(l^{\prime})\hat{\phi}(l-l^{\prime})-\int_{0}^{\infty}dl^{\prime}w(l+l^{\prime})\hat{\pi}(l^{\prime})l^{\prime}\\
 &  & -g_{\mathrm{s}}\int_{0}^{l}dl^{\prime}\hat{\phi}(l^{\prime})\hat{\phi}(l-l^{\prime})-g_{\mathrm{s}}\int_{0}^{\infty}dl^{\prime}\hat{\phi}(l+l^{\prime})\hat{\pi}(l^{\prime})l^{\prime}\,,
\end{eqnarray*}
and satisfies 
\begin{equation}
\left[\lim_{\tau\to\infty}\langle0|e^{-\tau\hat{H}_{\mathrm{FP}}}\right]l\hat{T}(l)=0\,.\label{eq:Virasoro}
\end{equation}
Eq.(\ref{eq:Virasoro}) is equivalent to the Virasoro constraints
\cite{Fukuma1991,Dijkgraaf1991}. $\hat{T}(l)$ satisfies an algebra
\begin{equation}
[l_{1}\hat{T}(l_{1}),l_{2}\hat{T}(l_{2})]=g_{\mathrm{s}}l_{1}l_{2}(l_{1}-l_{2})\hat{T}(l_{1}+l_{2})\,,\label{eq:Viralg}
\end{equation}
which serves as the integrability condition of (\ref{eq:Virasoro}).
Laplace transforming (\ref{eq:Viralg}), we obtain the Virasoro algebra.
We are not sure if $\hat{\mathcal{T}}^{I}$ satisfies a similar algebra.
Exploring the algebra of $\hat{\mathcal{T}}^{I}$ will be crucial
in understanding the structure of the theory. It may be also important
to point out that one can take the Fokker-Planck Hamiltonian to be
\[
\int_{0}^{\infty}dLf(L)\langle R|\mathcal{T}^{\alpha}(L)\rangle|\pi_{\alpha}(L)\rangle\,,
\]
instead of (\ref{eq:Hhat}). Here $f(L)$ is a function of $L$ satisfying
$f(L)\neq0$ for $L>0$. The recursion relation can be derived from
this modified Hamiltonian. Such a Hamiltonian was constructed in \cite{Ikehara:1995dd}
in the minimal string case. 

The classical equation of motion for string fields can be derived
in our formalism. It is possible to assign the target space ghost
number $g^{\mathrm{t}}$ such that \cite{Zwiebach1993}
\begin{eqnarray*}
g^{\mathrm{t}}(\phi^{I}) & = & \begin{cases}
4-n_{\varphi_{i}^{c}} & \alpha=+\\
2-n_{\varphi_{i}^{c}} & \alpha=-
\end{cases}\,,\\
g^{\mathrm{t}}(\pi_{I}) & = & \begin{cases}
2-n_{\varphi_{i}} & \alpha=+\\
4-n_{\varphi_{i}} & \alpha=-
\end{cases}\,.
\end{eqnarray*}
The fields with $g^{\mathrm{t}}=0$ can be considered as the classical
fields. Although the action $S[\phi^{I}]$ is not well-defined, eq.(\ref{eq:Seq})
implies that the equation
\[
L\phi^{I}-\frac{1}{2}g_{\mathrm{s}}D^{II^{\prime}I^{\prime\prime}}G_{I^{\prime}J^{\prime}}G_{I^{\prime\prime}J^{\prime\prime}}\phi^{J^{\prime\prime}}\phi^{J^{\prime}}=0\,.
\]
may be identified with the classical equation of motion for string
fields. In the BRST invariant formulation in section \ref{sec:BRST-invariant-formulation},
this equation coincides with
\begin{equation}
|\mathcal{T}^{\alpha}(\tau,L)\rangle=0\,,\label{eq:Talpha=00003D0}
\end{equation}
under the conditions
\begin{eqnarray}
 &  & |\pi_{\alpha}(\tau,L)\rangle=0\,,\nonumber \\
 &  & \partial_{\tau}|\phi^{\alpha}(\tau,L)\rangle=0\,.\label{eq:condition}
\end{eqnarray}
For BRST invariance, we may also have to impose 
\begin{equation}
|\mathcal{Q}^{\alpha}(\tau,L)\rangle=0\,.\label{eq:Q=00003D0}
\end{equation}
Indeed (\ref{eq:Talpha=00003D0}) (\ref{eq:condition}) and (\ref{eq:Talpha=00003D0})
solves the equation of motion derived from the action (\ref{eq:IBRST}),
if the auxiliary fields vanish.

\section*{Acknowledgments}

Part of this work was presented at the workshop ``Nonperturbative
Analysis of Quantum Field Theory and its Applications'' held at Osaka
University. The author would like to thank the organizers for their
warm hospitality and S. Aoki, S. Nishigaki and Y. Sumino for comments.
This work was supported in part by Grant-in-Aid for Scientific Research
(C) (18K03637) from MEXT.

\appendix

\section{Hyperbolic metric on the three holed sphere\label{sec:Hyperbolic-metric-on}}

Let us consider a hyperbolic pants whose boundaries are geodesics.
The pair of pants is conformally equivalent to $\mathbb{C}-\bigcup_{k=1}^{3}D_{k}$
where $D_{1},D_{2},D_{3}$ are disks around $z=0,1,\infty$ respectively.
We take the length of $\partial D_{k}$ to be $L_{k}=2\pi\lambda_{k}\ (k=1,2,3)$.
Around $\partial D_{k}$, a local coordinate $\rho_{k}$ is taken
so that the metric becomes
\[
ds^{2}=\frac{\lambda_{k}^{2}}{\left|\rho_{k}\right|\sin^{2}(\lambda_{k}\log\left|\rho_{k}\right|)}\left|d\rho_{k}\right|^{2}\,.
\]
The boundary $\partial D_{k}$ corresponds to 
\[
|\rho_{k}|=\exp\left[\frac{\pi}{\lambda_{k}}\left(\tilde{l}_{k}+\frac{1}{2}\right)\right]\,,
\]
where $\tilde{l}_{k}$ is an integer. $\rho_{k}$ can be expressed
as a function $\rho_{k}(z)$ of the complex coordinate $z$ on $\mathbb{C}$.
Although $\rho_{k}(z)$ have singularities in $\mathbb{C}-\bigcup_{k=1}^{3}D_{k}$,
it is well-defined around $\partial D_{k}$ and the three-holed sphere
corresponds to the region 
\[
\left|\rho_{k}(z)\right|>\exp\left[\frac{\pi}{\lambda_{k}}\left(\tilde{l}_{k}+\frac{1}{2}\right)\right]\,.
\]
The explicit forms of $\rho_{k}(z)$ are given by\cite{Hadasz2004,Firat2021}
\begin{eqnarray}
\rho_{1}(z) & = & e^{\frac{v(\lambda_{1},\lambda_{2},\lambda_{3})}{\lambda_{1}}}z(1-z)^{-\frac{\lambda_{2}}{\lambda_{1}}}\left[\frac{_{2}F_{1}\left(\frac{1+i\lambda_{1}-i\lambda_{2}+i\lambda_{3}}{2},\frac{1+i\lambda_{1}-i\lambda_{2}-i\lambda_{3}}{2};1+i\lambda_{1};z\right)}{_{2}F_{1}\left(\frac{1-i\lambda_{1}+i\lambda_{2}-i\lambda_{3}}{2},\frac{1-i\lambda_{1}+i\lambda_{2}+i\lambda_{3}}{2};1-i\lambda_{1};z\right)}\right]^{\frac{1}{i\lambda_{1}}}\,,\nonumber \\
\rho_{2}(z) & = & e^{\frac{v(\lambda_{2},\lambda_{1},\lambda_{3})}{\lambda_{2}}}(1-z)z{}^{-\frac{\lambda_{1}}{\lambda_{2}}}\left[\frac{_{2}F_{1}\left(\frac{1+i\lambda_{2}-i\lambda_{1}+i\lambda_{3}}{2},\frac{1+i\lambda_{2}-i\lambda_{1}-i\lambda_{3}}{2};1+i\lambda_{2};1-z\right)}{_{2}F_{1}\left(\frac{1-i\lambda_{2}+i\lambda_{1}-i\lambda_{3}}{2},\frac{1-i\lambda_{2}+i\lambda_{1}+i\lambda_{3}}{2};1-i\lambda_{2};1-z\right)}\right]^{\frac{1}{i\lambda_{2}}}\,,\nonumber \\
\rho_{3}(z) & = & e^{\frac{v(\lambda_{3},\lambda_{2},\lambda_{1})}{\lambda_{3}}}\frac{1}{z}(1-\frac{1}{z})^{-\frac{\lambda_{2}}{\lambda_{3}}}\left[\frac{_{2}F_{1}\left(\frac{1+i\lambda_{3}-i\lambda_{2}+i\lambda_{1}}{2},\frac{1+i\lambda_{3}-i\lambda_{2}-i\lambda_{1}}{2};1+i\lambda_{3};\frac{1}{z}\right)}{_{2}F_{1}\left(\frac{1-i\lambda_{3}+i\lambda_{2}-i\lambda_{1}}{2},\frac{1-i\lambda_{3}+i\lambda_{2}+i\lambda_{1}}{2};1-i\lambda_{3};\frac{1}{z}\right)}\right]^{\frac{1}{i\lambda_{3}}}\,,\label{eq:rhoz}
\end{eqnarray}
where 
\begin{equation}
e^{2iv(\lambda_{1},\lambda_{2},\lambda_{3})}=\frac{\Gamma(-i\lambda_{1})^{2}}{\Gamma(i\lambda_{1})^{2}}\frac{\gamma\left(\frac{1+i\lambda_{1}+i\lambda_{2}+i\lambda_{3}}{2}\right)\gamma\left(\frac{1+i\lambda_{1}-i\lambda_{2}+i\lambda_{3}}{2}\right)}{\gamma\left(\frac{1-i\lambda_{1}-i\lambda_{2}+i\lambda_{3}}{2}\right)\gamma\left(\frac{1-i\lambda_{1}+i\lambda_{2}+i\lambda_{3}}{2}\right)}\,,\label{eq:v}
\end{equation}
and 
\[
\gamma(x)=\frac{\Gamma(x)}{\Gamma(1-x)}\,.
\]
Notice that 
\begin{eqnarray}
\rho_{2}(z) & = & \left.\rho_{1}(1-z)\right|_{\lambda_{1}\leftrightarrow\lambda_{2}}\,,\nonumber \\
\rho_{3}(z) & = & \left.\rho_{1}(\frac{1}{z})\right|_{\lambda_{1}\leftrightarrow\lambda_{3}}\,,\nonumber \\
\left.\rho_{1}(z)\right|_{\lambda_{2}\leftrightarrow\lambda_{3}} & = & -\rho_{1}(\frac{z}{z-1})\,,\nonumber \\
\left.\rho_{2}(z)\right|_{\lambda_{1}\leftrightarrow\lambda_{3}} & = & -\rho_{2}(\frac{1}{z})\,,\nonumber \\
\left.\rho_{3}(z)\right|_{\lambda_{1}\leftrightarrow\lambda_{3}} & = & -\rho_{3}(1-z)\,,\label{eq:sl2}
\end{eqnarray}
hold. Attaching flat semi-infinite cylinders to the boundaries of
the three holed sphere, we get a surface conformally equivalent to
a three punctured sphere. The local coordinates on the cylinders are
given by 
\begin{equation}
W_{k}(z)=\exp\left[-\frac{\pi}{\lambda_{k}}\left(\tilde{l}_{k}+\frac{1}{2}\right)\right]\rho_{k}(z)\,,\label{eq:Wj}
\end{equation}
 up to a phase rotation and the metrics on the cylinders become
\[
ds^{2}=\lambda_{k}^{2}\frac{\left|dW_{k}\right|^{2}}{\left|W_{k}\right|^{2}}\,.
\]
Notice that $\mathrm{Im}z=0$ gives geodesics connecting the boundary
components and perpendicular to them. Therefore $W_{k}=\pm1$ become
the basepoints which are used to define the twist parameters of the
Fenchel-Nielsen coordinates \cite{Abikoff1980}. 

For studying various properties of the amplitudes, it is useful to
examine the limits $L_{k}\to0,\infty$ of the formula (\ref{eq:Wj}).
Since the behavior does not depend on $k$ because of (\ref{eq:sl2}),
we consider the case $k=1$. In the limit $L_{1}=2\pi\lambda_{1}\to0$,
(\ref{eq:v}) implies
\[
v(\lambda_{1},\lambda_{2},\lambda_{3})=n\pi+c\lambda_{1}+\mathcal{O}(\lambda_{1}^{2})\,,
\]
where 
\[
c=2\gamma+\frac{1}{2}\left[\psi\left(\frac{1+i\lambda_{2}+i\lambda_{3}}{2}\right)+\psi\left(\frac{1-i\lambda_{2}-i\lambda_{3}}{2}\right)+\psi\left(\frac{1-i\lambda_{2}+i\lambda_{3}}{2}\right)+\psi\left(\frac{1+i\lambda_{2}-i\lambda_{3}}{2}\right)\right]\in\mathbb{R}\,,
\]
and $n\in\mathbb{Z}$. From (\ref{eq:rhoz}) and (\ref{eq:Wj}), we
get
\[
z\sim e^{-c+\frac{\pi}{\lambda_{1}}(\tilde{l}_{1}+\frac{1}{2}-n)}W_{1}\,.
\]
$\tilde{l}_{1}$ should be taken \cite{Hadasz2004} so that $\tilde{l}_{1}-n=-1$
and we eventually obtain
\begin{equation}
z\sim e^{-c-\frac{\pi^{2}}{L_{1}}}W_{1}\,.\label{eq:L0}
\end{equation}
The limit $L_{1}=2\pi\lambda_{1}\to\infty$ can be obtained from the
Fuchsian equation. We get 
\begin{equation}
z\sim4W_{1}+\mathcal{O}(\frac{1}{\lambda_{1}})\,.\label{eq:Linfty}
\end{equation}

\section{BRST identity\label{sec:BRST-identity}}

In this appendix, we would like to prove 
\begin{eqnarray}
 &  & \langle\Sigma_{g,n,\mathbf{L}}|B_{6g-6+2n}\tilde{B}_{\alpha_{1}}^{1}\cdots\tilde{B}_{\alpha_{n}}^{n}\sum_{a=1}^{n}Q^{(a)}\nonumber \\
 &  & \quad=d\left[\langle\Sigma_{g,n,\mathbf{L}}|B_{6g-7+2n}\tilde{B}_{\alpha_{1}}^{1}\cdots\tilde{B}_{\alpha_{n}}^{n}\vphantom{\sum_{a=1}^{n}}\right.\nonumber \\
 &  & \hphantom{\quad=d\quad}\left.+\sum_{a=1}^{n}\langle\Sigma_{g,n,\mathbf{L}}|B_{6g-6+2n}\tilde{B}_{\alpha_{1}}^{1}\cdots\tilde{b}_{\alpha_{a}}^{a}\cdots\tilde{B}_{\alpha_{n}}^{n}\right]\,,\label{eq:BRST2}
\end{eqnarray}
where
\begin{eqnarray*}
B_{6g-7+2n} & = & \sum_{t=1}^{3g-3+n}\left[\prod_{s\ne t}\left(b(\partial_{l_{s}})b(\partial_{\tau_{s}})\right)\bigwedge_{s\ne t}\left(dl_{s}\wedge d\tau_{s}\right)\right]\left[b(\partial_{l_{t}})dl_{t}+b(\partial_{\tau_{t}})d\tau_{t}\right]\\
\tilde{B}_{\alpha_{a}}^{a} & = & \begin{cases}
1 & \alpha_{a}=+\\
b_{S}(\partial_{L_{a}})b(\partial_{\theta_{a}})e^{i\theta_{a}(L_{0}^{(a)}-\bar{L}_{0}^{(a)})}dL_{a}\wedge d\theta_{a} & \alpha_{a}=-
\end{cases}\,,\\
\tilde{b}_{\alpha_{a}}^{a} & = & \begin{cases}
0 & \alpha_{a}=+\\
b_{S}(\partial_{L_{a}})e^{i\theta_{a}(L_{0}^{(a)}-\bar{L}_{0}^{(a)})}dL_{a}+b(\partial_{\theta_{a}})e^{i\theta_{a}(L_{0}^{(a)}-\bar{L}_{0}^{(a)})}d\theta_{a} & \alpha_{a}=-
\end{cases}\,.
\end{eqnarray*}
It is straightforward to derive (\ref{eq:BRST}) and (\ref{eq:BRST1})
from (\ref{eq:BRST2}).

Decomposing $\Sigma_{g,n,\mathbf{L}}$ into pairs of pants, $\langle\Sigma_{g,n,\mathbf{L}}|$
can be expressed in terms of $\langle\Sigma_{0,3,\mathbf{L}}|$. In
order to prove (\ref{eq:BRST2}), we will study some of the properties
of $\langle\Sigma_{0,3,\mathbf{L}}|$. $\langle\Sigma_{0,3,\mathbf{L}}|$
satisfies
\[
\langle\Sigma_{0,3,\mathbf{L}}|\Psi_{1}\rangle|\Psi_{2}\rangle|\Psi_{3}\rangle=\left\langle W_{1}^{-1}\circ\mathcal{O}_{\Psi_{1}}(0)W_{2}^{-1}\circ\mathcal{O}_{\Psi_{2}}(0)W_{3}^{-1}\circ\mathcal{O}_{\Psi_{3}}(0)\right\rangle _{\mathbb{C}\cup\{\infty\}}\,.
\]
Here we take the local coordinate on $\mathbb{C}\cup\{\infty\}$ to
be the $z$ in appendix \ref{sec:Hyperbolic-metric-on} and $W_{k}(z)$
is given in (\ref{eq:Wj}). We introduce the twist angles $\theta_{a}$
by deforming $\langle\Sigma_{0,3,\mathbf{L}}|$ as 
\[
\langle\Sigma_{0,3,\mathbf{L}}|\to\langle\Sigma_{0,3,\mathbf{L}}|\prod_{a=1}^{3}e^{i\theta_{a}(L_{0}^{(a)}-\bar{L}_{0}^{(a)})}\,,
\]
so that we have
\begin{equation}
\langle\Sigma_{0,3,\mathbf{L}}|\prod_{a=1}^{3}e^{i\theta_{a}(L_{0}^{(a)}-\bar{L}_{0}^{(a)})}|\Psi_{1}\rangle|\Psi_{2}\rangle|\Psi_{3}\rangle=\left\langle f_{1}\circ\mathcal{O}_{\Psi_{1}}(0)f_{2}\circ\mathcal{O}_{\Psi_{2}}(0)f_{3}\circ\mathcal{O}_{\Psi_{3}}(0)\right\rangle _{\mathbb{C}\cup\{\infty\}}\,,\label{eq:Sigma03}
\end{equation}
where
\[
f_{a}(w_{a})=W_{a}^{-1}(e^{i\theta_{a}}w_{a})\,.
\]
(\ref{eq:Sigma03}) implies 
\[
\langle\Sigma_{0,3,\mathbf{L}}|\prod_{a=1}^{3}e^{i\theta_{a}(L_{0}^{(a)}-\bar{L}_{0}^{(a)})}=\sum_{i_{1},i_{2},i_{3}}\left\langle f_{1}\circ\mathcal{O}_{\varphi_{i_{1}}}(0)f_{2}\circ\mathcal{O}_{\varphi_{i_{2}}}(0)f_{3}\circ\mathcal{O}_{\varphi_{i_{3}}}(0)\right\rangle _{\mathbb{C}\cup\{\infty\}}\langle\varphi_{i_{3}}^{c}|\langle\varphi_{i_{2}}^{c}|\langle\varphi_{i_{1}}^{c}|\,,
\]
which can be regarded as the definition of the state $\langle\Sigma_{0,3,\mathbf{L}}|$. 

Following the formula (\ref{eq:partialxs}), we define
\begin{eqnarray*}
b(\partial_{L_{a}}) & = & \sum_{a^{\prime}=1}^{3}b^{(a^{\prime})}(\partial_{L_{a}})\,,\\
T(\partial_{L_{a}}) & \equiv & \sum_{a^{\prime}=1}^{3}T^{(a^{\prime})}(\partial_{L_{a}})\,,
\end{eqnarray*}
with
\begin{eqnarray*}
b^{(a^{\prime})}(\partial_{L_{a}}) & \equiv & -\oint_{0}\frac{dw_{a^{\prime}}}{2\pi i}\frac{\partial f_{a^{\prime}}}{\partial L_{a}}\frac{\partial w_{a^{\prime}}}{\partial z}b(w_{a^{\prime}})-\oint_{0}\frac{d\bar{w}_{a^{\prime}}}{2\pi i}\frac{\partial f_{a^{\prime}}}{\partial L_{a}}\frac{\partial\bar{w}_{a^{\prime}}}{\partial\bar{z}}\bar{b}(\bar{w}_{a^{\prime}})\,,\\
T^{(a^{\prime})}(\partial_{L_{a}}) & \equiv & -\oint_{0}\frac{dw_{a^{\prime}}}{2\pi i}\frac{\partial f_{a^{\prime}}}{\partial L_{a}}\frac{\partial w_{a^{\prime}}}{\partial z}T(w_{a^{\prime}})-\oint_{0}\frac{d\bar{w}_{a^{\prime}}}{2\pi i}\frac{\partial f_{a^{\prime}}}{\partial L_{a}}\frac{\partial\bar{w}_{a^{\prime}}}{\partial\bar{z}}\bar{T}(\bar{w}_{a^{\prime}})\,.
\end{eqnarray*}
Here $T(z),\bar{T}(\bar{z})$ are the stress tensors of the worldsheet
theory and we have 
\begin{equation}
\left\{ Q,b(\partial_{L_{a}})\right\} =T(\partial_{L_{a}})\,,\label{eq:QbL}
\end{equation}
with
\[
Q\equiv\sum_{a^{\prime}}Q^{(a^{\prime})}\,.
\]
It is possible to show that for any state $|\Psi\rangle$, 
\begin{eqnarray*}
 &  & \left[-\oint_{0}\frac{dw_{a^{\prime}}}{2\pi i}\frac{\partial f_{a^{\prime}}}{\partial L_{a}}\frac{\partial w_{a^{\prime}}}{\partial z}T(w_{a^{\prime}})-\oint_{0}\frac{d\bar{w}_{a^{\prime}}}{2\pi i}\frac{\partial f_{a^{\prime}}}{\partial L_{a}}\frac{\partial\bar{w}_{a^{\prime}}}{\partial\bar{z}}\bar{T}(\bar{w}_{a^{\prime}})\right]f_{a^{\prime}}\circ\mathcal{O}_{\Psi}(0)\\
 &  & \quad=-\partial_{L_{a}}\left[f_{a^{\prime}}\circ\mathcal{O}_{\Psi}(0)\right]\,,
\end{eqnarray*}
holds. Hence we obtain
\begin{eqnarray}
\langle\Sigma_{0,3,\mathbf{L}}|T(\partial_{L_{a}})\prod_{a=1}^{3}e^{i\theta_{a}(L_{0}^{(a)}-\bar{L}_{0}^{(a)})} & = & -\sum_{i_{1},i_{2},i_{3}}\langle\Sigma_{0,3,\mathbf{L}}|T(\partial_{L_{a}})\prod_{a=1}^{3}e^{i\theta_{a}(L_{0}^{(a)}-\bar{L}_{0}^{(a)})}|\varphi_{i_{1}}\rangle|\varphi_{i_{2}}\rangle|\varphi_{i_{3}}\rangle\langle\varphi_{i_{3}}^{c}|\langle\varphi_{i_{2}}^{c}|\langle\varphi_{i_{1}}^{c}|\nonumber \\
 &  & -\sum_{i_{1},i_{2},i_{3}}\partial_{L_{a}}\left\langle f_{1}\circ\mathcal{O}_{\Psi_{1}}(0)f_{2}\circ\mathcal{O}_{\Psi_{2}}(0)f_{3}\circ\mathcal{O}_{\Psi_{3}}(0)\right\rangle _{\mathbb{C}\cup\{\infty\}}\langle\varphi_{i_{3}}^{c}|\langle\varphi_{i_{2}}^{c}|\langle\varphi_{i_{1}}^{c}|\nonumber \\
 & = & -\partial_{L_{a}}\left[\langle\Sigma_{0,3,\mathbf{L}}|\prod_{a=1}^{3}e^{i\theta_{a}(L_{0}^{(a)}-\bar{L}_{0}^{(a)})}\right]\,.\label{eq:TL}
\end{eqnarray}
We can also define 
\begin{eqnarray*}
b(\partial_{\theta_{a}}) & = & -i(b_{0}^{(a)}-\bar{b}_{0}^{(a)})\,,\\
T(\partial_{\theta_{a}}) & = & -i(L_{0}^{(a)}-\bar{L}_{0}^{(a)})\,,
\end{eqnarray*}
and it is easy to prove
\begin{eqnarray}
 &  & \left\{ Q,b(\partial_{\theta_{a}})\right\} =T(\partial_{\theta_{a}})\,,\nonumber \\
 &  & \langle\Sigma_{0,3,\mathbf{L}}|T(\partial_{\theta_{a}})\prod_{a=1}^{3}e^{i\theta_{a}(L_{0}^{(a)}-\bar{L}_{0}^{(a)})}=-\partial_{\theta_{a}}\left[\langle\Sigma_{0,3,\mathbf{L}}|\prod_{a=1}^{3}e^{i\theta_{a}(L_{0}^{(a)}-\bar{L}_{0}^{(a)})}T(\partial_{\theta_{a}})\right]\,.\label{eq:bTtheta}
\end{eqnarray}
 $b(\partial_{L_{a}})\,,T(\partial_{L_{a}})\,,b(\partial_{\theta_{a}})$
and $T(\partial_{\theta_{a}})$ satisfy the following commutation
relations:
\begin{eqnarray}
[T(\partial_{L_{a}}),b(\partial_{L_{a^{\prime}}})] & = & \partial_{L_{a}}b(\partial_{L_{a^{\prime}}})-\partial_{L_{a^{\prime}}}b(\partial_{L_{a}})\,,\nonumber \\{}
[T(\partial_{L_{a}}),b(\partial_{\theta_{a^{\prime}}})] & = & -\partial_{\theta_{a^{\prime}}}b(\partial_{L_{a}})\,,\nonumber \\{}
[b(\partial_{L_{a}}),T(\partial_{\theta_{a^{\prime}}})] & = & -\partial_{\theta_{a^{\prime}}}b(\partial_{L_{a}})\,,\nonumber \\{}
[T(\partial_{\theta_{a}}),b(\partial_{\theta_{a^{\prime}}})] & = & 0\,.\label{eq:Tb}
\end{eqnarray}

Using (\ref{eq:QbL}), (\ref{eq:TL}), (\ref{eq:bTtheta}) and(\ref{eq:Tb}),
it is straightforward to show 
\begin{eqnarray}
 &  & \langle\Sigma_{0,3,\mathbf{L}}|\tilde{B}_{\alpha_{1}}^{1}\tilde{B}_{\alpha_{2}}^{2}\tilde{B}_{\alpha_{3}}^{3}\sum_{a=1}^{3}Q^{(a)}\nonumber \\
 &  & \quad=d\left[\langle\Sigma_{0,3,\mathbf{L}}|\left(\tilde{b}_{\alpha_{1}}^{1}\tilde{B}_{\alpha_{2}}^{2}\tilde{B}_{\alpha_{3}}^{3}+\tilde{B}_{\alpha_{1}}^{1}\tilde{b}_{\alpha_{2}}^{2}\tilde{B}_{\alpha_{3}}^{3}+\tilde{B}_{\alpha_{1}}^{1}\tilde{B}_{\alpha_{2}}^{2}\tilde{b}_{\alpha_{3}}^{3}\right)\right]\,,\label{eq:g0n3Q}
\end{eqnarray}
which is (\ref{eq:BRST2}) for $g=0,n=3$.

Other cases can be proved by using (\ref{eq:g0n3Q}). Let us consider
the next simplest case $g=1,n=1$. The surface state $\langle\Sigma_{1,1,L}|$
can be expressed as
\begin{eqnarray}
\langle\Sigma_{1,1,L}| & = & \sum_{i,j}{}_{123}\langle\Sigma_{0,3,(L,l_{\gamma},l_{\gamma})}|e^{i\theta_{\gamma}(L_{0}^{(2)}-\bar{L}_{0}^{(2)})}|\varphi_{i}\rangle_{2}|\varphi_{j}\rangle_{3}\langle\varphi_{i}^{c}|\varphi_{j}^{c}\rangle(-1)^{n_{\varphi_{j}}}\nonumber \\
 & = & _{123}\langle\Sigma_{0,3,(L,l_{\gamma},l_{\gamma})}|e^{i\theta_{\gamma}(L_{0}^{(2)}-\bar{L}_{0}^{(2)})}|R_{23}\rangle\,.\label{eq:g1n1f}
\end{eqnarray}
Using this, we obtain
\begin{eqnarray}
\langle\Sigma_{1,1,L}|B_{2}\tilde{B}_{\alpha}Q & = & _{123}\langle\Sigma_{0,3,(L,l_{\gamma},l_{\gamma})}|B_{2}\tilde{B}_{\alpha}^{1}Q^{(1)}e^{i\theta_{\gamma}(L_{0}^{(2)}-\bar{L}_{0}^{(2)})}|R_{23}\rangle\nonumber \\
 & = & \left._{123}\langle\Sigma_{0,3,(L,l_{2},l_{3})}|(b(\partial_{l_{2}})+b(\partial_{l_{3}}))b(\partial_{\theta_{\gamma}})\tilde{B}_{\alpha}^{1}\sum_{a=1}^{3}Q^{(a)}e^{i\theta_{\gamma}(L_{0}^{(2)}-\bar{L}_{0}^{(2)})}|R_{23}\rangle\right|_{l_{2}=l_{3}=l_{\gamma}}dl_{\gamma}\wedge d\theta_{\gamma}\nonumber \\
\label{eq:g1n1Q1}
\end{eqnarray}
In going from the first line to the second line, we have used 
\begin{equation}
(Q^{(2)}+Q^{(3)})|R_{23}\rangle=0\,.\label{eq:g1n12}
\end{equation}
Using (\ref{eq:g0n3Q}), we eventually get
\begin{eqnarray}
\langle\Sigma_{1,1,L}|B_{2}\tilde{B}_{\alpha}Q & = & \left.\partial_{l_{2}}\left[_{123}\langle\Sigma_{0,3,(L,l_{2},l_{3})}|b(\partial_{\theta_{\gamma}})\tilde{B}_{\alpha}^{1}e^{i\theta_{\gamma}(L_{0}^{(2)}-\bar{L}_{0}^{(2)})}|R_{23}\rangle\right]\right|_{l_{2}=l_{3}=l_{\gamma}}dl_{\gamma}\wedge d\theta_{\gamma}\nonumber \\
 &  & +\left.\partial_{l_{3}}\left[_{123}\langle\Sigma_{0,3,(L,l_{2},l_{3})}|b(\partial_{\theta_{\gamma}})\tilde{B}_{\alpha}^{1}e^{i\theta_{\gamma}(L_{0}^{(2)}-\bar{L}_{0}^{(2)})}|R_{23}\rangle\right]\right|_{l_{2}=l_{3}=l_{\gamma}}dl_{\gamma}\wedge d\theta_{\gamma}\nonumber \\
 &  & +\left.\partial_{\theta_{\gamma}}\left[_{123}\langle\Sigma_{0,3,(L,l_{2},l_{3})}|(b(\partial_{l_{2}})+b(\partial_{l_{3}}))\tilde{B}_{\alpha}^{1}e^{i\theta_{\gamma}(L_{0}^{(2)}-\bar{L}_{0}^{(2)})}|R_{23}\rangle\right]\right|_{l_{2}=l_{3}=l_{\gamma}}dl_{\gamma}\wedge d\theta_{\gamma}\nonumber \\
 &  & +\left._{123}\langle\Sigma_{0,3,(L,l_{2},l_{3})}|(b(\partial_{l_{2}})+b(\partial_{l_{3}}))b(\partial_{\theta_{\gamma}})\tilde{b}_{\alpha}^{1}e^{i\theta_{\gamma}(L_{0}^{(2)}-\bar{L}_{0}^{(2)})}|R_{23}\rangle\right|_{l_{2}=l_{3}=l_{\gamma}}dl_{\gamma}\wedge d\theta_{\gamma}\nonumber \\
 & = & d\left[\langle\Sigma_{1,1,L}|(B_{1}\tilde{B}_{\alpha}+B_{2}\tilde{b}_{\alpha})\right]\,.\label{eq:g1n13}
\end{eqnarray}

The proof for all the other cases goes in the same way. We use induction
with respect to $2g-2+n$. So far we have shown (\ref{eq:BRST2})
for $2g-2+n=1$. Assuming that (\ref{eq:BRST2}) is true for $2g-2+n=K>0$,
let us prove (\ref{eq:BRST2}) for $2g-2+n=K+1$. $\langle\Sigma_{g,n,\mathbf{L}}|$
can be expressed by $\langle\Sigma_{0,3,\mathbf{L}}|$ and surface
states with $2g-2+n\leq K$ by factorizing the surface as in Figure
\ref{fig:The-decomposition-of} or Figure \ref{fig:Factorizations-of-the}.
Using the induction hypothesis, we obtain (\ref{eq:BRST2}) for $\langle\Sigma_{g,n,\mathbf{L}}|$
in the same way as we did for $\langle\Sigma_{1,1,L}|$. 

\bibliographystyle{utphys}
\bibliography{stochastic}

\providecommand{\href}[2]{#2}\begingroup\raggedright\begin{thebibliography}{10}

\bibitem{Kaku1974}
M.~Kaku and K.~Kikkawa, ``The field theory of relativistic strings. i. trees,''
  \href{http://dx.doi.org/10.1103/PhysRevD.10.1110}{{\em Phys. Rev. D} {\bf 10}
  (1974)  1110}.

\bibitem{Witten1986}
E.~Witten, ``Noncommutative geometry and string field theory,''
  \href{http://dx.doi.org/10.1016/0550-3213(86)90155-0}{{\em Nucl. Phys. B}
  {\bf 268} (1986)  253--294}.

\bibitem{Kugo1992}
T.~Kugo and B.~Zwiebach, ``Target space duality as a symmetry of string field
  theory,'' \href{http://dx.doi.org/10.1143/ptp/87.4.801}{{\em Prog. Theor.
  Phys.} {\bf 87} (1992)  801--860},
  \href{http://arxiv.org/abs/hep-th/9201040}{{\tt arXiv:hep-th/9201040}}.

\bibitem{Zwiebach1993}
B.~Zwiebach, ``Closed string field theory: Quantum action and the b-v master
  equation,'' \href{http://dx.doi.org/10.1016/0550-3213(93)90388-6}{{\em Nucl.
  Phys. B} {\bf 390} (1993)  33--152},
  \href{http://arxiv.org/abs/hep-th/9206084}{{\tt arXiv:hep-th/9206084}}.

\bibitem{Lacroix2017}
C.~de~Lacroix, H.~Erbin, S.~P. Kashyap, A.~Sen, and M.~Verma, ``Closed
  superstring field theory and its applications,''
  \href{http://dx.doi.org/10.1142/S0217751X17300216}{{\em Int. J. Mod. Phys. A}
  {\bf 32} (2017) no.~28n29, 1730021},
  \href{http://arxiv.org/abs/1703.06410}{{\tt arXiv:1703.06410 [hep-th]}}.

\bibitem{Moosavian2019}
S.~F. Moosavian and R.~Pius, ``Hyperbolic geometry and closed bosonic string
  field theory. part i. the string vertices via hyperbolic riemann surfaces,''
  \href{http://dx.doi.org/10.1007/JHEP08(2019)157}{{\em JHEP} {\bf 08} (2019)
  157}, \href{http://arxiv.org/abs/1706.07366}{{\tt arXiv:1706.07366
  [hep-th]}}.

\bibitem{Moosavian2019a}
S.~F. Moosavian and R.~Pius, ``Hyperbolic geometry and closed bosonic string
  field theory. part ii. the rules for evaluating the quantum bv master
  action,'' \href{http://dx.doi.org/10.1007/JHEP08(2019)177}{{\em JHEP} {\bf
  08} (2019)  177}, \href{http://arxiv.org/abs/1708.04977}{{\tt
  arXiv:1708.04977 [hep-th]}}.

\bibitem{Costello2022}
K.~Costello and B.~Zwiebach, ``Hyperbolic string vertices,''
  \href{http://dx.doi.org/10.1007/JHEP02(2022)002}{{\em JHEP} {\bf 02} (2022)
  002}, \href{http://arxiv.org/abs/1909.00033}{{\tt arXiv:1909.00033
  [hep-th]}}.

\bibitem{DHoker1988}
E.~D'Hoker and D.~H. Phong, ``The geometry of string perturbation theory,''
  \href{http://dx.doi.org/10.1103/RevModPhys.60.917}{{\em Rev. Mod. Phys.} {\bf
  60} (1988)  917}.

\bibitem{Mirzakhani2006}
M.~Mirzakhani, ``Simple geodesics and weil-petersson volumes of moduli spaces
  of bordered riemann surfaces,''
  \href{http://dx.doi.org/10.1007/s00222-006-0013-2}{{\em Invent. Math.} {\bf
  167} (2006) no.~1, 179--222}.

\bibitem{Mirzakhani2007}
M.~Mirzakhani, ``Weil-petersson volumes and intersection theory on the moduli
  space of curves,''
  \href{http://dx.doi.org/10.1090/S0894-0347-06-00526-1}{{\em J. Am. Math.
  Soc.} {\bf 20} (2007) no.~01, 1--24}.

\bibitem{Eynard2007}
B.~Eynard and N.~Orantin, ``Weil-petersson volume of moduli spaces,
  mirzakhani's recursion and matrix models,''
  \href{http://arxiv.org/abs/0705.3600}{{\tt arXiv:0705.3600 [math-ph]}}.

\bibitem{Saad2019a}
P.~Saad, S.~H. Shenker, and D.~Stanford, ``Jt gravity as a matrix integral,''
  \href{http://arxiv.org/abs/1903.11115}{{\tt arXiv:1903.11115 [hep-th]}}.

\bibitem{Ishibashi:1993nq}
N.~Ishibashi and H.~Kawai, ``{String field theory of noncritical strings},''
  \href{http://dx.doi.org/10.1016/0370-2693(93)90448-Q}{{\em Phys. Lett. B}
  {\bf 314} (1993)  190--196}, \href{http://arxiv.org/abs/hep-th/9307045}{{\tt
  arXiv:hep-th/9307045}}.

\bibitem{Jevicki1994}
A.~Jevicki and J.~P. Rodrigues, ``{Loop space Hamiltonians and field theory of
  noncritical strings},''
  \href{http://dx.doi.org/10.1016/0550-3213(94)90329-8}{{\em Nucl. Phys. B}
  {\bf 421} (1994)  278--292}, \href{http://arxiv.org/abs/hep-th/9312118}{{\tt
  arXiv:hep-th/9312118}}.

\bibitem{Sen2015}
A.~Sen, ``Off-shell amplitudes in superstring theory,''
  \href{http://dx.doi.org/10.1002/prop.201500002}{{\em Fortsch. Phys.} {\bf 63}
  (2015)  149--188}, \href{http://arxiv.org/abs/1408.0571}{{\tt arXiv:1408.0571
  [hep-th]}}.

\bibitem{Erler2020}
T.~Erler, ``Four lectures on closed string field theory,''
  \href{http://dx.doi.org/10.1016/j.physrep.2020.01.003}{{\em Phys. Rept.} {\bf
  851} (2020)  1--36}, \href{http://arxiv.org/abs/1905.06785}{{\tt
  arXiv:1905.06785 [hep-th]}}.

\bibitem{Erbin2021}
H.~Erbin, \href{http://dx.doi.org/10.1007/978-3-030-65321-7}{{\em String Field
  Theory: A Modern Introduction}}, vol.~980 of {\em Lecture Notes in Physics}.
\newblock 3, 2021.

\bibitem{Polchinski2007}
J.~Polchinski, \href{http://dx.doi.org/10.1017/CBO9780511816079}{{\em String
  theory. Vol. 1: An introduction to the bosonic string}}.
\newblock Cambridge Monographs on Mathematical Physics. Cambridge University
  Press, 12, 2007.

\bibitem{Hadasz2004}
L.~Hadasz and Z.~Jaskolski, ``Classical liouville action on the sphere with
  three hyperbolic singularities,''
  \href{http://dx.doi.org/10.1016/j.nuclphysb.2004.03.012}{{\em Nucl. Phys. B}
  {\bf 694} (2004)  493--508}, \href{http://arxiv.org/abs/hep-th/0309267}{{\tt
  arXiv:hep-th/0309267}}.

\bibitem{Firat2021}
A.~H. F\i{}rat, ``Hyperbolic three-string vertex,''
  \href{http://dx.doi.org/10.1007/JHEP08(2021)035}{{\em JHEP} {\bf 08} (2021)
  035}, \href{http://arxiv.org/abs/2102.03936}{{\tt arXiv:2102.03936
  [hep-th]}}.

\bibitem{Cohen1986}
A.~G. Cohen, G.~W. Moore, P.~C. Nelson, and J.~Polchinski, ``An off-shell
  propagator for string theory,''
  \href{http://dx.doi.org/10.1016/0550-3213(86)90148-3}{{\em Nucl. Phys. B}
  {\bf 267} (1986)  143--157}.

\bibitem{Jaskolski1991}
Z.~Jaskolski, ``The polyakov path integral over bordered surfaces (the closed
  string off-shell amplitudes),''
  \href{http://dx.doi.org/10.1007/BF02352499}{{\em Commun. Math. Phys.} {\bf
  139} (1991)  353--376}.

\bibitem{Bolte1991}
J.~Bolte and F.~Steiner, ``The on-shell limit of bosonic off-shell string
  scattering amplitudes,''
  \href{http://dx.doi.org/10.1016/0550-3213(91)90249-W}{{\em Nucl. Phys. B}
  {\bf 361} (1991)  451--468}.

\bibitem{Do2011}
Do, ``Moduli spaces of hyperbolic surfaces and their weil-petersson volumes,''
  {\em arXiv:1103.4674 [math]} (2011)  .

\bibitem{Huang2016}
Y.~Huang, ``Mirzakhani's recursion formula on weil-petersson volume and
  applications,'' \href{http://dx.doi.org/10.4171/161-1/5}{{\em IRMA Lect.
  Math. Theor. Phys.} {\bf 27} (2016)  95--127},
  \href{http://arxiv.org/abs/1509.06880}{{\tt arXiv:1509.06880 [math.GT]}}.

\bibitem{McShane1991}
G.~McShane, ``Doctoral thesis.,'' {\em Ph. D. thesis, University of Warwick}
  (1991)  .

\bibitem{Parisi:1980ys}
G.~Parisi and Y.-s. Wu, ``{Perturbation Theory Without Gauge Fixing},'' {\em
  Sci.\ Sin.} {\bf 24} (1981)  483.

\bibitem{Ishibashi:1993nqz}
N.~Ishibashi and H.~Kawai, ``{String field theory of c <= 1 noncritical
  strings},'' \href{http://dx.doi.org/10.1016/0370-2693(94)90492-8}{{\em Phys.
  Lett. B} {\bf 322} (1994)  67--78},
  \href{http://arxiv.org/abs/hep-th/9312047}{{\tt arXiv:hep-th/9312047}}.

\bibitem{Ikehara1994}
M.~Ikehara, N.~Ishibashi, H.~Kawai, T.~Mogami, R.~Nakayama, and N.~Sasakura,
  ``String field theory in the temporal gauge,''
  \href{http://dx.doi.org/10.1103/PhysRevD.50.7467}{{\em Phys. Rev. D} {\bf 50}
  (1994)  7467--7478}, \href{http://arxiv.org/abs/hep-th/9406207}{{\tt
  arXiv:hep-th/9406207}}.

\bibitem{Ikehara1995}
M.~Ikehara, N.~Ishibashi, H.~Kawai, T.~Mogami, R.~Nakayama, and N.~Sasakura,
  ``A note on string field theory in the temporal gauge,''
  \href{http://dx.doi.org/10.1143/PTPS.118.241}{{\em Prog. Theor. Phys. Suppl.}
  {\bf 118} (1995)  241--258}, \href{http://arxiv.org/abs/hep-th/9409101}{{\tt
  arXiv:hep-th/9409101}}.

\bibitem{Sen2016}
A.~Sen, ``Reality of superstring field theory action,''
  \href{http://dx.doi.org/10.1007/JHEP11(2016)014}{{\em JHEP} {\bf 11} (2016)
  014}, \href{http://arxiv.org/abs/1606.03455}{{\tt arXiv:1606.03455
  [hep-th]}}.

\bibitem{Pius2014a}
R.~Pius, A.~Rudra, and A.~Sen, ``Mass renormalization in string theory: Special
  states,'' \href{http://dx.doi.org/10.1007/JHEP07(2014)058}{{\em JHEP} {\bf
  07} (2014)  058}, \href{http://arxiv.org/abs/1311.1257}{{\tt arXiv:1311.1257
  [hep-th]}}.

\bibitem{Pius2014}
R.~Pius, A.~Rudra, and A.~Sen, ``Mass renormalization in string theory: General
  states,'' \href{http://dx.doi.org/10.1007/JHEP07(2014)062}{{\em JHEP} {\bf
  07} (2014)  062}, \href{http://arxiv.org/abs/1401.7014}{{\tt arXiv:1401.7014
  [hep-th]}}.

\bibitem{Stanford2020}
D.~Stanford and E.~Witten, ``Jt gravity and the ensembles of random matrix
  theory,'' \href{http://dx.doi.org/10.4310/ATMP.2020.v24.n6.a4}{{\em Adv.
  Theor. Math. Phys.} {\bf 24} (2020) no.~6, 1475--1680},
  \href{http://arxiv.org/abs/1907.03363}{{\tt arXiv:1907.03363 [hep-th]}}.

\bibitem{Kugo1987}
T.~Kugo, ``Covariantized light cone string field theory,'' in {\em {2nd Meeting
  on Quantum Mechanics of Fundamental Systems (CECS)}}.
\newblock 10, 1987.

\bibitem{Fukuma1991}
M.~Fukuma, H.~Kawai, and R.~Nakayama, ``Continuum schwinger-dyson equations and
  universal structures in two-dimensional quantum gravity,''
  \href{http://dx.doi.org/10.1142/S0217751X91000733}{{\em Int. J. Mod. Phys. A}
  {\bf 6} (1991)  1385--1406}.

\bibitem{Dijkgraaf1991}
R.~Dijkgraaf, H.~L. Verlinde, and E.~P. Verlinde, ``Loop equations and virasoro
  constraints in nonperturbative 2-d quantum gravity,''
  \href{http://dx.doi.org/10.1016/0550-3213(91)90199-8}{{\em Nucl. Phys. B}
  {\bf 348} (1991)  435--456}.

\bibitem{Ikehara:1995dd}
M.~Ikehara, ``{String field theories from one matrix models},''
  \href{http://dx.doi.org/10.1143/PTP.93.1141}{{\em Prog. Theor. Phys.} {\bf
  93} (1995)  1141--1144}, \href{http://arxiv.org/abs/hep-th/9504094}{{\tt
  arXiv:hep-th/9504094}}.

\bibitem{Abikoff1980}
W.~Abikoff, {\em The real analytic theory of Teichm\"{u}ller space}.
\newblock Springer, Berlin, 1980.

\end{thebibliography}\endgroup

\end{document}